\documentclass{elsart}
\usepackage{graphicx}
\usepackage{dcolumn}
\usepackage{rotate}
\usepackage{latexsym}
\usepackage{placeins}

\def\met{\mbox{${\hbox{$E$\kern-0.6em\lower-.1ex\hbox{/}}}_T$}} 
\def\lsim{\mathrel{\rlap{\lower4pt\hbox{\hskip1pt$\sim$}}
    \raise1pt\hbox{$<$}}}         
\def\gsim{\mathrel{\rlap{\lower4pt\hbox{\hskip1pt$\sim$}}
    \raise1pt\hbox{$>$}}}         
\def\pbarp{\mbox{$p\overline{p}$}}
\def\ppbar{\mbox{$p\overline{p}$}}

\def\pbar{\mbox{$\overline{p}$}}

\begin{document}
\bibliographystyle{elsart-num}

\begin{frontmatter}

\title{The Upgraded D\O\ Detector}

\collab{D\O\ Collaboration}

\begin{flushleft}
\author{ \small
V.M.~Abazov,$^{ai}$
B.~Abbott,$^{bt}$
M.~Abolins,$^{bk}$
B.S.~Acharya,$^{ac}$
D.L.~Adams,$^{br}$
}
\author{ \small
M.~Adams,$^{ax}$
T.~Adams,$^{av}$
M.~Agelou,$^{r}$
J.-L.~Agram,$^{s}$
S.N.~Ahmed,$^{ah}$
S.H.~Ahn,$^{ae}$
}
\author{ \small
M.~Ahsan,$^{be}$
G.D.~Alexeev,$^{ai}$
G.~Alkhazov,$^{am}$
A.~Alton,$^{bj}$
G.~Alverson,$^{bi}$
}
\author{ \small
G.A.~Alves,$^{b}$
M.~Anastasoaie,$^{ah}$
T.~Andeen,$^{az}$
J.T.~Anderson,$^{aw}$
S.~Anderson,$^{ar}$
}
\author{ \small
B.~Andrieu,$^{q}$
R.~Angstadt,$^{aw}$
V.~Anosov,$^{ai}$
Y.~Arnoud,$^{n}$
M.~Arov,$^{ay}$
A.~Askew,$^{av}$
}
\author{ \small
B.~{\AA}sman,$^{an}$
A.C.S.~Assis Jesus,$^{c}$
O.~Atramentov,$^{bc}$
C.~Autermann,$^{u}$
C.~Avila,$^{h}$
}
\author{ \small
L.~Babukhadia,$^{bq}$
T.C.~Bacon,$^{ap}$
F.~Badaud,$^{m}$
A.~Baden,$^{bg}$
S.~Baffioni,$^{o}$
}
\author{ \small
L.~Bagby,$^{ay}$
B.~Baldin,$^{aw}$
P.W.~Balm,$^{ag}$
P.~Banerjee,$^{ac}$
S.~Banerjee,$^{ac}$
}
\author{ \small
E.~Barberis,$^{bi}$
O.~Bardon,$^{bi}$
W.~Barg,$^{d}$
P.~Bargassa,$^{bx}$
P.~Baringer,$^{bd}$
C.~Barnes,$^{ap}$
}
\author{ \small
J.~Barreto,$^{b}$
J.F.~Bartlett,$^{aw}$
U.~Bassler,$^{q}$
M.~Bhattacharjee,$^{bq}$
M.A.~Baturitsky,$^{ai}$
}
% ...
\address{$^{h}$Universidad de los Andes, Bogot\'{a}, Colombia}             
% ...
\address{$^{m}$Laboratoire de Physique Corpusculaire, IN2P3-CNRS,         
                 Universit\'e Blaise Pascal, Clermont-Ferrand, France}        
\address{$^{n}$Laboratoire de Physique Subatomique et de Cosmologie,      
                  IN2P3-CNRS, Universite de Grenoble 1, Grenoble, France}     
\address{$^{o}$CPPM, IN2P3-CNRS, Universit\'e de la M\'editerran\'ee,     
                  Marseille, France}                                          
% ...
\address{$^{q}$LPNHE, IN2P3-CNRS, Universit\'es Paris VI and VII,         
                  Paris, France}                                              
\address{$^{r}$DAPNIA/Service de Physique des Particules, CEA, Saclay,    
                  France}                                                     
\address{$^{s}$IReS, IN2P3-CNRS, Universit\'e Louis Pasteur, Strasbourg,  
                  France and Universit\'e de Haute Alsace, Mulhouse, France}  
% ...
\address{$^{u}$III.\ Physikalisches Institut A, RWTH Aachen,               
                   Aachen, Germany}                                           
% ...
\address{$^{ac}$Tata Institute of Fundamental Research, Mumbai, India}     
% ...
\address{$^{ae}$Korea Detector Laboratory, Korea University,               
                   Seoul, Korea}                                              
% ...
\address{$^{ag}$FOM-Institute NIKHEF and University of                     
                  Amsterdam/NIKHEF, Amsterdam, The Netherlands}               
\address{$^{ah}$Radboud University Nijmegen/NIKHEF, Nijmegen, The               
                  Netherlands}                                                
\address{$^{ai}$Joint Institute for Nuclear Research, Dubna, Russia}       
% ...
\address{$^{am}$Petersburg Nuclear Physics Institute,                      
                   St. Petersburg, Russia}                                    
\address{$^{an}$Lund University, Lund, Sweden, Royal Institute of          
                   Technology and Stockholm University, Stockholm,            
                   Sweden and}                                                
% ...
\address{$^{ap}$Imperial College, London, United Kingdom}                  
% ...
\address{$^{ar}$University of Arizona, Tucson, Arizona 85721, USA}         
% ...
\address{$^{av}$Florida State University, Tallahassee, Florida 32306, USA} 
\address{$^{aw}$Fermi National Accelerator Laboratory, Batavia,            
                   Illinois 60510, USA}                                       
\address{$^{ax}$University of Illinois at Chicago, Chicago,                
                   Illinois 60607, USA}                                       
\address{$^{ay}$Northern Illinois University, DeKalb, Illinois 60115, USA} 
\address{$^{az}$Northwestern University, Evanston, Illinois 60208, USA}    
% ...
\address{$^{bc}$Iowa State University, Ames, Iowa 50011, USA}              
\address{$^{bd}$University of Kansas, Lawrence, Kansas 66045, USA}         
\address{$^{be}$Kansas State University, Manhattan, Kansas 66506, USA}     
% ...
\address{$^{bg}$University of Maryland, College Park, Maryland 20742, USA} 
% ...
\address{$^{bi}$Northeastern University, Boston, Massachusetts 02115, USA} 
\address{$^{bj}$University of Michigan, Ann Arbor, Michigan 48109, USA}    
\address{$^{bk}$Michigan State University, East Lansing, Michigan 48824,   
                   USA}                                                       
% ...
\address{$^{bq}$State University of New York, Stony Brook,                 
                   New York 11794, USA}                                       
\address{$^{br}$Brookhaven National Laboratory, Upton, New York 11973, USA}
% ...
\address{$^{bt}$University of Oklahoma, Norman, Oklahoma 73019, USA}       
% ...
\address{$^{bx}$Rice University, Houston, Texas 77005, USA}                
% ...
\end{flushleft}

\begin{abstract}

The D\O\ experiment enjoyed a very successful data-collection run at the
Fermilab Tevatron collider between 1992 and 1996.  Since then, the 
detector has been upgraded to take advantage of improvements to the Tevatron 
and to enhance its physics capabilities.  We describe the new elements of 
the detector, including the silicon 
microstrip tracker, central fiber tracker, solenoidal magnet, preshower 
detectors, forward muon detector, and forward proton detector.  
The uranium/liquid-argon calorimeters and central muon detector,
remaining from Run~I, are discussed briefly.  We also present 
the associated electronics, triggering, and data acquisition systems,  
along with the design and implementation of software specific to D\O.

\end{abstract}


\end{frontmatter}

\FloatBarrier

\section{Introduction}
\label{sec:intro}

The D\O\ experiment was proposed in 1983 to study proton-antiproton collisions
at a center-of-mass energy of 1.8~TeV at the Fermilab Tevatron collider.  
The focus of the
experiment was the study of high mass states and large $p_T$ phenomena.  The 
detector performed very well during Run I of the Tevatron, 1992--1996, 
leading to the discovery of the top quark \cite{top_discovery} and measurement
of its mass \cite{top_mass1,top_mass2,top_mass3,top_mass4,top_mass5}, a 
precision measurement of the mass of the $W$ boson
\cite{W_mass1,W_mass2,W_mass3,W_mass4,W_mass5,W_mass6,W_mass7}, 
detailed analysis
of gauge boson couplings \cite{gc3,gc4,gc5,gc6,gc7,gc8,gc9,gc10,gc11}, studies
of jet production \cite{jet1,jet2,jet3,jet4}, 
and greatly improved limits on
the production of new phenomena such as leptoquarks
\cite{lq1,lq2,lq3,lq4,lq5,lq6,lq7} and 
supersymmetric particles \cite{susy1,susy2,susy3,susy4,susy5,susy6,susy7,susy8},
among many other accomplishments
\cite{d0_pub_list}.

During Run~I, the Tevatron operated using six bunches each of protons and 
antiprotons, with 3500~ns between bunch crossings and a center-of-mass energy 
of 1.8~TeV.  The peak luminosity was typically  
1--2$\times 10^{31}\ {\rm cm^{-2} s^{-1}}$ and approximately 120~pb$^{-1}$ 
of data were recorded by D\O.  Following the completion of the new 
Main Injector and associated Tevatron upgrades \cite{tev1,tev2}, 
the collider began running again in 2001.  In Run~II, which began in March 
2001, the Tevatron is operated with 36 bunches of protons 
and antiprotons with a bunch spacing of 396~ns and at an increased 
center-of-mass energy of 1.96~TeV.  The instantaneous luminosity will increase 
by more than a factor of ten to greater than $10^{32}\ {\rm cm^{-2} s^{-1}}$ 
and more than 4~fb$^{-1}$ of data are expected to be recorded. 

\begin{figure}
\centerline{\includegraphics[width=6.in]{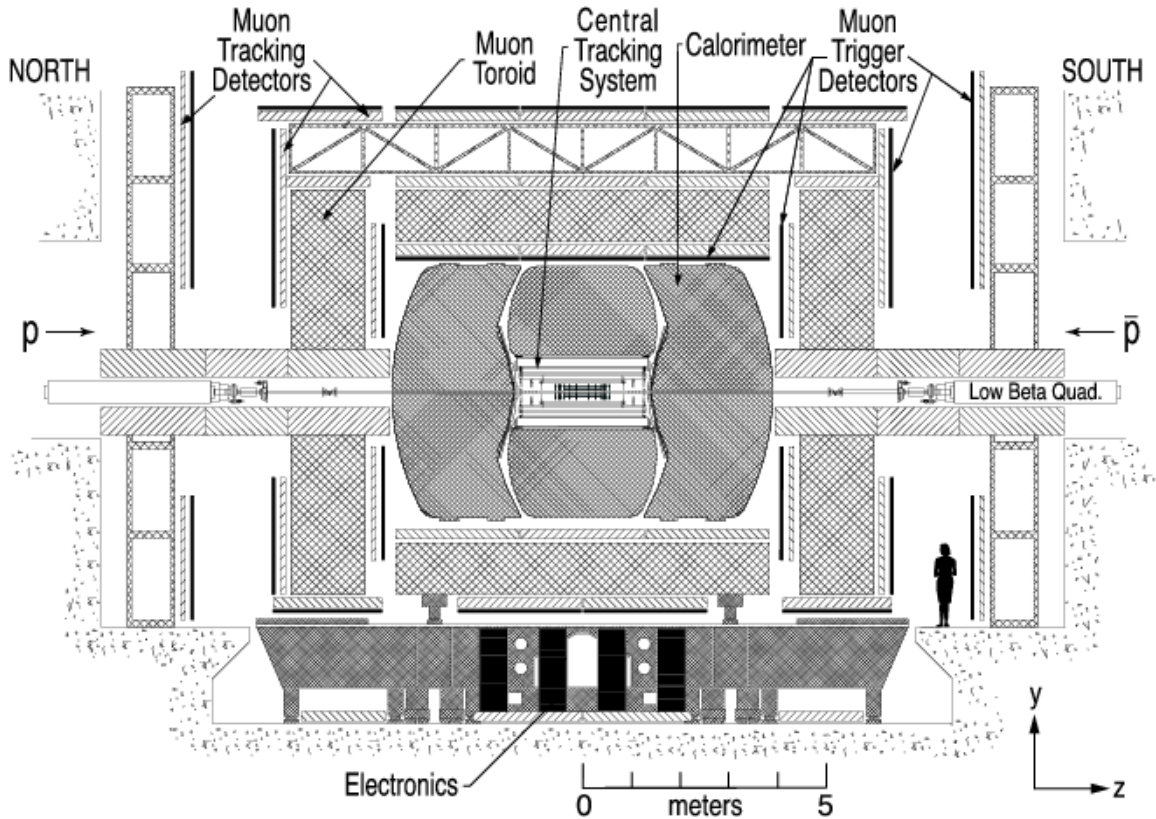}}
\caption{Diagram of the upgraded D\O\ detector, as installed in the collision
hall and viewed from inside the Tevatron ring. The forward proton detector is
not shown.  The detectors in the central region of the detector are shown in 
Fig.~\ref{fig:tracker}.}
\label{fig:detector}
\end{figure}

To take advantage of these improvements in the Tevatron and to enhance the
physics reach of the experiment, we have significantly upgraded the D\O\ 
detector.  The detector consists of three major subsystems: central tracking
detectors, uranium/liquid-argon calorimeters, and a muon spectrometer.  
The original D\O\ detector is described in detail 
in Ref.~\cite{d0_nim}.  The central tracking system has been completely 
replaced.  The old system lacked a magnetic field and suffered from radiation 
damage, and improved tracking 
technologies are now available.  The new system includes a silicon microstrip 
tracker and a scintillating-fiber tracker located within a 2 T solenoidal 
magnet.  The silicon microstrip tracker is able to identify displaced vertices 
for $b$-quark tagging.  The magnetic field enables measurement of the 
energy to momentum ratio ($E/p$) 
for electron identification and calorimeter calibration,
opens new capabilities for tau lepton identification and hadron spectroscopy, 
and allows precision muon
momentum measurement.  Between the solenoidal magnet and the central 
calorimeter and in front of the forward calorimeters, preshower detectors have 
been added for improved electron identification.  In the forward muon system,
proportional drift chambers have been replaced by mini drift tubes and trigger 
scintillation counters that can withstand the 
harsh radiation environment and additional shielding has been added.  In the
central region, scintillation counters have been added for improved muon
triggering. 
We have also added a forward proton detector for the study of diffractive
physics.  A side view of the upgraded D\O\ detector is shown in 
Figure~\ref{fig:detector}. 

The large reduction in the bunch spacing required the improvement of the 
read-out electronics and the implementation of pipelining for the front-end
signals from the tracking, calorimeter, and muon systems.  The calorimeter
preamplifiers and signal-shaping electronics have been replaced, as have
all of the electronics for the muon system.  The trigger system has been
significantly upgraded, providing three full trigger levels 
to cope with the higher collision rate and new hardware to identify displaced 
secondary vertices for $b$-quark tagging.  
Muon triggering has been enhanced by the addition of scintillation counters in
the central and forward regions.

A significant improvement to the detector's performance 
resulted from the removal of the old 
Main Ring beam pipe from the calorimeters.  During Run~I, the Main Ring
was used to accelerate protons for antiproton production while the Tevatron
operated in collider mode.  Losses from the Main Ring produced spurious
energy deposits in the calorimeters and muon system, and most triggers were 
not accepted while Main Ring protons passed through the detector.  Removal of 
the Main Ring increased the livetime of the detector by approximately 10\%, 
depending on the trigger. 

In the following sections of this paper, we describe the design and performance
of the upgraded D\O\ detector.  The new central tracking system and 
solenoidal magnet are presented in Sections~\ref{sec:central_tracking} and 
\ref{sec:solenoid}, respectively.  The preshower detectors are described
in Section~\ref{sec:preshower}.  The calorimeters are briefly described in
Section~\ref{sec:calorimeters} along with the new calorimeter electronics.
The muon system is discussed in Section~\ref{sec:muon}.  The new forward
proton detector is presented in Section~\ref{sec:fpd}.  The luminosity monitor
is described in Section~\ref{sec:lum-monitor}.  The triggering
and data acquisition systems are described in Sections~\ref{sec:trigger} and
\ref{sec:daq}.  Section~\ref{sec:controls} covers detector 
controls and monitoring and Section~\ref{sec:software} contains an overview 
of the software components of the experiment.

In the detector description and data analysis, we use a right-handed coordinate 
system in which the $z$-axis is along the 
proton direction and the $y$-axis is upward 
(Figure~\ref{fig:detector}).  The angles $\phi$ and $\theta$ 
are the azimuthal and polar angles, respectively.  The $r$ coordinate denotes 
the perpendicular 
distance from the $z$ axis.  The pseudorapidity, $\eta = -\ln[\tan(\theta/2)]$,
approximates the true rapidity, $y=1/2 \ln[(E+p_zc)/(E-p_zc)]$, for finite 
angles in the limit that $(mc^2/E) \to 0$.   We use the term ``forward'' to
describe the regions at large $|\eta|$. 

\FloatBarrier

\section{Central tracking}
\label{sec:central_tracking}

Excellent tracking in the central region is necessary for studies of  
top quark, electroweak, and $b$  physics and to search for new phenomena,
including the Higgs boson.  The central tracking system consists of the silicon
microstrip tracker (SMT) and the central fiber tracker (CFT) surrounded by a 
solenoidal magnet.  It surrounds the D\O\ beryllium beam pipe, which has a wall 
thickness of 0.508~mm and an outer diameter of 38.1~mm, and is 2.37~m long.  
The two tracking detectors locate the primary interaction 
vertex with a resolution of about 35~$\mu$m along the beamline.  They can tag 
$b$-quark jets with an impact parameter resolution of better than 15~$\mu$m 
in $r-\phi$ for particles with transverse momentum $p_T > 10$~GeV/$c$ at 
$|\eta| = 0$.  The high resolution of the
vertex position allows good measurement of lepton $p_T$, jet transverse energy 
($E_T$), and missing transverse energy \met.  Calibration of the 
electromagnetic calorimeter using $E/p$ for electrons is now possible.   
 
Both the SMT and CFT provide tracking information to the trigger.  The SMT 
provides signals to the Level~2 and 3 trigger systems and 
is used to trigger on displaced vertices from $b$-quark decay.
The CFT provides a fast and continuous readout of discriminator signals to the 
Level~1 trigger system; upon a Level~1 trigger accept, track information based
on these signals is sent to Level~2.  The Level~3 trigger receives 
a slower readout of the CFT's digitized analog signals, in addition to the
discriminator information available at Level~1 and Level~2.
 
\begin{figure}
\centerline{\includegraphics[width=6.in]{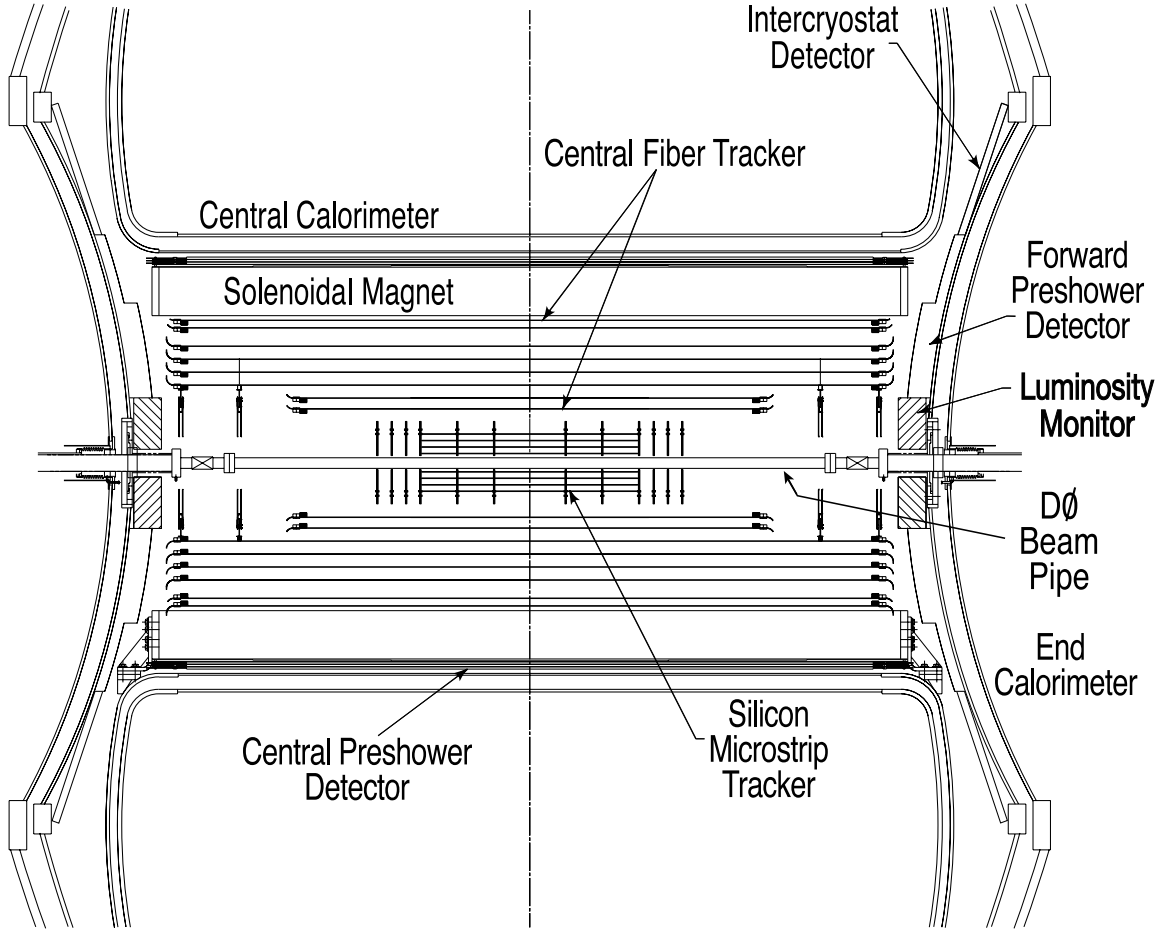}}
\caption{Cross-sectional view of the new central tracking system in the $x-z$
plane.  Also shown are the locations of the solenoid, the preshower detectors, 
luminosity monitor, and the calorimeters.}
\label{fig:tracker}
\end{figure}

A schematic view of the central tracking system is shown in
Figure~\ref{fig:tracker}.  

\subsection{Silicon microstrip tracker}
\label{sec:smt}

The SMT provides both tracking and vertexing over 
nearly the full $\eta$ coverage of the calorimeter and muon 
systems.  Design of the detector, electronics, and cooling are, in large 
part, dictated by the accelerator environment.  The length of the interaction 
region
($\sigma \approx 25$~cm) sets the length scale of the device.  With 
a long interaction region, it is a challenge to deploy detectors such that 
the tracks are generally perpendicular to detector surfaces for all $\eta $.  
This led us to a design of barrel modules interspersed with disks in the 
center and assemblies of disks in the forward regions.  
The barrel detectors primarily measure the $r-\phi$ coordinate and the disk 
detectors measure $r-z$ as well as $r-\phi $.  Thus vertices for particles at 
high $\eta$ are reconstructed in three dimensions by 
the disks, and vertices of particles at small values of  $\eta$ are measured 
in the barrels and central fiber tracker.  This design poses difficult 
mechanical challenges in arranging the detector components and minimizing
dead areas while providing sufficient space for cooling and cables.

\begin{figure}
\centerline{\includegraphics[width=6.in]{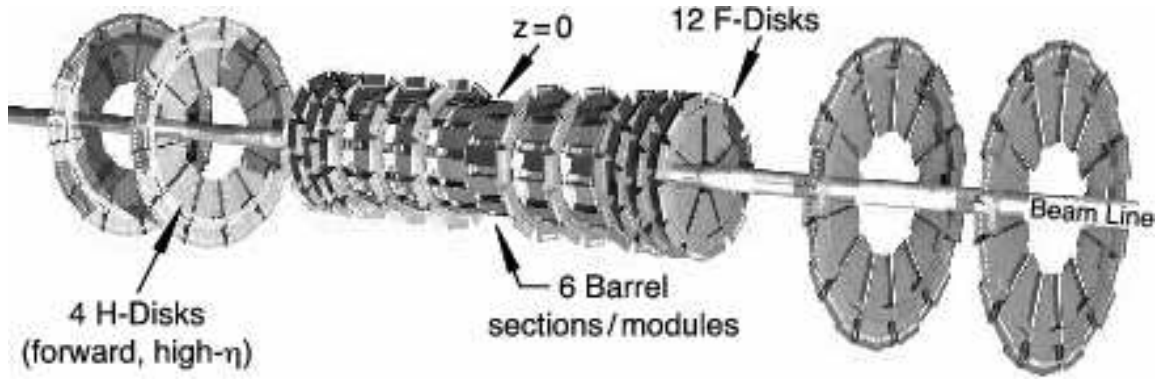}}
\caption{The disk/barrel design of the silicon microstrip tracker.} 
\label{fig:smt}
\end{figure}

An isometric view of the SMT is shown in Figure~\ref{fig:smt}.
The detector has six barrels in the central region.  Each barrel has four 
silicon readout layers.  The silicon modules installed in the barrels are 
called ``ladders.'' 
Layers 1 and 2 have twelve ladders each; layers 3 and 4 have twenty-four 
ladders each, for a total of 432 ladders.
Each barrel is capped at high $\vert z\vert$ with a disk of twelve double-sided 
wedge detectors, called an ``F-disk.'' Forward of the three disk/barrel
assemblies on each side is 
a unit consisting of three F-disks.  In the far forward regions, 
two large-diameter disks, ``H-disks,'' provide tracking at high 
$\vert \eta \vert$.  Twenty-four full wedges, each consisting of two 
back-to-back single-sided ``half'' wedges, are mounted on each H-disk.
There are 144 F-wedges and 96 full H-wedges in the tracker; each side of a 
wedge (upstream and downstream) is read out independently. 
There is a grand total of 912 readout modules, with 792,576 channels.  The
centers of the H-disks are located at $\vert z\vert = 100.4,~121.0$~cm; 
the F-disks are at $\vert z\vert = 12.5,~25.3,~38.2,~43.1,~48.1,$ 
and $53.1$~cm.  The centers of the barrels are at $\vert z\vert =
6.2,~19.0,~31.8$~cm.
The SMT is read out by custom-made 128-channel SVXIIe readout chips.

\subsubsection{Sensors}
\paragraph{Sensor types}
The SMT uses a combination of single-sided (SS), double-sided (DS), and 
double-sided double-metal (DSDM) technologies.  Silicon sensors were obtained 
from three manufacturers.  All barrel sensors were produced by Micron 
Semiconductor LTD~\cite{Micron}.  The wedges for the F-disks were produced by 
Micron Semiconductor LTD and Canberra Eurisys Mesures~\cite{EurisysMesures}.  
The wedge sensors for the H-disks were manufactured by ELMA~\cite{elma}; these 
sensors use intermediate strips for 
charge interpolation.  Single-sided and double-sided devices were 
produced from high resistivity 4'' silicon wafers, with crystal orientation 
$<$111$>$ and $<$100$>$.  The 90$^\circ$ stereo sensors used in layers 1 and 3 
of the four centermost barrels are DSDM sensors, 
manufactured using $<$100$>$ 6'' wafers.  Isolation on the n-side of all 
double-sided sensors is provided by p-stop implants.  All traces are biased 
using polysilicon resistors.  Table~\ref{tab:silicon} shows the sensor types 
used in the SMT and their locations.

Disk sensors are trapezoids with readout strips arranged parallel to the 
long edge of the devices.  This provides an effective 30$^\circ$ stereo angle 
for the double-sided F-disks.  A wedge for the H-disks consists of a pair of 
single-sided half-wedges mounted back-to-back, giving an effective stereo 
angle of 15$^\circ$.  This arrangement means that the strip length varies for 
strips which originate past the base of the trapezoid.  There are three types 
of sensors in the central barrels.  The second and fourth layers use 
double-sided stereo sensors with the n-side implants at a 2$^\circ$ angle with 
respect to the p-side axial strips.  Two of these sensors are bonded together 
to form a 12-cm-long ladder.  The first and third layers of the outer barrels 
use single-sided sensors with axial strips and, again, two sensors are bonded
together to make one 12-cm ladder.  The inner four barrels use single 
12-cm-long 90$^\circ$ stereo sensors.  Ninety degree readout was achieved by 
using a second metal layer on the n-side insulated from the first metal by 
3~$\mu$m of PECVD (plasma enhanced chemical vapor deposition) silicon oxide.  
Two readout strips on this side are multiplexed to a single readout channel.  
Implants on the n-side are isolated by individual p-stop frames in addition 
to a common p-stop enclosure.

\begin{table}
\begin{center}
\caption{Characteristics and deployment of various sensor types in the SMT. 
{\it i} indicates the length of the inner H-disk sensor; 
{\it o} is the length of the outer H-disk sensor.}
\label{tab:silicon}
\begin{tabular}{ l c c p{55pt} p{36pt} p{36pt} p{36pt}}
\hline
Module & Type & Layer & Pitch ($\mu$m) \par p/n & 
Length \par (cm) & Inner \par radius (cm) & Outer \par radius (cm) \\
\hline
F-disks & DS & -- & 50/62.5 & 7.93 & 2.57 & 9.96  \\
\hline
H-disks & SS & -- & 40 \par 80 readout 
         & 7.63$^i$ \par 6.33$^o$  & 9.5 & 26 \\
\hline
\raisebox{-1.ex}[0pt]{Central}  
                           & DSDM & 1, 3 & 50/153.5 & 12.0 & 2.715 & 7.582 \\
\cline{2-7} 
\raisebox{1.ex}[0pt]{barrels (4)} 
        & DS & 2, 4 & 50/62.5 & 6.0 & 4.55 & 10.51  \\
\hline
\raisebox{-1.ex}[0pt]{Outer} & SS & 1, 3 & 50 & 6.0 & 2.715 & 7.582 \\
\cline{2-7} 
\raisebox{1.ex}[0pt]{barrels (2)} 
             & DS & 2, 4 & 50/62.5 & 6.0 & 4.55 & 10.51  \\
\hline
\end{tabular}
\end{center}
\end{table}

\paragraph{Sensor testing}
All sensors were required to exhibit leakage current below 260~nA/cm$^2$, 
and each channel was tested for leakage current, AC coupling 
capacitance, and AC coupling leakage to 80~V.  The AC coupling leakage test is 
especially important for the double-sided sensors since these capacitors are 
required to stand off the depletion voltage.  Several classes of problems 
were identified during the testing:

\begin{itemize}
\item Areas of low interstrip resistance ---  Washing bad sensors in pure 
water at the vendor sometimes cured this problem.
\item Isolation-implant shorts ---  These occur in all sensor types but the 
effects are particularly severe in the DSDM sensors where a single isolation 
short can affect a region of approximately twenty channels.  Strip-by-strip DC 
scans were performed at the manufacturer to identify problems early in 
processing.
\item Poorly controlled bias resistance ---  Sensors with bias resistors as 
high as 10~M$\Omega$ ($5\times$ nominal) were eventually accepted.
\item Microdischarge breakdown --- This breakdown was observed with modest 
negative voltage applied to the p-side of double-sided sensors.  The onset 
voltage of breakdown was measured for each sensor and used to 
set the split between p($-$)- and n($+$)-side bias voltage.
\end{itemize}

\begin{figure}
\centerline{\includegraphics[width=3.in]{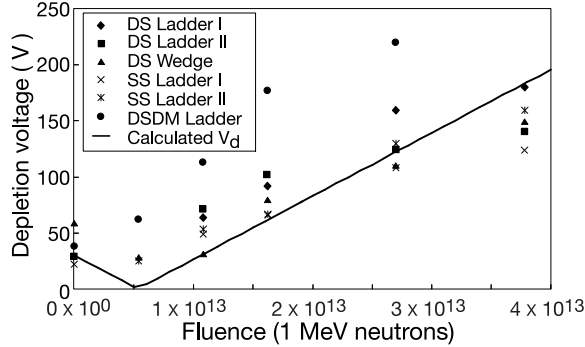}}
\caption{Depletion voltage as a function of fluence as measured 
for SMT modules in the Fermilab Booster.} 
\label{fig:vdepl}
\end{figure}

\paragraph{Radiation damage studies}
A set of radiation damage qualification studies was performed at the 
Fermilab Booster and the Radiation Laboratory at the University of 
Massachusetts  Lowell.  In general, the results of these studies conformed to 
behavior expected from RD48~\cite{RD48} parameterizations.  The exception was 
the depletion voltage ($V_d$) behavior of the DSDM devices, 
shown in Figure~\ref{fig:vdepl}.  These sensors exhibit a rise in depletion 
voltage almost twice as fast as that of the single-sided sensors.  To 
determine if the anomalous behavior is due to the bulk silicon properties, 
photodiode test structures from the same wafer were irradiated in the booster.  
These all showed normal behavior.  It is likely that the rapid rise in 
depletion voltage is related to the PECVD inter-metal isolation layer.

Microdischarge breakdown is caused by large fields near the junction 
implant.  This high-field region is expected to move with the junction from 
the p-side to the n-side upon type inversion.  In addition, fixed positive 
charge in the oxide insulation layers will tend to increase the field on the 
p-side and reduce it on the n-side.  This is indeed what we observe.  Breakdown 
moves from the p-side to the n-side after type inversion and the threshold 
increases to approximately 150~V after about 2~Mrad.  Microdischarge in the 
DSDM sensors will limit the lifetime of the SMT to an integrated luminosity 
of 3.5--6 fb$^{-1}$.

\subsubsection{Mechanical design}
\paragraph{Ladders}
Sensor ladder assemblies were designed for low mass, precise alignment, 
and good thermal performance.  With the exception of the DSDM ladders, which 
use a single 12-cm sensor, ladders were constructed using two 6-cm 
sensors.  All ladder assembly was done under the control of a coordinate 
measuring machine (CMM).  The two sensors were aligned using the fiducials on 
the sensors.  Alignment was achieved during assembly using locating notches in 
precisely machined beryllium pieces on both the ``active'' (SVXIIe-readout-chip 
carrying, see Section~\ref{sec:SVXII}) and ``passive'' sides of the ladder.  
These notches correspond to posts on the support bulkheads and are aligned to 
fiducials on the sensor.  Carbon-boron fiber/Rohacell rails bridge the 
beryllium pieces and maintain the flatness of the ladder assemblies.

\begin{figure}
\centerline{\includegraphics[width=4.in]{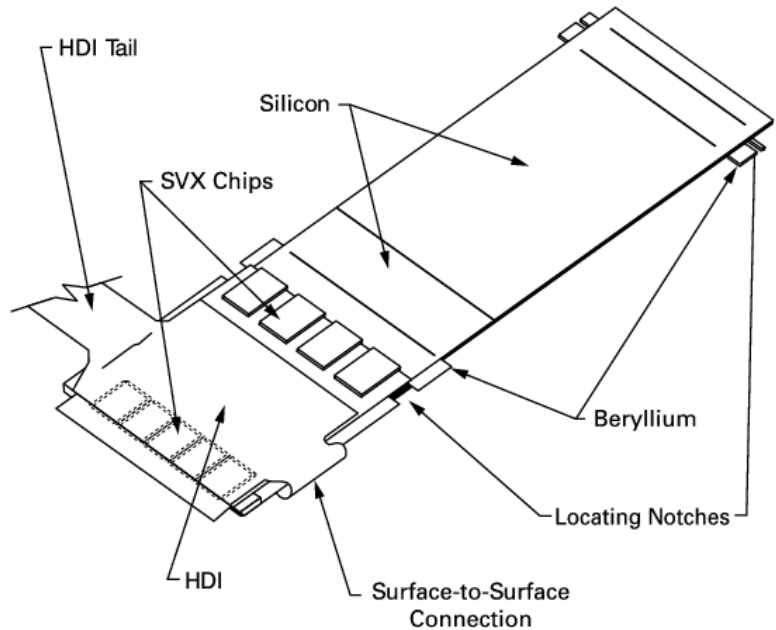}}
\caption{Double-sided ladder design, n-side.  The SVXIIe readout chips shown 
as dashed lines are located on the p-side of the ladder.} 
\label{fig:ldr_design}
\end{figure}

\begin{figure}
\centerline{\includegraphics[width=4.in]{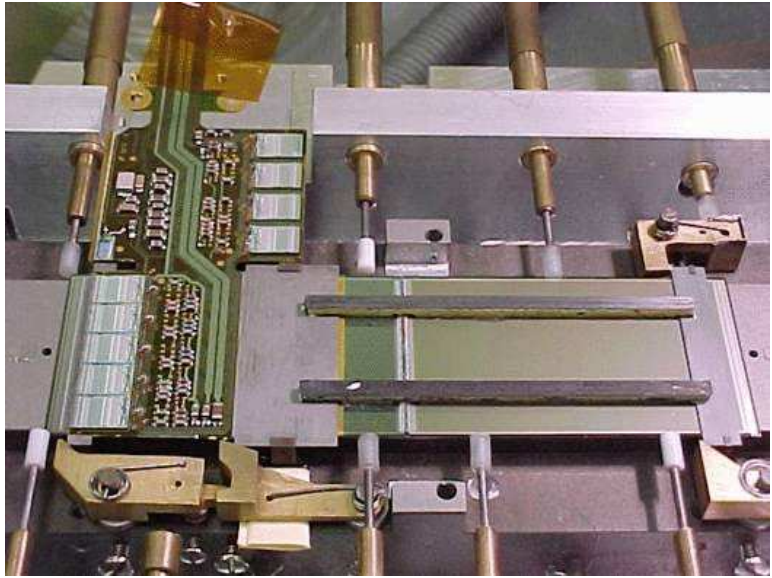}}
\caption{Double-sided ladder during assembly with the flex hybrid unfolded.  
The spring-loaded ``pushers'' and support rails are also visible. } 
\label{fig:ldr_assy}
\end{figure}

A sketch of a double-sided 2$^\circ$ ladder with nine SVXIIe readout chips 
is shown in Figure~\ref{fig:ldr_design}.  Figure~\ref{fig:ldr_assy} shows 
a similar ladder on an assembly fixture with the high density interconnect 
(HDI, Section~\ref{sec:SVXII}) unfolded, spring-loaded 
``pushers,'' and support rails.  Each ladder and wedge was surveyed using an 
automatic optical CMM after assembly.

\paragraph{Wedges}
In contrast to the barrel assemblies, F- and H-disks are planar 
modules, allowing for simple optical alignment during assembly.  A major 
design constraint was to provide minimum dead space between the disk/barrel 
assemblies.  F-wedges use double-sided sensors with separate 8- and 6-chip 
HDIs for each side.  The larger strip 
pitch (62.5~$\mu$m vs 50~$\mu$m) on the n-side requires an adapter to match 
the SVXIIe readout pitch.  The 50~$\mu$m-thick copper-clad Kapton~\cite{kapton}
pitch adaptors are part of the cooling path to the p-side 
and allow the six n-side SVXIIe chips to be located outboard of the cooling 
channel.  The double-sided hybrid assembly required a complex 12-layer vacuum 
lamination with pitch adaptor, Tedlar~\cite{tedlar}, Ablefilm
adhesive~\cite{ablefilm}, Kapton~\cite{kapton}, beryllium, and HDI layers.  

H-disk wedges are made from back-to-back single-sided sensors.  In 
this case, the precise front-to-back alignment was provided in the wedge 
assembly process.  A special dual-camera assembly fixture was developed to 
simultaneously image fiducials on the top and bottom sensors.  This 
allowed front-to-back alignment of the two-sensor wedges of better than 
15~$\mu$m.  An aluminized silicon pitch adapter, fabricated on the same wafer 
as the sensor, was used to match the 80~$\mu$m readout pitch to the SVXIIe
input.

\paragraph{Supports and assembly}
Barrel ladders are supported by beryllium bulkheads machined with posts and 
pinholes for ladder support.  Each barrel has a thick ``active'' bulkhead 
containing the cooling channels and connections to the outer support 
cylinder, and a thin ``passive'' bulkhead that serves to set the spacing 
of the ends of the ladders without readout chips.  Ladders were lowered onto 
the posts using a special insertion fixture and fixed into place with pins.  
Thermal contact to the cooled bulkhead edge is made with thermal grease.  Each 
ladder is grounded to the bulkheads with conducting epoxy.  Edges of the 
beryllium pieces were measured with the touch probe of the assembly CMM, and 
these measurements were correlated to the beryllium and sensor fiducial 
mark measurements taken with the optical CMM to provide the final ladder 
position.  Cables (HDI ``tails'') are routed between barrel sublayers inboard 
of ladders, so no inter-module space is taken by the HDI tails.  The tails 
are coupled to ``card edge'' style Hirose connectors
\cite{hirose} on the low-mass, flexible, Kapton cables on the outer surface of 
the support structure.

Disks are supported by beryllium rings.  Wedges are located on alternate sides 
of the ring with sufficient overlap to eliminate dead regions.  Wedges were 
manually aligned under a CMM and secured with  screws.  Finished F-disks were 
then assembled into a disk/barrel central module or one of the three-disk
modules at the end of the central disk/barrel section. 

\begin{figure}
\centerline{\includegraphics[width=3.in]{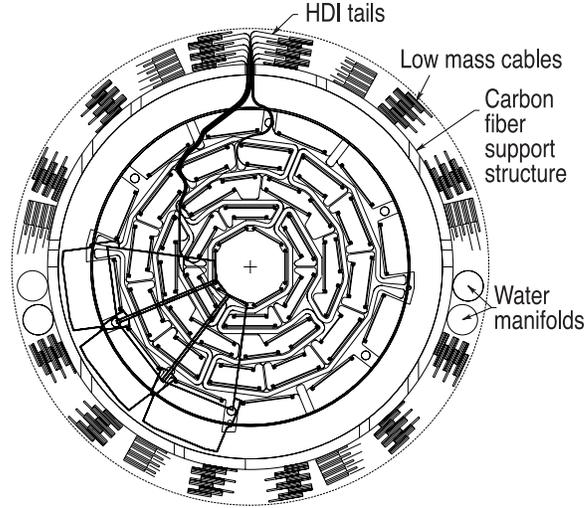}}
\caption{Cross section of the SMT disk/barrel module showing 
ladders mounted on the beryllium bulkhead, sample cable paths, three of twelve 
F-disk wedges, carbon fiber support structure, and the low-mass cable stack.} 
\label{fig:barrel_disk}
\end{figure}

Overall support of the SMT (exclusive of the H-disks) is provided by two 
double-walled carbon fiber cylinders spaced by carbon fiber ribs 
to eliminate differential contraction.  North and south half-cylinders are 
independent structures.  This limits the size of the units, allowing
installation of the SMT in the limited space available in the 
collision hall.  The central upper section of each half-cylinder is removed 
for placement of the disk/barrel modules.  Each module is supported by 
adjustable kinematic mounts.  Cables and services are accessed through holes 
in the cylinder whose outer surface is used for routing the low-mass cables 
and water manifolds.  Final alignment is provided by sapphire balls mounted 
on the bulkheads, which are accessed with touch probes through additional 
holes in the support cylinder.  The disk/barrel half cylinders are supported 
from the inner central fiber tracker barrel using mounts glued into place.  
Figure~\ref{fig:barrel_disk} shows the disk/barrel module within its support 
cylinder.  H-disks are located on separate mounts suspended from the third
layer of the CFT.

\paragraph{Cooling}
The maximum operating temperature of the detector is limited to reduce bulk 
damage to the silicon, which leads to type inversion and increasing 
depletion voltage.  Heat generated by 
the SVXIIe chips must flow through the HDI, epoxy, 
beryllium heat spreaders, and, for the top-side chips, through the silicon 
and the bottom-side readout structure.  

Coolant is a water/30\% ethylene glycol mixture supplied at $-10^{\circ}$~C 
by two redundant 4.4 kW chillers.  The coolant flows through channels 
machined in the active beryllium bulkheads and ring supports and is maintained 
below atmospheric pressure to minimize the effect of any possible leaks.  
Pressure drop across detector elements is below 5~psi with a temperature 
rise in the channel of 1--1.5$^{\circ}$~C.  A manifold is used for each side of 
the half-cylinder with flows through the modules controlled by restricting 
apertures.  The hottest point of the silicon at 
the tip of the ladder is typically 10--15$^\circ$~C above the coolant 
temperature, maintaining the silicon temperature below 5$^\circ$~C.

\subsubsection{Electronics}
\label{sec:SVXII}
\paragraph{SVXIIe chip and HDIs}
The SMT is read out using the 128-channel SVXIIe chip~\cite{SVXIIe}.  The chip 
includes preamp, analog delay, digitization, and data sparsification.  Input 
charge is integrated on the preamplifier for a train of beam crossings 
(typically twelve) and reset during inter-bunch gaps.  This charge is 
delivered to a 32-cell analog pipeline.  Upon a Level~1 trigger accept, 
double-correlated sampling is performed on the appropriate 
cells\footnote{Double correlated sampling is the process of subtracting the 
analog baseline pedestal value from the value being measured.} and this 
analog information is fed to a parallel set of Wilkinson ADCs.  Digitization 
utilizes both edges of the 53~MHz main accelerator clock, providing 8~bits of 
analog information in 2.4~$\mu$s.  Readout is half as fast.  Typical noise 
performance for a rise time setting of 100~ns is $490e + 50e$/pF.  Fabrication 
was done using the 1.2~$\mu$m UTMC radiation-hard process \cite{UTMC}
with yields of approximately 60\%.  

To make the HDIs, chips were mounted on Kapton flex circuits laminated to 
beryllium heat spreaders.  Eight different types of HDI are necessary to 
accommodate the various detector and readout geometries.  A readout cable 
``tail'' is part of each HDI.  In the case of double-sided ladders, a 
single HDI contains the readout for both the p- and n-sides with the 
Kapton folded over to sandwich the ladder.  The flex circuits are two-layer 
50~$\mu$m Kapton with 125~$\mu$m line spacing and 50~$\mu$m plated-through 
holes.

\paragraph{Readout}

\begin{figure}
\centerline{\includegraphics[width=6.in]{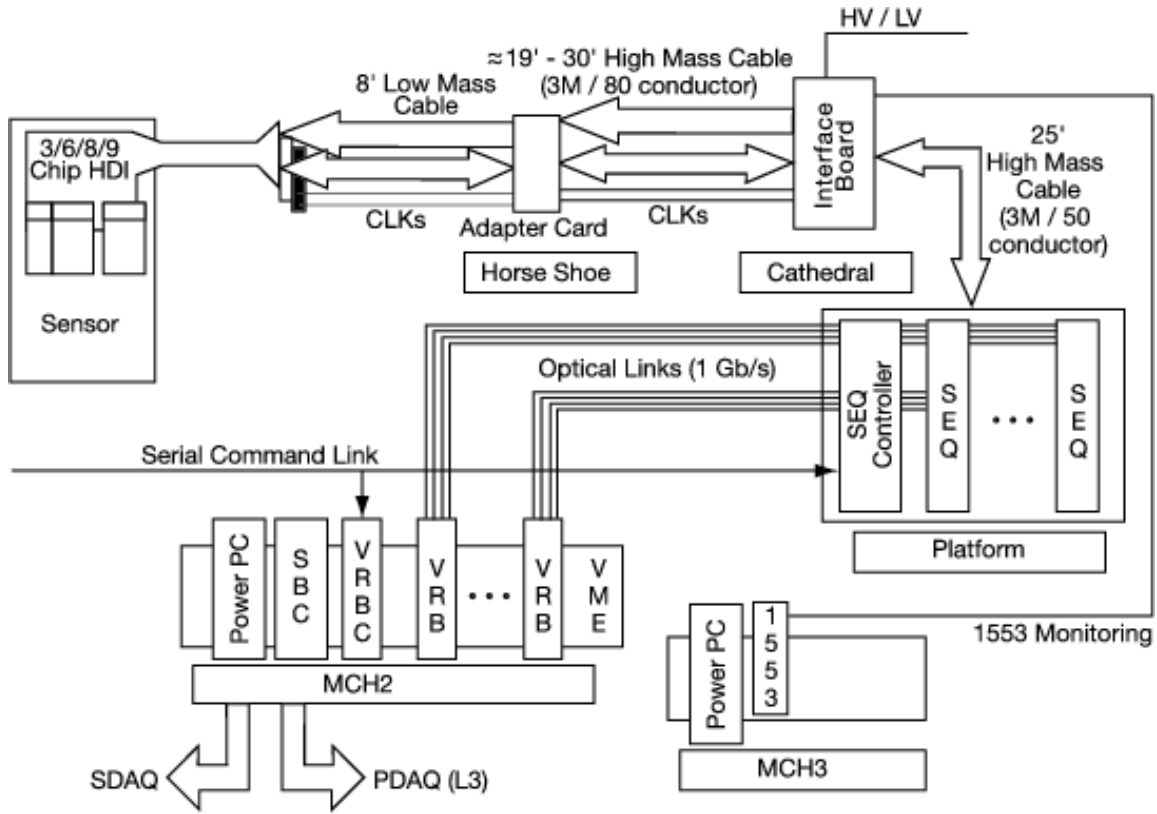}}
\caption{Diagram of the major components of the SMT readout 
system.  The data acquisition (DAQ) system is discussed in detail in
Section~\ref{sec:daq}.  The horse shoe, cathedral, platform, MCH2, and MCH3 are
places where various pieces of electronics are located.} 
\label{fig:readout}
\end{figure}

Figure~\ref{fig:readout} shows the SMT readout chain.  Trigger information is 
received via the serial command link (SCL) by the sequencer crate 
controller.  The SVXIIe sequencer provides timing and control signals for the 
SVXIIe chips on eight HDIs.  These signals are regenerated by interface boards 
that also control power and bias for the SVXIIe chips and provide interfaces 
to the monitoring systems and individual HDI temperature and current 
trips.

Data from the ladders and wedges are sent from the sequencers to VME readout 
buffer (VRB) memories via optical link fibers.  The VRBC (VRB controller) 
receives trigger data from the SCL and uses that information to control the 
operation of the VRBs.  Two single-board computers are resident in each 
readout crate.  The primary purpose of the Power PC is slow monitoring and 
calibration.  The second computer (SBC)
collects data from the VRBs upon Level~2 accepts and then sends the data 
to Level~3 for readout.  Downloads and slow controls are provided by a 
MIL-STD-1553B \cite{mil-1553} control system.  

\paragraph{Cabling}
Clock, power, and signal quality and timing are critical to proper operation 
of the SVXIIe chip.  The SMT is read out using low-mass, flexible, Kapton 
cables within the detector volume followed by high-mass 80- and 50-conductor 
``pleated-foil'' cables on the outside.  The cables carry both power (except 
the 50-conductor high-mass cables)
and digital signals.  A pair of coaxial cables carries the differential clock.  
Low-mass cables are routed along the half cylinder and 
coupled to the 80-conductor pleated-foil cables at a ring of adapter cards 
mounted on the end of the central calorimeter cryostat.

\subsubsection{Production and testing}
Production of ladders and wedges and assembly into barrels and disks were 
performed at the Silicon Detector Facility (SiDet) at Fermilab; the work took
slightly more than one year.  Components 
such as chips, HDIs, and cables were tested before shipment to Fermilab.  A 
number of sensor- and HDI-related problems were identified that necessitated 
extensive quality control testing.  Tests included:

\begin{itemize}
\item HDI burn-in --- Each HDI was operated at room temperature for 24 hours.  
Current, pedestal, and gain were monitored.  
\item Ladder/wedge debugging --- Pedestal and noise were measured for each 
assembled module.  Pinholes (shorted AC coupling capacitors) generated in the 
wirebonding process were identified in 0.8\% of the channels and the 
associated wirebonds pulled.  The onset voltage for microdischarge was 
determined on the n-side of double-sided sensors.  Resistance between the 
HDI ground and beryllium was measured.  Grounds were remade if the resistance 
was greater than 10~$\Omega$.
\item Ladder/wedge burn-in --- Assembled and tested ladders and wedges were 
operated at $5^{\circ}$~C for 72 hours.  Currents, pedestals, gain, data 
integrity, and noise were monitored.  
\item Laser test --- Modules were tested using a 1064~nm laser on a 
computer-controlled $x-y$ table.  This test allowed us to measure depletion 
voltage and tabulate broken and noisy channels.
\item Assembly test --- Noise and readout were tested for each ladder (wedge) 
as it was installed in a barrel (disk).  Coherent noise was evident in badly 
grounded ladders and grounds were remade.
\item Final cabling test --- Readout was verified after the low-mass cables 
were attached.  Upon leaving SiDet, 99.5\% of the sensors were functional.
\end{itemize}
This extensive testing was made possible by the development of a ``Stand 
Alone Sequencer'' (SASEQ), a 3U VME module which could be read out simply 
using a PC-VME interface and Microsoft EXCEL.  This allowed for easy 
replication of testing systems at Fermilab and universities.

A large scale ``10\%'' test was organized to test major detector 
components with pre-production versions of final readout components.  This 
test was crucial for debugging readout hardware, testing termination 
schemes, and adjusting sequencer timings.  Features were identified 
in the SVXIIe chip that caused pedestal jumps and readout errors.  
Additional initialization states were added to the SVXIIe control sequence 
to address these problems.  An 
assembled barrel and F- and H-disks were tested in the 10\% test to check 
system integration and search for noise problems.  Cosmic ray data were taken 
with the barrel as a final check of system performance.

\subsubsection{Operation}
\paragraph{Operating experience}
\label{sec:smt-operation}
The SMT has been included in physics data taking since the start of Run II. 
Full electronics debugging was completed in October 2001, 
when 94\% of the sensors were functional.  As of May 2005, 90\% of sensors 
were functional (note that problems can be anywhere along the readout
chain).  Most operational difficulties have been peripheral 
to the silicon detector itself.  These include latchup of operational 
amplifiers on the 
interface boards, low voltage power supply failures, and high leakage 
currents in high voltage distribution boxes.  

\begin{figure}
\centerline{\includegraphics[width=3.in]{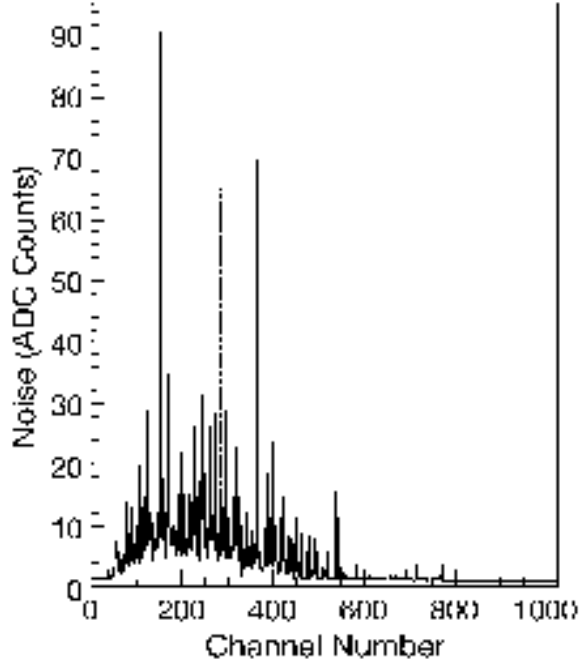}}
\caption{Example of the grassy noise seen in the Micron-supplied F-disk 
detectors.  Ideally, the entire plot would look like the region above channel
800.} 
\label{fig:grassy-noise} 
\end{figure}

The most serious detector feature is ``grassy noise,'' shown in
Figure~\ref{fig:grassy-noise}, which is confined to the Micron-supplied 
F-disk detectors (75\% of the F-disk sensors).  This noise is 
characterized by large charge spikes which cover 10--20 strips and occur in 
about 20\% of the events for affected devices.  Leakage currents 
typically rise to greater than 100~$\mu$A within one hour of 
turn-on at the beginning of a store.  
Both the SMT and CFT observe pedestal shifts that depend on the phase of the 
beam crossing with respect to the SVXIIe reset pulse.  These shifts are 
typically 1--2 counts (27 counts/MIP) in most ladders but can be as large as 
10 counts in a few that presumably have poor ground connections.  Digital 
voltage  
supply current in the SVXIIe chip rises steadily if the chip is not properly 
initialized or not read out steadily and can cause individual HDIs to trip 
off in 5--10 minutes.  These currents are constantly displayed and a special 
pulser is turned on if there is an extended interruption in data 
taking.

\begin{figure}
\centerline{\includegraphics[width=3.in]{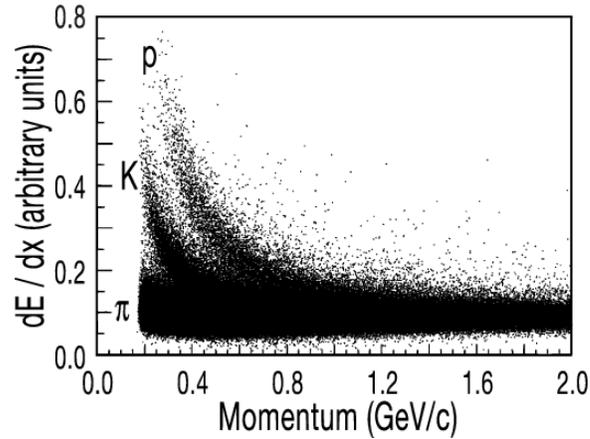}}
\caption{Distribution of energy loss for a kaon-enriched sample 
of tracks showing $\pi$, $K$, and proton bands.} 
\label{fig:smt-dedx} 
\end{figure}

\paragraph{Alignment and calibration}
Signal/noise performance varies with detector type from 12:1 to 18:1.  
Coherent noise is typically one-third of the random noise but varies due to 
the crossing-dependent pedestal shifts described in
Section~\ref{sec:smt-operation}.  Gains vary among detector 
types with the n-sides 5--15\% lower than the p-sides due to the larger 
load capacitance.  Pulse height information from the SVXIIe is used to 
calculate cluster centroids and can also be used for d$E$/d$x$ tagging of 
low momentum tracks.  Figure~\ref{fig:smt-dedx} shows d$E$/d$x$ distributions
after corrections for gain and incident angle are made.  

\begin{figure}
\centerline{\includegraphics[width=3.in]{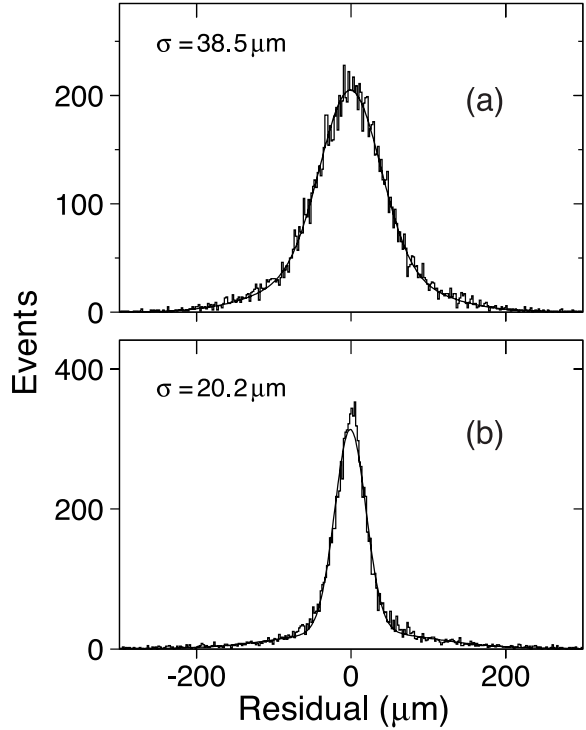}}
\caption{Axial residual distribution using tracks with 
$p_T > 3$~GeV/$c$ (a) upon initial installation and (b) after software 
alignment of the central barrel detectors.  
The residual is the distance between the SMT hit and the track.  The track 
fit was done using all SMT and CFT hits except the SMT hit in question.  
The simulated resolution of the residual distribution for a perfectly aligned 
detector is about 16~$\mu$m.}
\label{fig:smt-align} 
\end{figure}

Detector alignment was transferred from optical CMM measurements of detector 
fiducials to ladder beryllium features to barrel sapphire balls to 
half-cylinder targets to the D\O\ coordinate system.  
Final alignment for the combined SMT-CFT tracking system 
is better than 10~$\mu$m (Figure~\ref{fig:smt-align}).

\subsection{Central fiber tracker}
\label{sec:CFT}

The CFT consists of scintillating
fibers mounted on eight concentric support cylinders and
occupies the radial space from 20 to 52~cm from the center of the
beampipe.  To accomodate the forward SMT H-disks, the two innermost cylinders 
are 1.66~m long; the outer six cylinders are 2.52~m long.  The outer cylinder 
provides coverage for $|\eta| \lsim 1.7$.  Each cylinder supports one doublet 
layer of fibers oriented along the beam direction ($z$) and a second doublet 
layer at a stereo angle in $\phi$ of $+3^\circ$ ($u$) or $-3^\circ$ ($v$).  
Doublet layers with fibers oriented along the beam  axis are referred to as 
axial layers, while the doublet layers oriented at small angles are referred 
to as stereo layers. From the smallest cylinder outward, the fiber doublet 
orientation is $zu-zv-zu-zv-zu-zv-zu-zv$.  
The scintillating fibers are coupled to clear fiber waveguides which carry the 
scintillation light to visible light photon counters (VLPCs,
Section~\ref{sec:VLPC}) for read out.  The small fiber diameter (835~$\mu$m) 
gives the CFT an inherent doublet layer resolution of about 100~$\mu$m as 
long as the location of the individual fibers is known to better than 
50~$\mu$m.  Scintillating fiber detectors are discussed in detail in 
Ref.~\cite{scint_fiber_ann_rev}.

Discriminator signals from the axial doublet layers are used to form a fast 
Level~1 hardware trigger based upon the number of track candidates above  
specified $p_T$ thresholds (with a minimum threshold of 1.5 GeV/$c$).  Level~1 
track candidates are used by the Level~2 trigger, while the Level~3 trigger 
uses the full CFT readout information.

\subsubsection{Fibers}
\label{sec:fibers}

The scintillating fibers, including the cladding, are 835~$\mu$m in diameter 
and 1.66 or 2.52~m in length. 
They are optically connected to clear fiber waveguides of identical diameter
which are 7.8 to 11.9~m long.  The fibers were manufactured by 
Kuraray \cite{Kuraray} and have a multi-clad structure consisting of a core 
surrounded by two claddings.  The scintillating fiber is structurally and 
chemically similar to
the clear fiber, but contains fluorescent dyes.  The CFT uses
about 200~km of scintillating fiber and 800~km of clear fiber.  

Light production in the fibers is a multistep process.  The base core material
is polystyrene (PS).  The PS is doped with the organic fluorescent dye
paraterphenyl (pT) to about 1\% by weight.  Excitations in the PS are rapidly
transferred to the pT via a non-radiative dipole-dipole interaction.  pT has
a rapid fluorescence decay (a few nanoseconds) and a short emission wavelength
($\approx 340$~nm).  The mean free path of the emitted light is only a few
hundred microns in the PS.  To get the light out of the detector, a secondary
wave-shifter dye, 3-hydroxyflavone (3HF), is added at a low concentration 
(1500~ppm).  The 3HF is 
spectrally matched to the pT but has minimal optical self-absorption.  The 3HF
absorbs the 340~nm radiation from the pT and re-emits it at 530~nm which is
well-transmitted in PS.

Surrounding the PS core, whose refractive index is $n=1.59$, are two claddings,
each approximately 25~$\mu$m thick: an inner layer of polymethylmethacrylate 
(PMMA) with $n=1.49$, and an outer layer of fluoro-acrylic with $n=1.42$.  
The PMMA inner cladding serves as a mechanical
interface between the core and the outer cladding, which are mechanically
incompatible.  The multiclad fiber is both mechanically and optically superior
to single-clad fiber and typical values of the attenuation length are
about 5~m for the scintillating fiber and about 8~m for the clear fiber. 

We observe the light from only one end of each scintillating fiber.  
The opposite end of each of the scintillating fibers was mirrored with a 
sputtered aluminum coating that provides a reflectivity of about 
90\%.  

\begin{figure}
\centerline{\includegraphics[width=3.in]{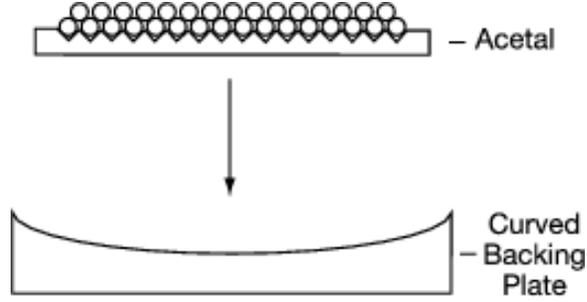}}
\caption{Technique for curved scintillating fiber ribbon fabrication.}
\label{fig:CFT_ribbon_fabrication}
\end{figure}

The scintillating fibers were assembled into ribbons consisting
of 256 fibers in two layers of 128 fibers each.  Precisely spaced grooves were 
machined into a long, 1/16"-thick piece of acetal.  The spacing between 
the grooves varies between 928 and 993~$\mu$m and depends on the 
radius of the corresponding support cylinder.  The grooved plastic was 
inserted into a rigid, curved backing plate of the desired 
radius, and the scintillating fibers were laid in and glued together to form the
doublet ribbons; the two layers of fiber are offset by one-half of the fiber
spacing.  The technique is illustrated in 
Figure~\ref{fig:CFT_ribbon_fabrication}.
It enables curved ribbons to match the curvature of each
support cylinder without machining precisely-spaced grooves into a curved
surface.  Details on fiber lengths and spacings are provided in
Table~\ref{tab:CFT_params}.

\begin{table}
\begin{center}
\caption{Design parameters of the CFT; $u = +3^\circ$, $v = -3^\circ$.  A
through H correspond to the eight axial layers of the CFT.}
\label{tab:CFT_params}
\begin{tabular}{lc c c c c}
\hline 
Layer & Radius (cm) & Fibers/layer & Fiber separation ($\mu$m) 
& Active length (m)\\
\hline
A    & 20.04 & $1280 \times 2$ & 982.4 & 1.66 \\
A$u$ & 20.22 & $1280 \times 2$ & 990.3 & 1.66  \\
B    & 24.93 & $1600 \times 2$ & 978.3 & 1.66 \\
B$v$ & 25.13 & $1600 \times 2$ & 985.1 & 1.66  \\
C    & 29.87 & $1920 \times 2$ & 976.1 & 2.52 \\
C$u$ & 30.05 & $1920 \times 2$ & 980.9 & 2.52  \\
D    & 34.77 & $2240 \times 2$ & 974.4 & 2.52  \\
D$v$ & 34.95 & $2240 \times 2$ & 979.3 & 2.52 \\
E    & 39.66 & $2560 \times 2$ & 971.7 & 2.52 \\
E$u$ & 39.86 & $2560 \times 2$ & 976.3 & 2.52 \\
F    & 44.56 & $2880 \times 2$ & 970.0 & 2.52 \\
F$v$ & 44.74 & $2880 \times 2$ & 974.3 & 2.52 \\
G    & 49.49 & $3200 \times 2$ & 969.8 & 2.52 \\
G$u$ & 49.67 & $3200 \times 2$ & 973.3 & 2.52 \\
H    & 51.97 & $3520 \times 2$ & 926.1 & 2.52 \\
H$v$ & 52.15 & $3520 \times 2$ & 927.8 & 2.52 \\
\hline
\end{tabular}
\end{center}
\end{table}

\begin{figure}
\centerline{\includegraphics[width=6.in]{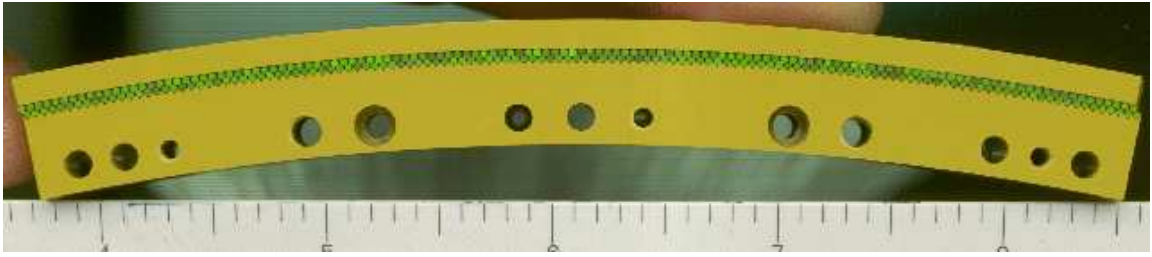}}
\caption{A CFT fiber ribbon mass terminated via a v-groove connector.}
\label{fig:CFT_v-groove-connector}
\end{figure}

The readout ends of the fibers were carefully positioned and adhesively
bonded into v-groove connectors, which are located around the outer
perimeter of the detector, and then the mass-terminated ribbon and
connector were polished to facilitate high efficiency light transmission
across the connector joint.  A polished curved connector is shown in
Figure~\ref{fig:CFT_v-groove-connector}.  Each 256-fiber
waveguide bundle terminates in a matching curved connector.  
The connectors for
each doublet fiber layer are different since the 
connectors must have the proper
curvature for each layer.  The light transmission through the v-groove
connectors, with optical grease between the fiber ends, is approximately 95\%. 

After the fiber ribbons were fabricated, a $^{57}$Co x-ray source was used to
verify the accuracy of the fiber placement, the responses of the fibers, and 
the transmission efficiency of the connectors.  The position of 
each fiber within a ribbon was determined with an accuracy of better than 
25~$\mu$m rms.

\subsubsection{Mechanical support structure}

The eight support cylinders are each double-walled with a 0.25"-thick core of 
Rohacell~\cite{Rohacell}.  
The walls are constructed from linear carbon fibers impregnated with about 
40\% resin.  To minimize sagging of the cylinders, the carbon fibers were 
applied in layers in the following pattern:  
$0^\circ/60^\circ/-60^\circ/{\rm core}/-60^\circ/60^\circ/0^\circ$ 
where the angles are those of the carbon fibers with respect 
to the axis of the cylinder. 
Each carbon fiber layer is about 0.0025" thick; the total 
thickness of carbon per cylinder is 0.020".  After fabrication, the outside
diameters of the cylinders were measured and compared to a perfect cylinder.  
The radial deviation for all cylinders is approximately 100~$\mu$m rms.

The requirements of the Level~1 hardware trigger constrain the design of the 
CFT, particularly in the placement of the scintillating fiber ribbons.  
Offline reconstruction of the CFT data can incorporate corrections for 
misalignment of fibers, but this is not possible in the hardware trigger;
the axial fibers had to be placed on the support cylinders such that any skew 
from parallel to the beam axis is less than 200~$\mu$m from end to end.  
Consequently, the individual fiber ribbons were precisely positioned and 
bonded to the outer surface of each cylinder prior to nesting the cylinders.
The placement precision is approximately 35~$\mu$m rms for both axial and
stereo layers.   
The connectors for the axial ribbons are all on one end of the cylinder,
while the connectors for the corresponding stereo ribbons are all on the
opposite end of the cylinder.

Successive cylinders are nested together by thin carbon-fiber annular 
rings that connect the inner surface of the end ring of one cylinder to a 
carbon-fiber ring mounted on the outer fiber surface of the cylinder 
immediately inside.  The fiber connectors and interlocking rings seal the 
fiber volume.

For tracks traversing the detector at normal incidence, the thickness of each
cylinder can be described as follows:  0.28\% of a radiation length for the
scintillating fibers, 0.32\% for the carbon fiber support cylinder, 0.13\% for
the glue used to make ribbons out of fibers, and 0.17\% for the glue used to
attach the ribbons to the support cylinders.

\subsubsection{Waveguides}
\label{sec:waveguides}

The clear fiber waveguides range in length from 7.8 to 11.9~m, and
use the clear fibers described in Section~\ref{sec:fibers}.  The waveguides
generally contain 256 clear fibers inside a flexible protective 
plastic light shield.  One end is mass terminated in a curved 
acetal connector machined to mate to the corresponding connector at
the end of the scintillating fiber ribbon. The mass-terminated
curved connector end was polished using a diamond fly cutter.  
About 40~cm from the other end of the waveguide, the 256 fibers are 
separated in $\phi$ into two groups of 128 fibers.
Each group is covered with a flexible plastic protective light shield.
These fibers are individually routed into the appropriate locations
in two rectangular connectors
that are designed to mate to the VLPC cassettes (Section~\ref{sec:VLPC}).
These molded plastic rectangular connectors are composed of Noryl N190
\cite{noryl} with Celogen RA foam \cite{celogen} to
minimize distortions.
After the fibers were potted into the rectangular connectors, the 
fiber-connector assembly was polished and the routing of the fibers was 
verified.

\begin{figure}
\centerline{\includegraphics[width=6.in]{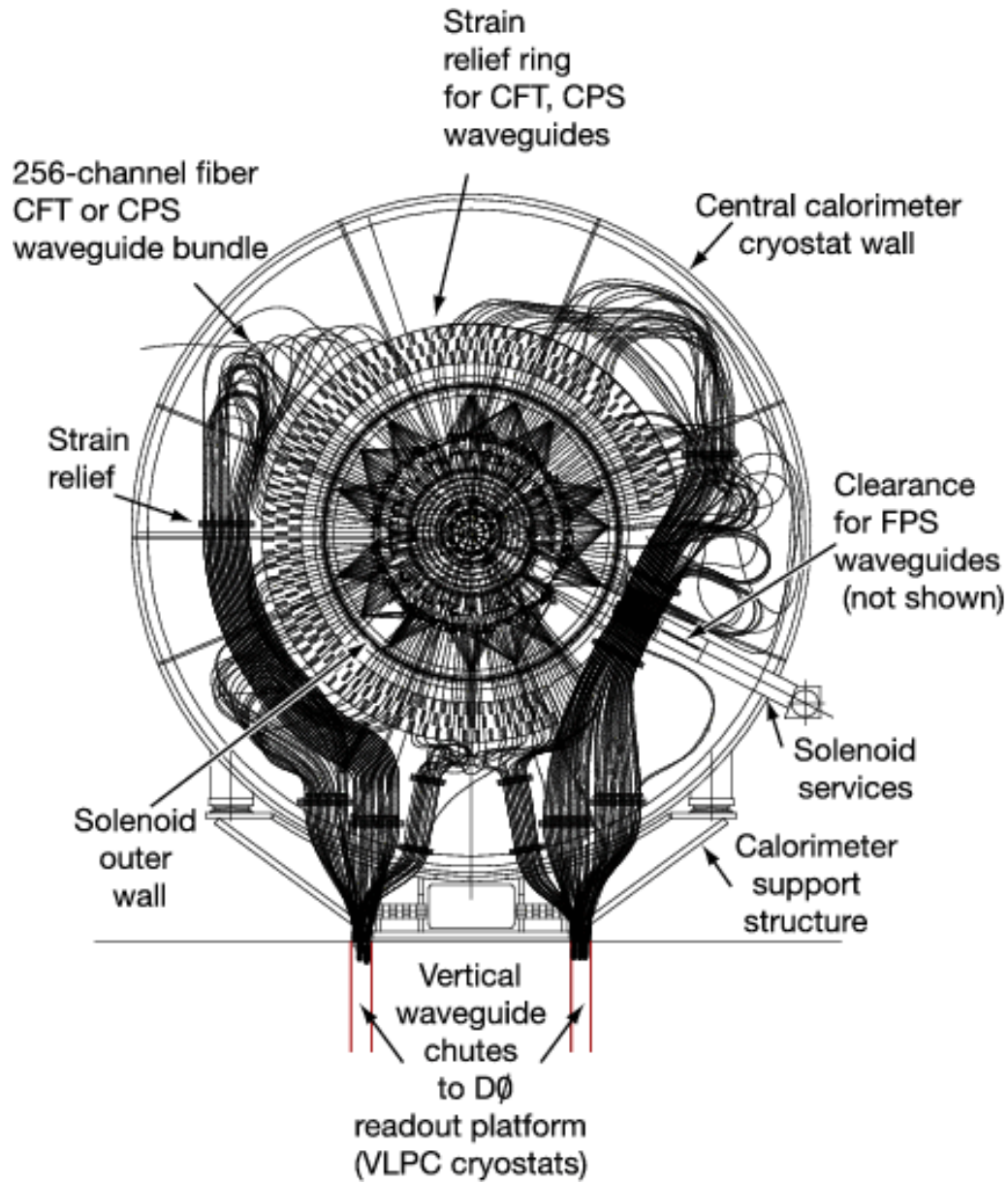}}
\caption{Routing of the clear waveguide fibers on the south face of the central
cryostat.}
\label{fig:CFT_fiber_routing}
\end{figure}

The clear fiber waveguides are routed from the ends of the CFT through the
small gaps between the central and end calorimeter cryostats
to the VLPC cassettes located about 6~m below the central calorimeter cryostat.  
These gaps also contain the forward preshower detectors 
(Section~\ref{sec:fps}), the waveguides for the forward and central 
(Section~\ref{sec:cps}) preshower detectors, and the readout cables for the 
SMT.  The narrowest region, between the forward preshower 
detector and the solenoidal magnet, is about 1.5" wide in $z$, requiring the
waveguides to follow complex paths along the 
central calorimeter cryostat.  Moreover, the waveguide bundles must have the
correct lengths to provide the proper timing for signals at the 
VLPCs.  The waveguide routing is illustrated 
in Figure~\ref{fig:CFT_fiber_routing}.  

\subsubsection{Visible light photon counter cassettes}
\label{sec:VLPC}

The light generated by the passage of charged particles through the
scintillating fibers of the CFT is converted
into electrical signals by the VLPCs housed in 
the VLPC cassettes.  VLPCs are impurity-band silicon avalanche photodetectors
that operate nominally at 9~K, and are capable of detecting
single photons \cite{VLPC1,VLPC2}.  They provide fast response, excellent 
quantum efficiency ($\geq$75\%), high gain (22,000 to 65,000), low gain
dispersion, and the capability of functioning in a high background
environment.  

VLPCs are fabricated by growing a series of doped and undoped
silicon layers on silicon substrate wafers.
Individual wafers yield a maximum of 176 VLPC chips after dicing, and each chip 
contains a two by four array of 1-mm-diameter pixels.  Each 
eight-pixel chip is soldered to an aluminum nitride substrate, and the outputs
from individual pixels are wirebonded to individual contact pads on the 
substrate.
Non-uniformities in the production process result in variations
in characteristics such as gain, quantum efficiency, and thermal
noise rates among and across VLPC wafers~\cite{Bross}.  Due to these variations,
the bias voltage at which the VLPCs operate at optimal signal-to-noise ratio 
varies between 6 and 8 volts.  To reduce the complexity of the bias voltage and 
threshold implementations in the readout electronics, 
the VLPC chips are carefully 
sorted and assigned to specific cassettes to allow for optimal performance.

\begin{figure}
\centerline{\includegraphics[angle=180,width=6.in]{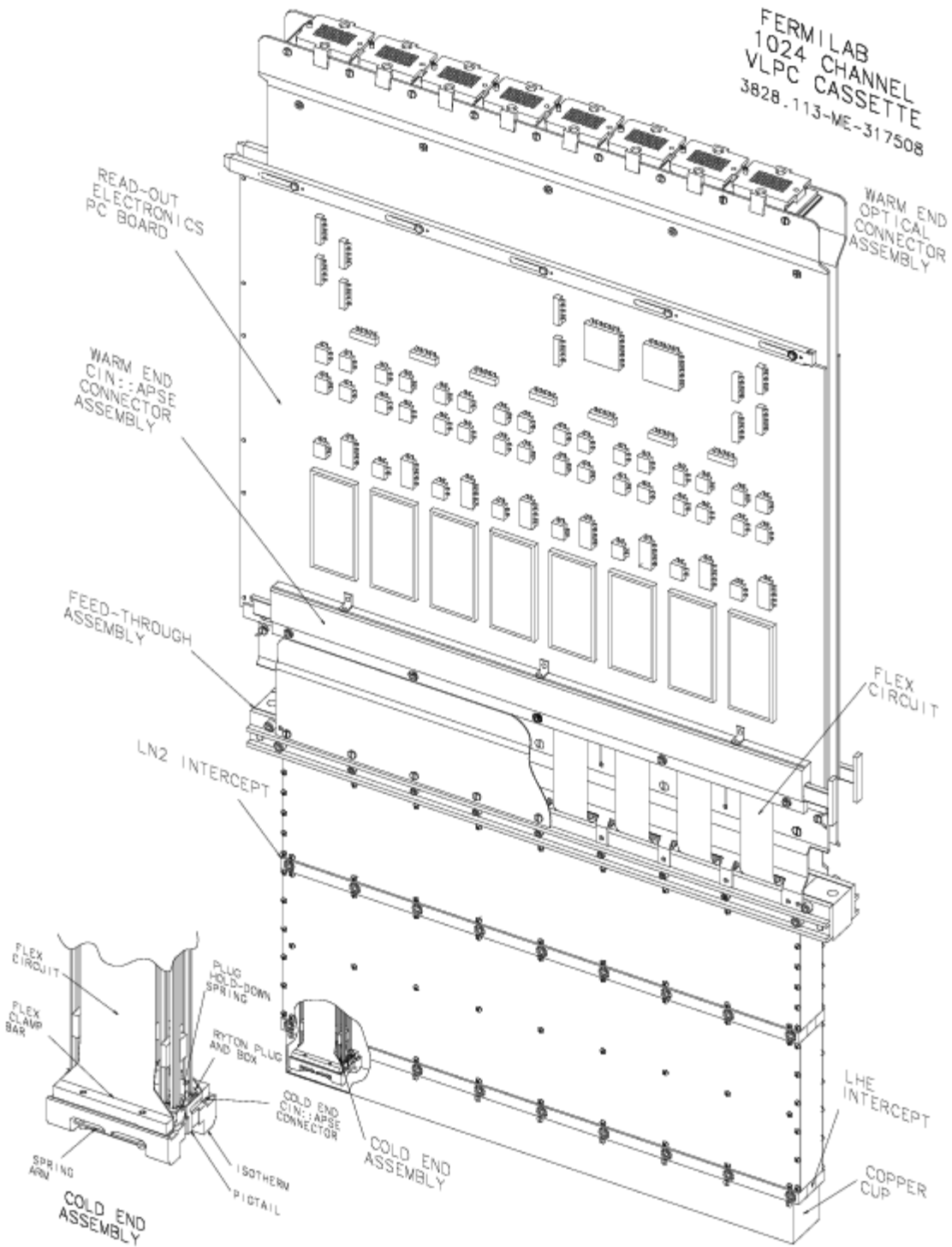}}
\caption{A VLPC cassette supporting AFE readout boards as viewed 
from the left side.  The VLPC hybrids are located on the isotherms housed 
inside the copper cup shown at the bottom of the figure.}
\label{fig:cass_overview}
\end{figure}

VLPC cassettes
mounted in cryostat slots provide the 
mechanical support, optical alignment, and appropriate operating environment 
for the VLPCs.  
Figure~\ref{fig:cass_overview} shows an outside view of a cassette with a 
readout board attached.
Each VLPC cassette houses 128 VLPC chips, and thus provides
1024 individual pixels of light-sensitive detector. 
Details of the specially-designed VLPC cryostats are 
available in Ref.~\cite{Rucinski}.
Cassettes are approximately 88~cm tall, 48~cm wide, and 4.4~cm thick.
Individual 0.965-mm-diameter fibers inside each cassette guide light from the 
clear fiber waveguides to individual VLPC pixels, and flex circuits provide 
paths for the electrical signals from the VLPCs to the preamplifiers on the 
analog front-end boards (AFEs) that are mounted on the cassette body.  
In addition to preamplifiers, the 
AFEs also provide trigger discriminator signals, temperature control, and 
bias-voltage control electronics.  
Details on the AFEs are provided in
Section~\ref{sec:AFE}.

\begin{figure}
\centerline{\includegraphics[angle=-90,width=4.5in]{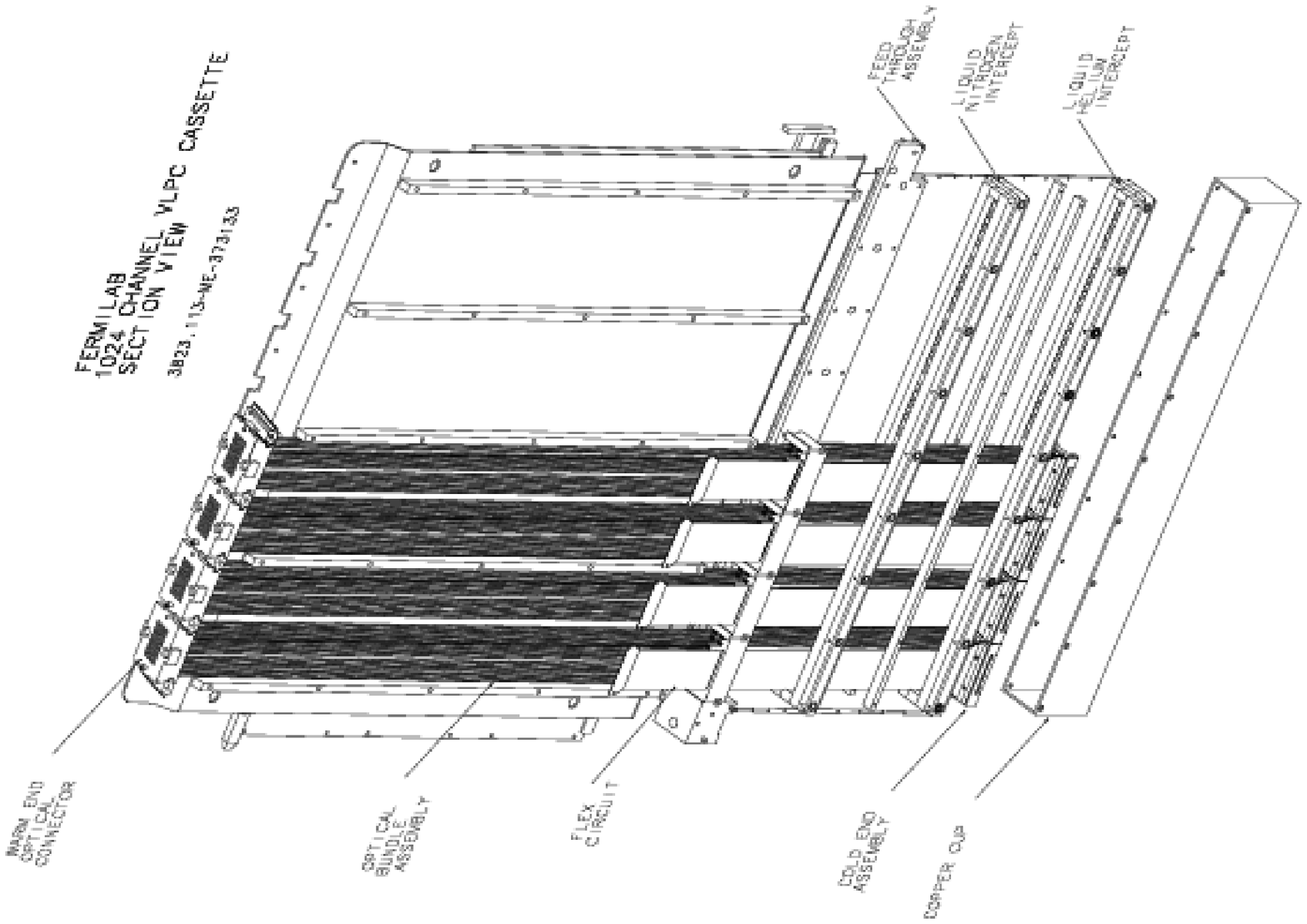}}
\caption{VLPC cassette body with left side body panel and
side panels removed to show four installed modules.}
\label{fig:cass_section}
\end{figure}

Each VLPC cassette consists of a cassette body housing eight modules.
A view of the internal structure of a partially assembled cassette is shown 
in Figure~\ref{fig:cass_section}.
Each module is composed of a 128-fiber bundle and a cold-end assembly.  The
cassettes are designed so that modules can be readily extracted to allow repair 
of the cold-end assemblies. The 128-fiber bundle 
terminates at the top end in a rectangular
molded optical connector (referred to as the warm-end optical connector), and 
at the bottom (cold) end in sixteen groups of eight fibers.  The ends of the
fibers are polished, and each group of eight fibers is glued into a 
molded plug.  The plugs and mating boxes are precision-molded
parts made of carbon-fiber-loaded polyphenylene sulfide (PPS) plastic.
A polyurethane feedthrough block is
cast around the 128 fibers and the flex circuits to form the barrier between 
the warm  and the cold ends of the cassette.  The fibers accept light from the 
clear fiber waveguides which are connected to the warm-end optical 
connectors at 
the top of the cassette and pipe the light to the VLPCs mounted in the cold-end 
assemblies.

\begin{figure}
\centerline{
\includegraphics[angle=270,width=5.5in]{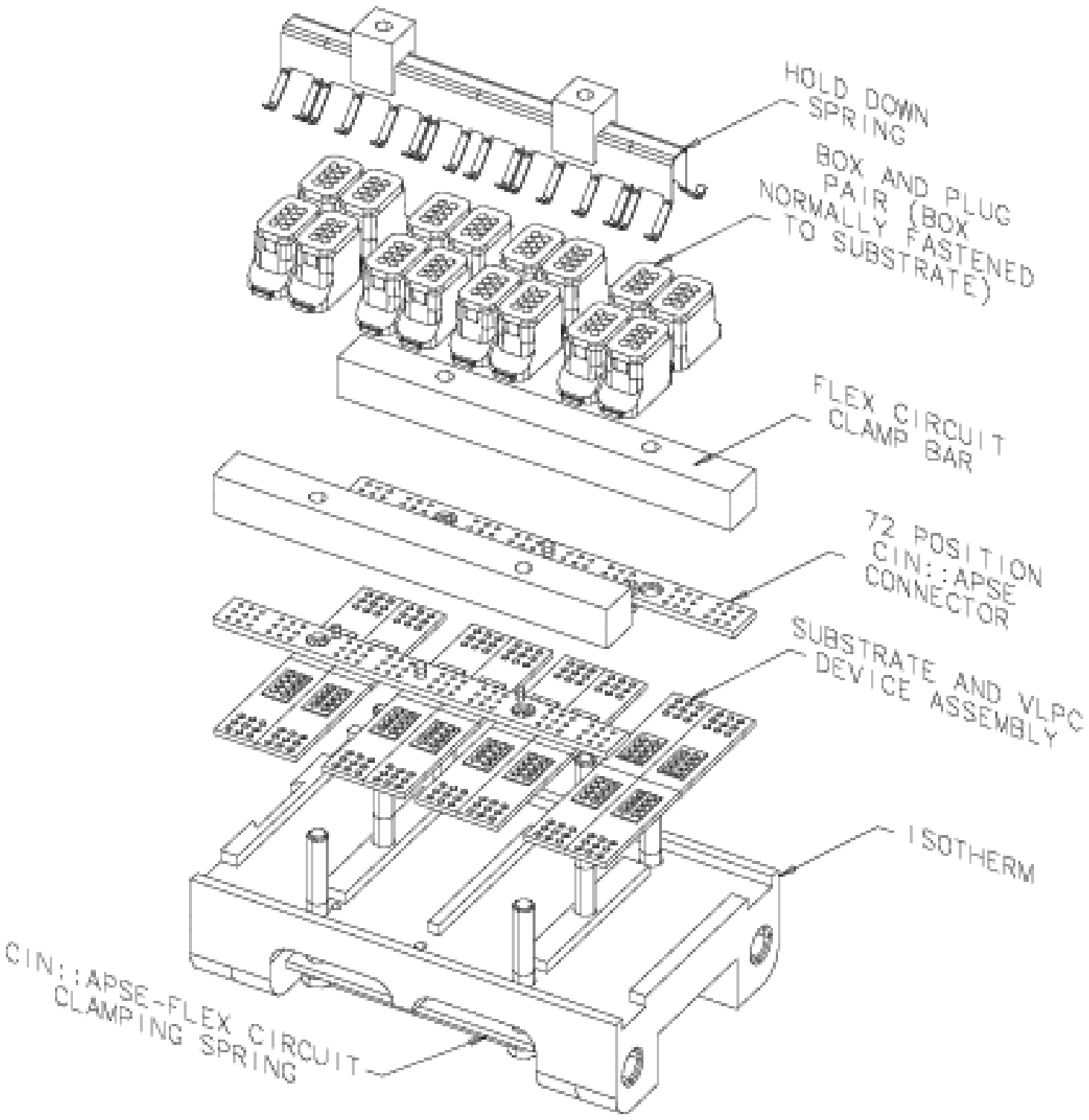}}
\caption{Cold-end assembly for one 128-channel module of a VLPC cassette.}
\label{fig:cass_ceexpl}
\end{figure}

The cold-end assembly hangs from the PPS plugs and consists of sixteen 
8-channel VLPC hybrid assemblies supported on an isotherm.  The inset in the 
lower left corner of Figure~\ref{fig:cass_overview} shows a view of a 
cold-end assembly supported on a 128-fiber bundle,
and Figure~\ref{fig:cass_ceexpl}
shows an exploded view of the components of
a cold-end assembly.

A VLPC hybrid assembly is composed of a PPS box adhesively bonded to the
aluminum nitride substrate upon which a VLPC chip is mounted.  
The molded PPS box is precisely positioned (maintaining 25~$\mu$m tolerances)
so that the eight polished fibers in the PPS plug are aligned over the
individual VLPC pixels when the plug and box are mated, thus achieving good
light collection efficiency while minimizing optical crosstalk between 
neighboring channels.  Connectors are clamped across the contact pads on the 
aluminum nitride substrates and mated to flex circuits which transmit 
single-ended electrical signals between 
the cold-end and the AFE.  Each module contains two flex 
circuits, and these flexible printed circuits are realized on 5-$\mu$m-thick 
adhesiveless copper coatings on a 51-$\mu$m-thick 
polyamide base material.  The individual traces are 76~$\mu$m 
wide with minimum 89~$\mu$m 
spacings between circuit features.  Each flex circuit is 41.1~cm
long.  
The isotherm also supports a calibrated carbon resistor that
serves as a temperature sensor, 
resistors used as heaters, and the required springs and 
fasteners.  The temperature sensor and heaters are employed to control the
temperature of the VLPCs to within 0.1~K.

The cassette body can be viewed as composed of cold-end and warm-end mechanical
structures.  The cold-end structure (that portion of the cassette that is 
inserted in the cryostat) comprises several sub-assemblies: the 
feedthrough assembly, the G-10 side walls, the heat intercept assemblies, and 
the cold-end copper cup.  The cassette is mounted to the top plate of the 
cryostat at the feedthrough assembly.  This assembly provides the gas-tight 
seal for the cold, $\leq2$~psig, stagnant helium gas 
volume within the cryostat.   
Side walls composed of G-10 support the top (or liquid nitrogen) 
intercept which serves to cut off the flow of heat from the warm end. 
Another set of G-10 side walls support the bottom (or liquid helium) intercept.
The bottom intercept supports the copper cup which surrounds the isotherms at 
the ends of the modules.

The warm-end structure is made of parallel tin-plated  
aluminum panels separated by spacer 
bars that form a protective box for the optical fibers.  Rails mounted on the
aluminum panels support the two AFEs.  
The warm-end Cin::apse \cite{Cinapse} connector assembly 
enables the connection between the flex circuits and the AFEs.
The AFEs mate to 
backplane connectors mounted on a backplane support structure via card-edge 
connectors.  The backplane support structures are bolted to the VLPC cryostats,
and thus the combination of cassettes and 
cryostat serve as the crates for the AFEs which are mounted 
on the cassettes.  This design, with connections on two orthogonal edges 
of the AFE, allows the AFEs to be
removed for service without extracting the cassette from the cryostat.

The CFT requires 76,800 channels of VLPC readout,
and the central and forward preshower detectors (Section~\ref{sec:preshower}) 
are instrumented with
an additional 22,564 channels of VLPC readout.  The ninety-nine cassettes 
necessary to provide this readout are mounted in two custom-designed cryostats 
located on the platform beneath the D\O\ detector.
Over 99.8\% of the individual VLPC channels in these cassettes met or 
exceeded the desired performance specifications
during cryogenic qualification tests performed prior to installation
at D\O.

\subsubsection{CFT readout electronics}
\label{sec:AFE}

The CFT and the central and forward preshower detectors
(Section~\ref{sec:preshower}) share a VLPC-based readout as well as similar 
Level~1 and Level~2 trigger electronics, and therefore all three use 
the same front-end electronics to process the signals from the VLPCs. 
The front-end electronics are custom printed circuit boards (the AFEs) 
approximately 14" tall (9U) and 18" long which are mounted on the VLPC 
cassettes inserted into cryostats 
as described in Section~\ref{sec:VLPC}. 

The AFE is a large and complex board that must perform a number of 
functions with competing requirements.  It has charge-sensitive 
amplifiers to deal with the very small signals from the VLPCs. It is 
part of the Level~3 readout, part of the Level~1 and Level~2 triggers, and part 
of the slow control and monitoring system.  It must also control the bias and 
temperature of the VLPCs. This functionality must be embedded in the 
AFE because it is the only piece of electronics that interfaces to the VLPCs.
It must serve three different sub-detectors with different dynamic range 
requirements and comes in two types --- left-handed and right-handed, 
depending on which side of the cassette body the board is mounted. 

A number of features make it possible for the AFE to fulfill all 
of the above requirements. First, only a single printed circuit board 
was designed and laid out. The ``handedness'' of the board is determined by 
the way a few key components are installed. So, for example, there 
are mounting points and traces for the hard metric connectors on both 
ends of the board. When the connector is mounted on one end of the 
board, the board becomes left-handed and mounts on the left side of 
the cassette.  When the connector is mounted on the other end of the board,
the AFE is 
right-handed.  The great majority of the components are mounted exactly 
the same way on both left- and right-handed AFEs. 
Second, the most noise sensitive parts of the board, 
including the front-end amplifier chips, are mounted on separate multi-chip 
modules (MCMs), each with its own regulator for the power and 
separate ground planes.  The amplifier chips are wire bonded directly to
these much smaller (3.5" by 1.5") subassemblies.  This allows 
the very fine pitch required for wire bonding to be confined to only 
the MCM substrate. Otherwise the AFE boards would not be manufacturable. 
This also allows the MCM subassemblies to be tested separately from the 
AFEs and to be removed as required for repair or replacement. There are 
eight MCMs on each AFE, each serving sixty-four VLPC channels, to match the 
construction of the cassette modules as described in Section~\ref{sec:VLPC}. 
The MCMs are intended to be identical. To accommodate the different dynamic 
ranges of the CFT, the central preshower detector, and the forward preshower
detector, capacitive charge division is used. 
By properly sizing the AC coupling capacitors on the AFE, the
charge seen by the amplifiers is adjusted to accommodate the 
required dynamic range 
with the same electronics.

\begin{figure}
\centerline{\includegraphics[width=6.in]{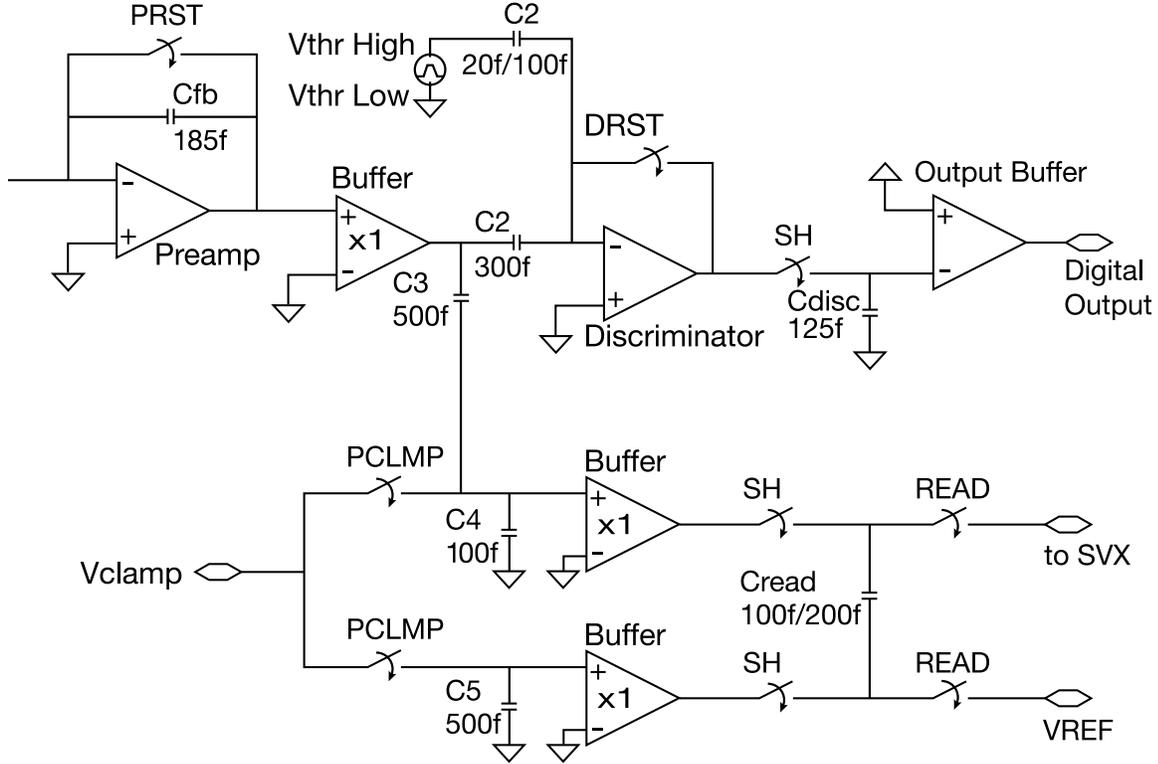}}
\caption{Simplified schematic diagram of the SIFT chip.  
Vclamp, Vthr, and VREF are externally-supplied DC voltage levels.}
\label{fig:sift-schematic}
\end{figure}

To further reduce the cost and simplify the system, the front-end 
amplifier and digitizer is the same chip used for the SMT. 
This also facilitates commonality further downstream in the 
readout chain. The SVXIIe and its readout are described in
Section~\ref{sec:SVXII}.  On the AFE, the SVXIIe provides for the integration 
of the charge signals from the VLPCs, a pipeline for storing the signals 
while the trigger is formed, and digitization of the signals and 
sparsification of the digitized data for readout. There are eight SVXIIe chips
on every AFE, one on each MCM. The SVXIIe chips share a single 8-bit bus for 
readout. However, because the signals from the AFE are also needed for 
the trigger system and because the SVXIIe digitization speed is too slow to
generate a trigger decision, a second chip, the SIFT chip, 
provides a trigger pickoff.  A simplified schematic is shown in 
Figure~\ref{fig:sift-schematic}. 
Each SIFT chip has eighteen channels so there are four SIFT chips on every 
MCM (eight channels are unused). The SIFT chips receive the signals from the 
VLPCs before the SVXIIe does. For each channel, the SIFT has a preamplifier 
that integrates the incoming charge and switched capacitors that are used to 
split the amplified signal and send it along two paths:  one to the SVXIIe, 
for subsequent digitization and readout, the other to a 
discriminator which fires if the charge exceeds a preset threshold 
(typically 10 to 15~fC).  

There are five clock signals that control the operation of the SIFT, 
as can be seen from the schematic. One is the reset of the integrator, 
PRST, another is the reset for the discriminator, DRST. The third, SH 
(sample and hold) samples the discriminator output and the analog value 
of the integrator at the end of the cycle. READ transfers the analog 
charge to the SVXIIe. PCLMP resets the bias for the followers in the analog 
chain. All of these signals are derived from the main accelerator clock 
(53~MHz).

The digital signals from the discriminators serve as the  
inputs to the Level~1 central track trigger (L1CTT, Section~\ref{sec:l1ctt})) 
and a bit is sent to the trigger system for every axial fiber every 132~ns 
(the time originally anticipated between beam crossings). This means that each
AFE board is transmitting data at about 4.1 Gbits/s. 
The formatting and pipelining
of this data is accomplished by sixteen CPLDs (complex programmable logic 
devices) in conjunction with four fast FIFO (first-in, first-out) memories. 
The trigger data is sent over four LVDS (low-voltage differential signal) 
links. Each LVDS link is driven by Texas Instruments SN65LVDS95 transmitter 
chips~\cite{cft-transmitter-chip}.  Each chip has 21 bits of parallel-load, 
serial-out shift registers driven by a $7\times$ clock synthesizer 
and drives four LVDS lines.  Three line pairs are used to transmit 21 bits, 
and the fourth carries the clock. With the data loaded into the transmitter 
chips with the 53~MHz clock, the AFE can send $21\times7\times4= 588$\ 
bits every 132~ns. One bit on each link is used to carry the 53~MHz clock to 
help synchronize the data frames at the receiver end. This leaves
$20\times7\times4= 560$\ bits for trigger information.  The actual 
discriminator information is carried by 512 bits. The other bits
are used to send control information to the L1CTT. 
These include indication of which crossings had beam present, when a 
Level~1 trigger accept was received from the trigger system, when the AFE is 
in a test mode or in reset, and so on. The control bits are duplicated on each 
of the four links and this redundancy is used to provide diagnostic 
information about the integrity of the links from the AFE to the L1CTT.

There are twenty CPLDs on each AFE. Besides the sixteen used to handle 
the discriminator data, one CPLD is used as a ``virtual'' SVXIIe chip to allow 
stored discriminator data to be read out on the same bus as the SVXIIe chips. 
Another CPLD is used to generate all the signals required to operate the SIFT 
chips and other timing signals for the board. The last two CPLDs work in 
conjunction with a microcontroller on the AFE to implement communications 
with the monitoring and slow control system via the MIL-STD-1553B bus. The 
microcontroller is a mixed signal controller PIC14000 from Microchip
\cite{Microchip}. 
It has an integrated single-slope 16-bit ADC which is used to monitor 
analog values on the board such as VLPC bias, VLPC current, and VLPC 
temperature. The temperature of the VLPCs is measured via calibrated 
carbon resistors mounted near the VLPC chips on each module. The value 
of the resistance must be measured to better than one part in a thousand 
and this drives the requirement for the ADC. The microcontroller 
implements closed-loop control for individual heater resistors on each 
VLPC module which allows the temperature at the VLPCs to be kept constant 
to $9.00\pm0.05$~K despite larger cryostat temperature fluctuations.  The
VLPC temperature monitoring and control is performed by the right-handed
AFE boards.  The VLPC bias voltage (which varies between 6 and 8 volts
from module to module) must be controlled to approximately 50~mV to achieve
optimal detector performance.

The extensive use of programmable logic on the AFE (twenty CPLDs and a 
microcontroller) greatly eased the tasks of designing, building, and 
operating it.  The design of the AFE could be completed before all details of 
the interconnected detector systems were finalized.  A single printed circuit
board is flexible enough to instrument three very different detectors. 
Moreover, this intricate board operates with little external control:  
once set points and operating parameters are 
downloaded, firmware on the board controls the system using on-board DACs 
and ADCs.  We are also able to modify the firmware to optimize the 
operation of the electronics as running conditions change. 

\begin{figure}
\centerline{\includegraphics[width=3.in]{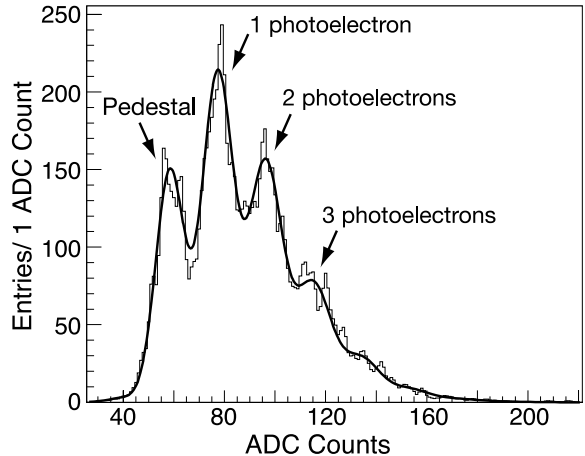}}
\caption{A typical LED spectrum for a single VLPC for an axial CFT fiber.  
Every channel is fit automatically and the parameters of the fit are extracted 
and used for monitoring.  Typically, more than 97\% of the axial 
channels are fit successfully.  The solid histogram is the data; the smooth 
curve is the fit.}
\label{fig:LED-spectrum}
\end{figure}

The performance of the electronics is dictated by the physics requirements 
of the detector.  The small-diameter scintillating fibers used in the 
CFT coupled with the long waveguides necessary
to carry the signals to the platform cause the signals generated by the
VLPCs to be small ($\approx$10 photoelectrons incident upon the VLPCs).
To assure acceptable efficiency for triggers and tracking, the 
individual channel thresholds must therefore be set between 1.5 and 
2~photoelectrons (pe).  To maintain a low and stable threshold and
to distinguish individual photoelectron peaks
during calibration, the analog signal must be digitized with a noise of less 
than 0.4 pe, which is equivalent to about 2~fC. 
In fact, the front-end electronics achieve or exceed all requirements. 
The mean pedestal width from fits to LED calibration spectra
(Section~\ref{sec:CFT-LED}) for all axial fibers is 0.24~pe or 1.6~fC,
and discriminator thresholds have 
similar noise and offsets of less than 2~fC so that it is possible to set
discriminator thresholds below 10~fC for most channels. A sample spectrum 
from calibration data 
is shown in Figure~\ref{fig:LED-spectrum}. A summary of the important 
parameters for the AFE is given in Table~\ref{tab:AFT-parameters}.

\begin{table}
\begin{center}
\caption{Summary of AFE board characteristics}
\label{tab:AFT-parameters}
\begin{tabular}{l p{7.cm} r}
\hline 
\multicolumn{2}{l}{AFE overview} & \\
\hline
\phantom{X} 
 & Number of readout channels per AFE &  512 \\
 & Number of MCMs per AFE             &   8 \\
 & Total number of individual components on AFE       & 2300 \\
 & \raggedright Threshold settings  \\  
   (analog V generated by onboard DACs) & 4 per MCM \tabularnewline
 & \raggedright Pedestal settings  \\
   (analog V generated by onboard DACs) & 1 per MCM \tabularnewline
 & Output bandwidth (L1 trigger)         & 4200 Mbit/s \\
 & Output bandwidth (SVXIIe bus)         & 159 Mbit/s \\
\multicolumn{2}{l}{AFE performance} & \\ 
\hline
 & Noise (input referred) --- analog readout & 1.6 fC \\
 & Noise (input referred) --- discriminator  & 1.2 fC \\
 & Threshold spread (rms)                  & 2 fC \\
 & Pedestal spread (rms)                   & 3 fC \\
\multicolumn{2}{l}{AFE stability and reliability} & \\
\hline
 & Mean time between failures (estimated)       & 35000 hrs/board  \\
 & Calibration stability  & Better than 1\% \\    
\hline
\end{tabular}
\end{center}
\end{table}     
    

The readout of the SVXIIe chips mounted on the AFE boards is coordinated 
via signals from the sequencers and sequencer controllers.
The sparsified data is transferred
from the sequencers to VME transition modules (VTM) 
via fiber optics cables.  The VTMs transfer the data to buffers in
VRBs in VME VIPA \cite{VIPA} crates.  A VRBC mounted in the same crate 
controls the
assignment of buffers.  A single-board computer mounted in that crate 
serves as the interface to the higher level readout.  The readout of the 
CFT and the preshower detectors is buffered in four VME crates.
Details of the higher level readout system are provided in 
Sections~\ref{sec:SVXII} and \ref{sec:daq}.

\subsubsection{Calibration}
\label{sec:CFT-LED}

The CFT and the preshower detectors (Section~\ref{sec:preshower}) are each 
equipped~\cite{Baumbaugh} with fast Nichia \cite{Nichia} blue-emitting LED 
pulsers that can be used to perform several functions.  The pulsers certify 
the proper connectivity and yes-no operation of the fiber channels and 
attached readout chain, provide a means for monitoring the stability of the 
VLPC readout over time, and calibrate the response of individual channels or, 
more specifically, provide a relative channel-to-channel energy calibration 
of the fiber-VLPC system.  This monitoring and calibration is done 
during detector commissioning or between collider stores.  Although the 
function of the LED system is the same for all three detectors, the method 
of light injection differs for each.

The limited space at the ends of the CFT cylinders complicates
injection of light directly into the ends of the scintillating fibers.  
However, a method has been devised to introduce light through the
cladding walls and into the active core of the fiber.  The method takes 
advantage of the fact that the 3HF dye in the CFT fibers has a small 
absorption in the blue wavelength range ($\approx 450$~nm).  
Blue light incident on several layers of the scintillating fiber 
can be absorbed in the top layer, exciting the 3HF dye and 
producing scintillation photons detectable by the VLPC.  
To distribute the light over the cylindrical geometry of 
the fiber tracker, flat optical panels consisting of a single  
thin ribbon of 100 (typically) clear fibers, each 
approximately 500~$\mu$m in diameter, are degraded in certain local 
regions so that light is emitted through their cladding walls. 
The fibers are bundled together at one end of the ribbon, potted, and 
finished to allow coupling with an LED.  The flat 
optical panels are then mounted around every 
fiber cylinder near each end of the cylinder and driven by pulsers.
The use of two bands provides not 
only a margin of safety against failure of an individual optical 
element but also allows comparison of the light from the two ends of the
tracker and monitoring of any changes in the quality
of the scintillating fibers over time.

Although similar flat optical panels could have been adapted to the preshower 
detectors, the CPS and FPS use different designs.  For the preshower detectors, 
injection of LED light directly into the scintillator strips
complicates the design appreciably, so instead, light is injected into the
wavelength shifting (WLS) fibers.  
Since the WLS fibers are grouped in sets of sixteen channels 
and routed from the scintillator strip to connectors located at the perimeter 
of the detectors, one LED  
illuminates a group of sixteen WLS fibers.  The mechanism, however, for
delivering light into these sixteen fibers differs for CPS and FPS.  

In the CPS, two LEDs are mounted inside a machined Lexan light block, 
directly before the outer perimeter of the detector, one on either side of 
the connector holding the WLS fibers.  Ten LEDs mounted on one side 
of the connector are chained in series, allowing for an RG-178 cable to supply 
power to each LED string.  The series of grouping ten 
LEDs continues across the perimeter of the CPS detector.

LEDs for the FPS system~\cite{patwa-thesis} are also mounted on the outer 
perimeter of the detector.  
However, due to the tight spatial constraints inside the FPS
scintillating wedges, a 0.835-mm-diameter clear fiber transports the
blue light to a group of sixteen WLS fibers, systematically arranged
within a hollow cylindrical cavity.  The clear fiber is
inserted directly into the central hole of the cavity, allowing for
light to be delivered uniformly over a full solid angle to the sixteen
WLS fibers.  The cavity holds the 
same group of WLS fibers that are routed from their respective 
scintillator strips to optical connectors coupling the clear fiber 
waveguides. The LEDs themselves are mounted on two PC boards,
each providing the relevant electrical circuitry to drive the device.

PC boards located on the readout platform supply power to the LED system of 
each detector.  The entire system is 
computer-controlled and can be operated directly from the control room, 
allowing users to program the pulse height and width as well as 
operate one LED in a detector at a time or the complete array.  

\FloatBarrier

\section{Solenoidal magnet}
\label{sec:solenoid}

The superconducting solenoidal magnet was designed \cite{magnet_cdr,solenoid} 
to optimize the momentum resolution, 
$\delta p_T/p_T$, and tracking pattern recognition within the constraints
imposed by the Run~I detector.  
The overall physical size of the magnet was determined by the available
space within the central calorimeter vacuum vessel:  2.73 m in length and
1.42 m in diameter.  We selected a central field of 2 T after considering
the momentum resolution and tracking pattern recognition, the available space, 
and the thickness of the cryostat which depends on the thicknesses of the 
conductor and support cylinder.  In addition, the magnet is required 
{\it i}) to operate safely and stably at either polarity, {\it ii}) 
to have a uniform field over
as large a percentage of the volume as practical, {\it iii}) to be as thin as 
possible to make the tracking volume as large as possible, {\it iv}) to 
have an overall thickness of approximately $1X_0$ at $\eta = 0$ to optimize the
performance of the central preshower detector mounted on the outside of the 
solenoid cryostat, and {\it v}) to quench safely without 
a protection resistor (although one is installed to reduce the recool time
after an inadvertent fast dump).  Services such as  
cryogens, magnet current buses, and vacuum pumpout and relief must
reach the magnet from the control dewar through the narrow space (7.6~cm) 
between the central and end 
calorimeter vacuum vessels.  The magnet system is controlled
remotely, including cool down, energization, de-energization for field
reversal, quench recovery, and warmup, without access to the magnet cryostat,
service chimney, or control dewar.

\begin{figure}
\centerline{\includegraphics[width=3.in]{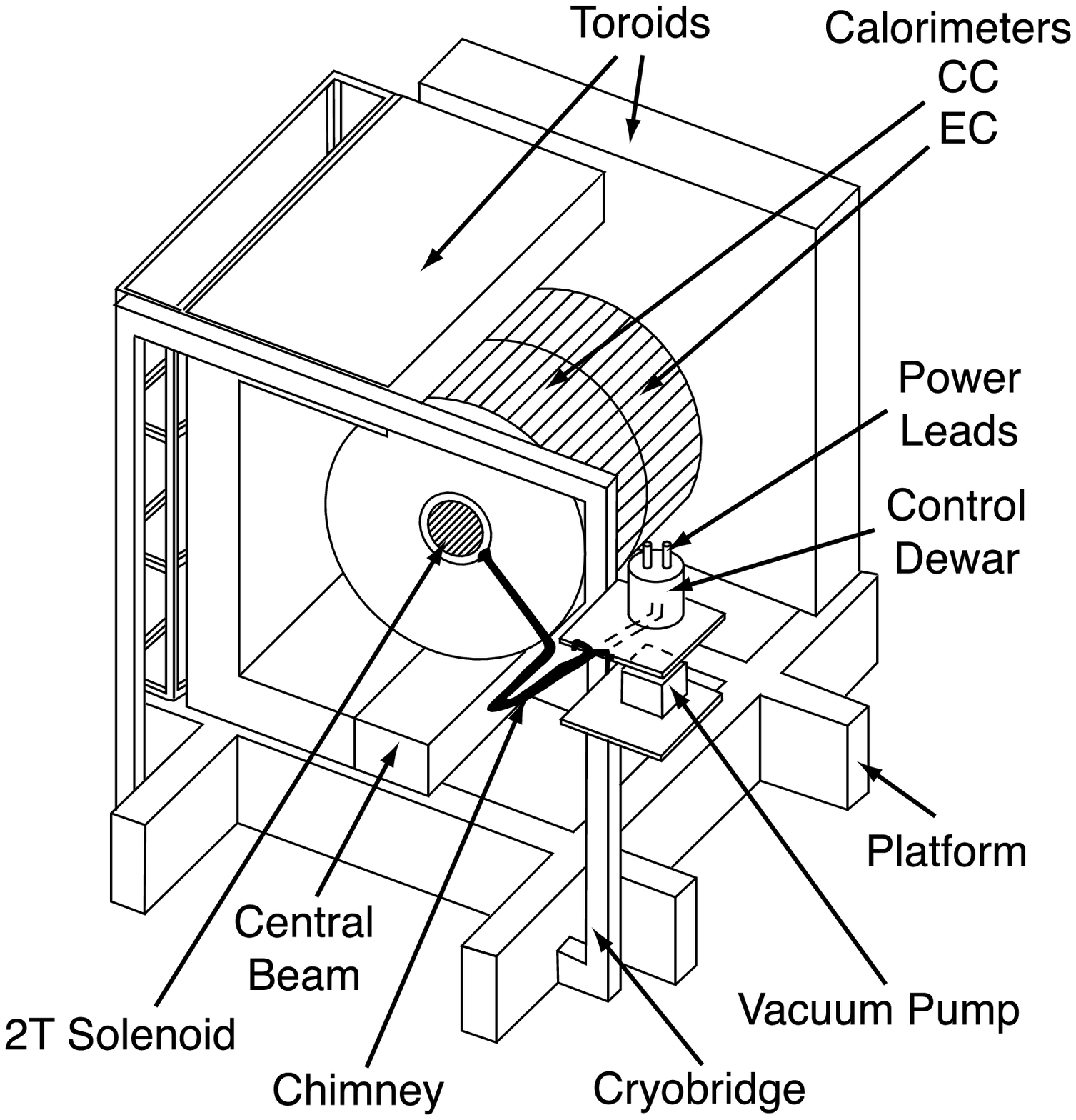}}
\caption{Perspective view of the solenoid inside the central calorimeter.
One end calorimeter, several muon chambers, and parts of the toroids have been 
omitted for clarity.  Also shown are the service chimney and control dewar.}
\label{fig:solenoid_perspective}
\end{figure}

The major parameters of the solenoid design are listed in Table
\ref{tab:solenoid_params}.  A perspective view of the solenoid inside the
central calorimeter with its chimney and control dewar is shown in Figure
\ref{fig:solenoid_perspective}. 

\begin{table}
\begin{center}
\caption{Major parameters of the solenoid}
\label{tab:solenoid_params}
\begin{tabular}{ll}
\hline 
Central field 			& 2.0 T \\
Operating current 		& 4749 A \\
Cryostat warm bore diameter 	& 1.067 m \\
Cryostat length			& 2.729 m \\
Stored energy			& 5.3 MJ \\
Inductance			& 0.47 H \\ 
Cooling				& Indirect, 2-phase forced flow helium \\
Cold mass			& 1460 kg \\
Conductor			& 18-strand Cu:NbTi, cabled \\
Conductor stabilizer		& High purity aluminum \\
Thickness			& 0.87 X$_0$ \\
Cooldown time			& $\le 40$ hours \\
Magnet charging time		& 15 minutes \\
Fast discharge time constant	& 11 seconds \\
Slow discharge time constant	& 310 seconds \\
Total operating heat load	& 15 W plus 0.8 g/s liquefaction \\
Operating helium mass flow	& 1.5 g/s \\
\hline
\end{tabular}
\end{center}
\end{table}

The solenoid, along with its cryostat, control dewar, and connecting service
chimney, was manufactured by Toshiba Corp.~\cite{Toshiba} in Yokohama, Japan.
The system was specified to operate safely and reliably over a twenty-year
lifetime with up to 150 cool-down cycles, 2500 energization cycles, and 400
fast dumps.

\subsection{Magnet construction}

\begin{figure}
\centerline{\includegraphics[width=3.in]{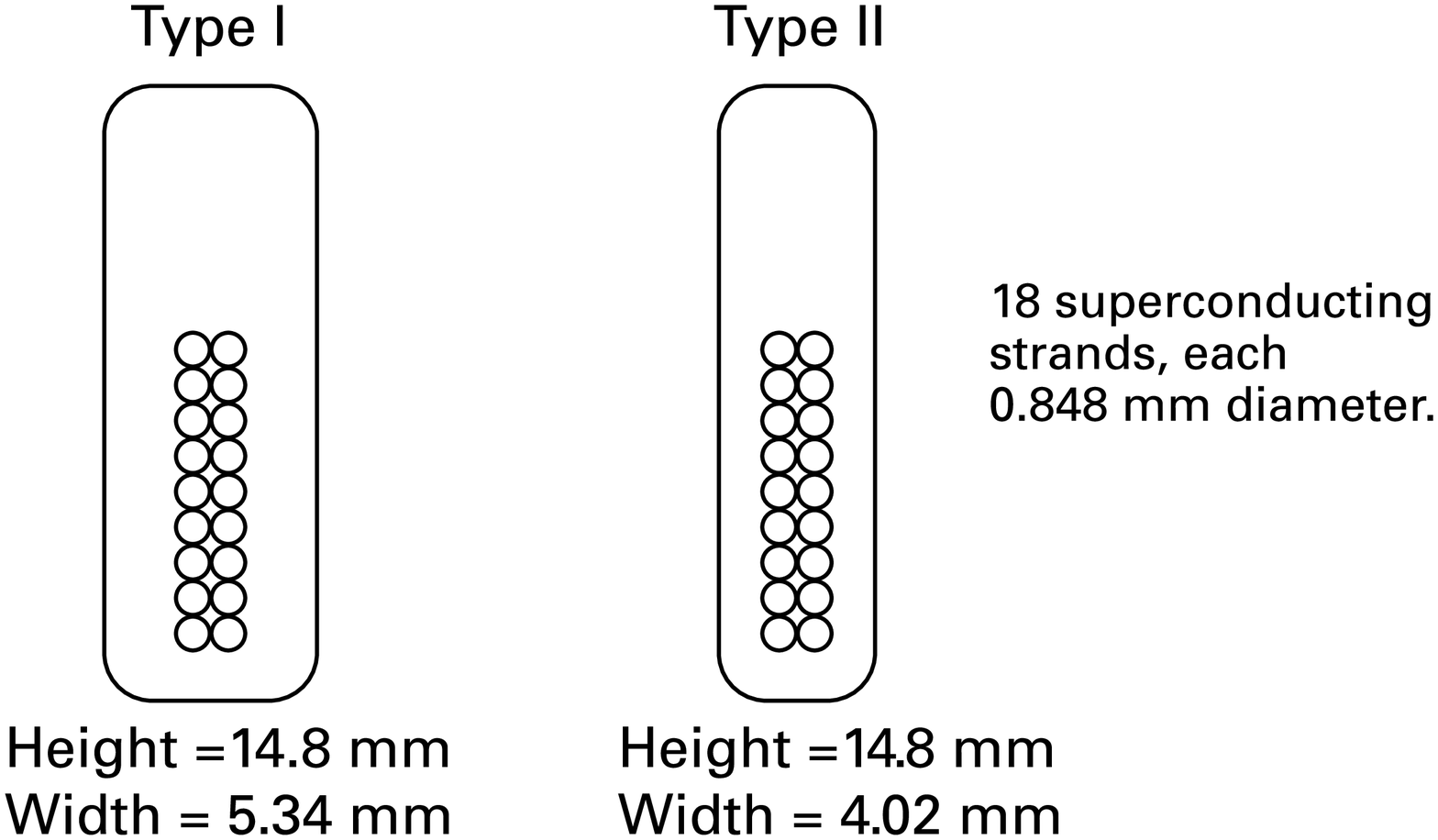}}
\caption{Cross sections of the two conductors used in construction of
the solenoidal magnet.}
\label{fig:solenoid_conductor}
\end{figure}

The solenoid is wound with two layers of superconductor to achieve the required
linear current density for a 2 T central field.  The support cylinder is located
on the outside of the windings to support the radial Lorentz forces on the
conductor, to provide axial rigidity to the finished coil, to provide a high
thermal conductivity path to the helium coolant piping, and to ensure reliable
quenchback for quench safety.  To maximize the
field uniformity inside the bore of the magnet, the current density in the 
windings is larger at the ends of the coil.  This is accomplished by the use 
of a narrower conductor at the ends of the coil.
Cross sectional views of the conductors are shown in
Figure~\ref{fig:solenoid_conductor}.  Both conductors are made with a 
superconducting Rutherford-type cable of multifilamentary 
Cu:NbTi strands stabilized with pure aluminum.  The basic strand has a 
Cu:NbTi ratio of 1.34:1 and a diameter of 0.848~mm; eighteen strands are used 
in each conductor.  

Both conductors are used in both layers.  The middle section of each
layer is wound with the wider conductor and the end sections with the narrower
conductor.  The transition point between the two conductors in the inner layer
occurs at $z=\pm 0.941$~m, and at $z=\pm 0.644$~m in the outer
layer.  At the four locations where the conductor width changes, the two types
of conductor are joined with a soldered lap joint one full turn in length.  
There are no other joints in the coil.

\subsection{Magnet cryostat}

The magnet cryostat consists of four major components:  the vacuum vessel, the
liquid-nitrogen-cooled radiation shield, the cold mass support system with 
liquid-nitrogen-cooled intercepts, and the helium cooling tube on the outer
support cylinder of the superconducting coil.  

The vacuum vessel consists of inner and outer coaxial aluminum shells with flat 
annular
bulkheads welded to each end.  The superconducting buses from the coil and the
cryogen pipes from the outer support cylinder and the radiation shields leave
the vacuum vessel through the service chimney nozzle welded in the bulkhead at 
the south end of the cryostat.  The cryostat is designed for full internal
vacuum and for an internal relieving pressure of 0.44 MPa (6.4 psig).  

The magnet cold mass, the superconducting coil and outer support cylinder,
weighs 1.46 metric tons.  The cold mass support system consists of axial members
which locate the coil axially and support it against axial thermal, decentering,
and seismic forces, and nearly tangential members that locate the coil radially
and provide radial support against thermal, gravitational, seismic, and
decentering forces.  The support members connect the outer support cylinder of
the coil to the flat annular bulkheads of the vacuum vessel.  All cold mass
supports have thermal intercepts that operate near 85~K.  The magnet cryostat 
is attached to the central calorimeter by support brackets that carry the 
weight of the cryostat and the tracking detectors that are attached to it, 
plus any residual magnetic decentering loads from the muon steel.

\subsection{Control dewar and refrigeration system}

The control dewar is the interface between the fixed cryogenic piping 
and the detector, which must be moved from the assembly area into the collision
hall. It has bayonet connections for cryogenic lines
and contains the vapor-cooled current leads and a liquid-helium reservoir,
helium supply subcooler, helium flow control valve, and other valving and
instrumentation.  The current leads carry the full operating current of
the magnet at the optimum cooling gas flow rate and are designed for safe 
operation without cooling gas flow for at
least the full slow-discharge time constant of the magnet (approximately 310~s)
in the event that cooling flow is lost while the magnet is de-energized.  

Cryogenic and electrical services are carried from the control dewar to the 
magnet cryostat via the service chimney, which also 
serves as the relief line for the solenoid vacuum space and provides a 
path for pumpout of the magnet cryostat and control dewar vacuum spaces. 

A cryogenic refrigeration system supplies liquid nitrogen and liquid helium to
the magnet and the VLPC cryostats for the CFT and central and
forward preshower detector readouts (Section~\ref{sec:CFT}).  
The refrigeration requirements of the magnet and VLPC systems,
as well as operation of the detector in the collision hall and adjacent 
assembly hall, were considered when the refrigeration system was designed. 

A standard Fermilab satellite stand alone refrigerator (STAR) provides helium
refrigeration.  The capacity of the STAR is sufficient for non-simultaneous
cooldown and simultaneous operation of both the solenoid and VLPC systems.
Liquid helium is stored in a 3000~L dewar which supplies the magnet control
dewar and the VLPC cryostats via separate transfer lines.

\subsection{Magnet energization and control}

\begin{figure}
\includegraphics[width=6.in]{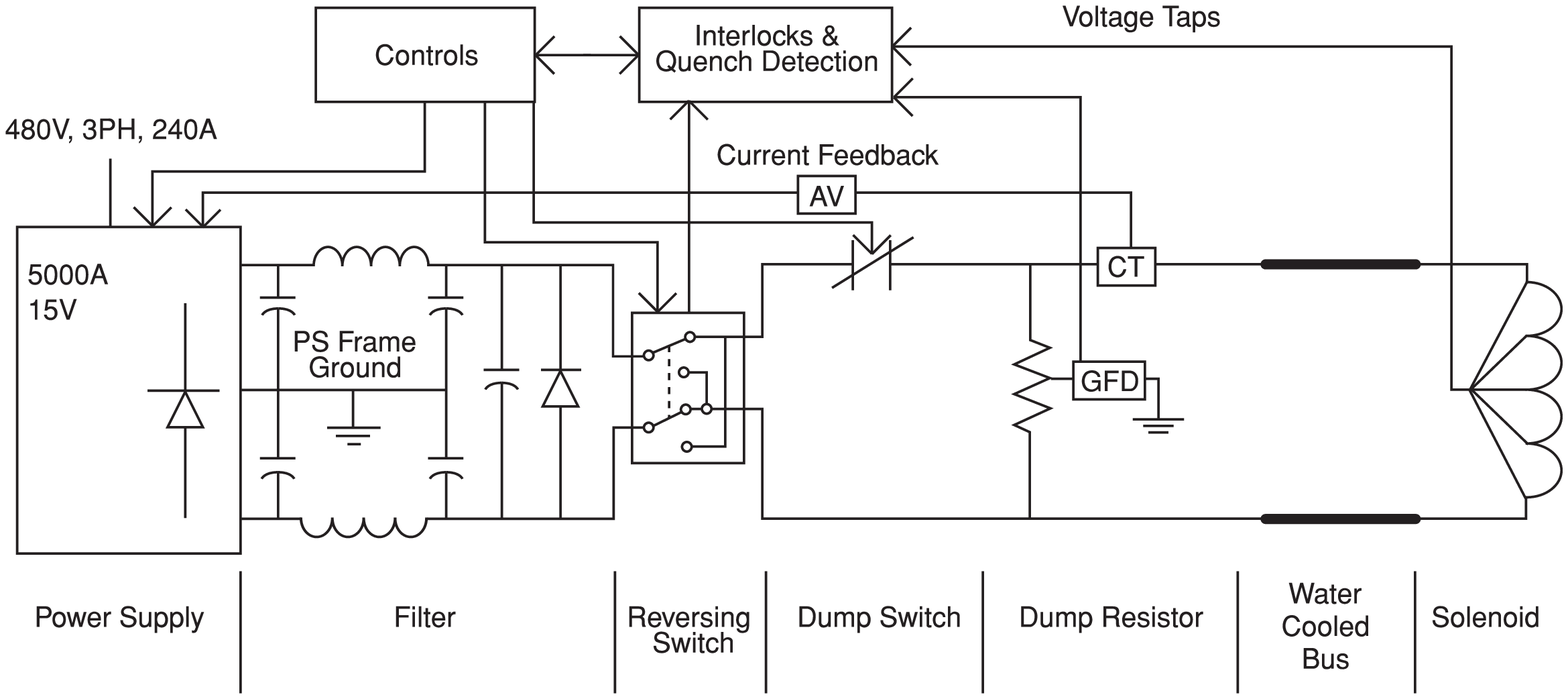}
\caption{Block diagram of the DC energization system for the solenoid.}
\label{fig:solenoid_energization}
\end{figure}

A block diagram of the DC energization system is shown in Figure
\ref{fig:solenoid_energization}.  The power supply is a special
Fermilab unit designed for superconducting loads.  It is a 
twelve-phase-thyristor water-cooled rectifier unit with precision feedback 
current regulation.  
The power supply taps are set at 15 V/5000 A for efficient
operation and reduced AC loading and DC ripple.   
The power supply regulates the current to within 0.01\%
using an external precision Holec~\cite{holec} 
5000~A direct current current transducer (also known as a DCCT or transductor) 
installed downstream of the ripple filter and dump resistor.

Magnet polarity is reversed using a 5000~A DC mechanical motorized polarity
reversing switch and a switch controller.  Polarity reversal requires 
complete discharge of the solenoid followed by recharging at the opposite
polarity.  The controller confirms that the
polarity reversal occurs at zero current load and that the power supply is
turned off.  Polarity reversal takes about forty minutes.

The Texas Instruments TI565T Programmable Logic Controller (PLC) (originally
installed to operate the liquid argon calorimeter refrigeration system, the 
muon chamber gas systems, and various building utility and safety systems) 
has been expanded to control and monitor the VLPC cryostats, solenoid, and 
helium refrigeration systems.  The PLC is a stand-alone system with an internal
battery-backed program memory that requires no host system to download the
control programs in the event of a power failure.  It consists of two
independent TI565T processors operated in a ``hot backup'' configuration.  One
PLC is online and actively controlling the system; the second is in standby,
running step-for-step with the online unit but with its input/output
communication disabled.  Each PLC runs continuous internal diagnostics for
errors.  When a fatal error is registered by the active PLC, it is taken offline
and the standby PLC is put online without intervention or disruption to the 
system.

Dedicated control and monitoring of the magnet 
energization and protection system is done by a new Texas Instruments TI555 
PLC.  A quench protection monitor (QPM) hardwired chassis and interlock logic 
unit provide primary quench protection for the magnet.  The QPM contains
filtering, signal averaging, and delay circuitry for the voltage taps and
thermometry which are used in the quench detection logic, vapor-cooled lead
fault detection, or power supply failure logic.  It contains preset limits for
selected sensors which trigger fast or slow discharge of the magnet and it 
preserves the time ordering of the detected fault signals which trigger magnet 
discharge.  The QPM defines a set of interlock conditions that must be 
satisfied before the magnet can be energized or the reversing switch operated.

The cryogenic controls and PLCs are powered by a 10~kW uninterruptable power 
supply that is backed up by an automatic diesel power generator.  The total 
power consumption is approximately 6~kW. 

\subsection{Magnetic field}

\begin{figure}
\centerline{\includegraphics[width=6.in]{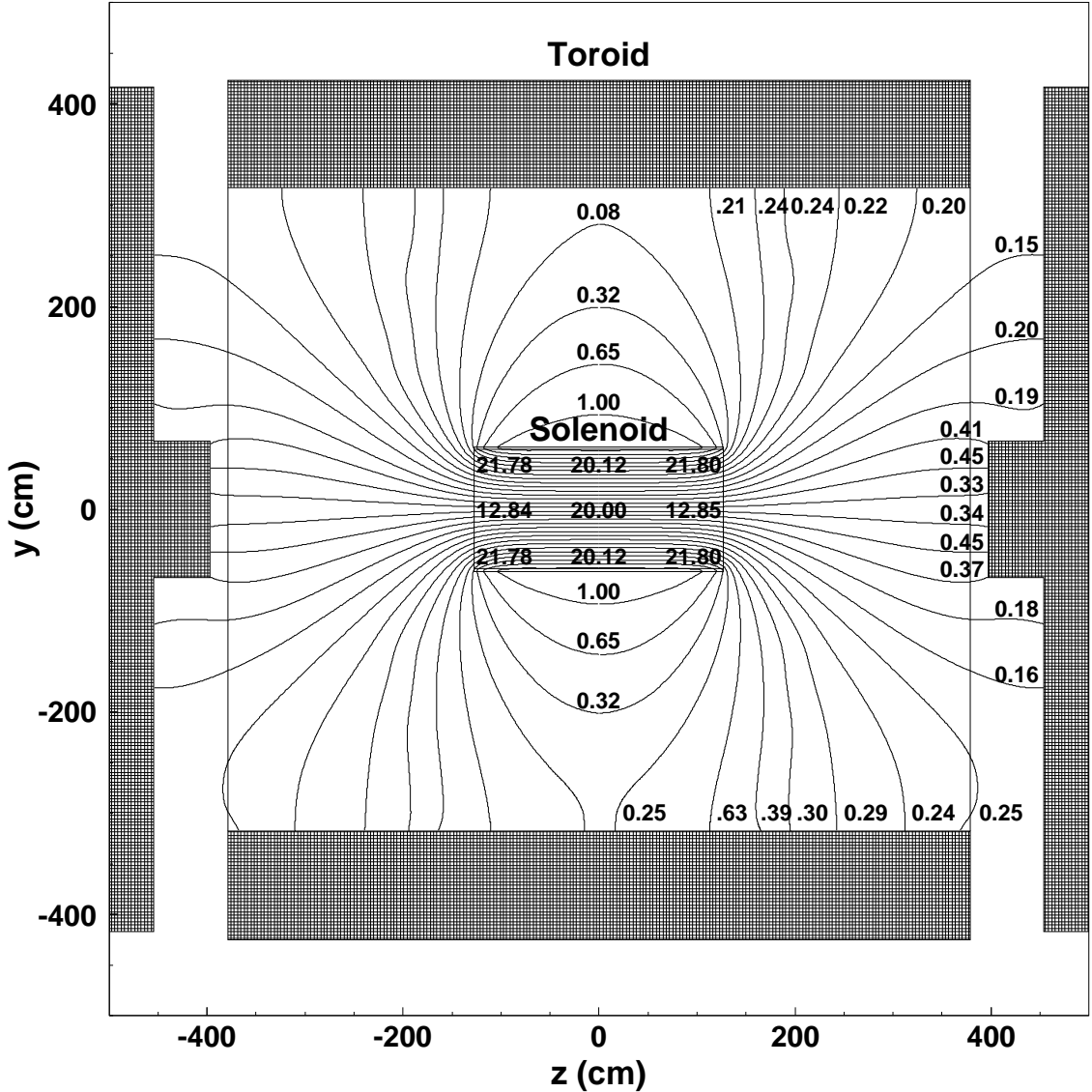}}
\caption{The $y-z$ view of the D\O\ magnetic field (in kG) with both the 
toroidal and solenoidal magnets at full current (1500~A and 4749~A, 
respectively).  The field in the central toroid is approximately 1.8~T; that in
the end toroids is about 1.9~T.  The field lines are projections
onto the $y-z$ plane; the left and right line ends differ by up to 2.5~m in 
$x$.}
\label{fig:solenoid-map}
\end{figure}

The magnetic field of the full magnet system is modeled using the TOSCA
\cite{tosca} program.  The calculated field map was compared with the measured
field in two locations:  near the internal radius of the solenoid cryostat 
($r\approx 54$~cm) at $z=4$~cm and in the gap (at $(x,y,z) = (0,372,105)$~cm)
at the top of the central muon toroid steel (Section~\ref{sec:toroids}).  
Within the solenoid (operated at 
4749~A), the measured field is $20.141\pm0.005$~kG; 
the calculated field at this location is 20.158~kG.  The calculated magnetic
field is scaled by 0.09\% to agree with the measurement.  With full 
operating current (1500~A) in the toroid coils, there is a 4.3\% difference 
between the calculated and measured field at the central toroid gap, 
requiring an adjustment in the calculated field for this magnet.   
The $y-z$ view of the magnetic field with both the toroid and solenoid magnets 
at full current is shown in Figure~\ref{fig:solenoid-map}.

\FloatBarrier

\section{Preshower detectors}
\label{sec:preshower}

The preshower detectors aid in electron identification and background rejection 
during both triggering and off\-line reconstruction.  They function as 
calorimeters as well as tracking detectors, enhancing the spatial matching 
between tracks and calorimeter showers~\cite{PS_cosmic_test}.  The detectors 
can be used offline to correct the electromagnetic energy measurement of the 
central and end calorimeters for losses in the solenoid and upstream material, 
such as cables and supports.   Their fast energy and position measurements 
allow preshower information to be included in the Level 1 trigger.  The 
central preshower detector (CPS) covers the region $|\eta| < 1.3$ and is
located between the solenoid and the central calorimeter with an inner radius of
28.25'' and an outer radius of 29.21''.  The two
forward preshower detectors (FPS)~\cite{patwa-thesis} 
cover $1.5 < |\eta| < 2.5$ and are attached to 
the faces of the end calorimeters.  The preshower detectors
can be seen in Figure~\ref{fig:tracker}.

\subsection{Common properties of the preshower detectors}

\begin{figure}
\centerline{\includegraphics[width=6.in]{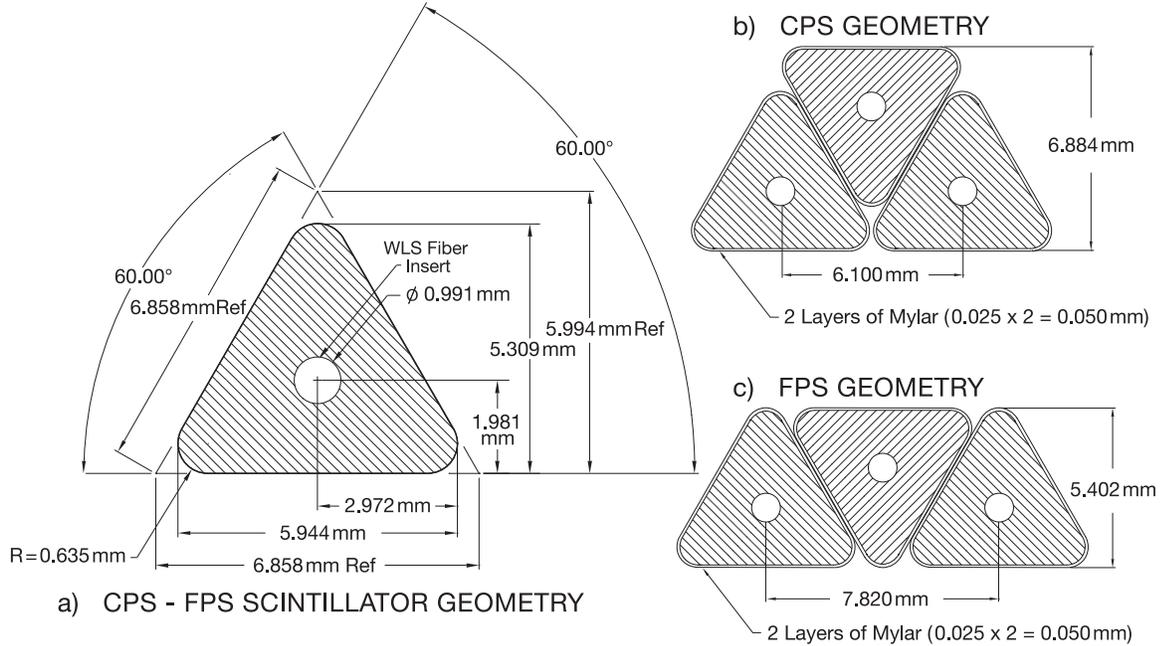}}
\caption{Cross section and layout geometry of the CPS and FPS scintillator 
strips.  The circles show the location of the embedded wavelength-shifting 
fibers.  Design dimensions are shown.}
\label{fig:ps-cross-section}
\end{figure}

Both preshower detectors are made from triangular strips of 
scintillator, as shown in Figure~\ref{fig:ps-cross-section}.  Since the
triangles are interleaved, there is no dead space between strips and most tracks
traverse more than one strip, allowing for strip-to-strip interpolations and
improved position measurement.  The strips are
made by extruding polystyrene plastic doped with 1\% p-terphenyl and
150~ppm diphenyl stilbene, with a light yield similar to that of commercial 
Bicron BC-404 scintillator.  Each scintillator strip is machine-wrapped in 
aluminized mylar for optical isolation, and the ends are painted 
white to enhance reflectivity.  The
packing density is different for the CPS and the FPS modules, resulting in
different layer thicknesses and strip pitches.  Because of the nesting process, 
which requires epoxying the strips together to form a layer, the  
measured pitch can differ by up to 20\% from the design dimensions shown in 
Figure~\ref{fig:ps-cross-section}.
After extrusion and wrapping, the triangular strips have a tendency to bow.  To
relieve stress in the plastic, making them easier to handle, the strips 
were slumped to the required shapes by heating to about $180^\circ$~F for two 
hours followed by gradual cooling at room temperature.

Embedded at the center of each triangular strip is a wavelength-shifting 
fiber that collects and carries the light to the end of the detector.  
The non-readout ends of the WLS fibers are diamond-polished and silvered.  At 
the readout end, fibers are grouped into bunches of sixteen and potted into 
connectors for transition to clear light-guide fibers.  Light is transmitted 
via the clear fibers to VLPC cassettes (Section~\ref{sec:VLPC}) for readout.  
Both the WLS and clear fibers are 835-$\mu$m-diameter Kuraray multiclad fibers.  

The preshower detectors share common elements with the central fiber tracker,
beginning with the waveguides and continuing through the entire readout 
electronics system.  The last elements which are unique to the preshower
detectors are the connections between the wavelength shifting fibers and the
waveguides.  The waveguides, VLPC cassettes, and readout electronics are 
described in Sections~\ref{sec:waveguides} -- 
\ref{sec:AFE}.  The calibration systems for the preshower detectors are also
similar to that of the CFT and are described in Section~\ref{sec:CFT-LED}.

\subsection{Central preshower detector}
\label{sec:cps}

The CPS consists of three concentric cylindrical layers of triangular
scintillator strips and is located in the nominal 5~cm gap between the 
solenoid and the central calorimeter.  Between the solenoid and the CPS is a  
lead radiator 7/32" thick (approximately 1 radiation length ($X_0$)) and 96"
long, covering $|\eta| < 1.31$.  The lead is covered by stainless steel skins 
1/32" thick and 103" long.  The solenoid itself is $0.9X_0$ thick, 
providing a total of about two radiation lengths of material for particles at 
normal incidence, increasing to about four radiation lengths at the largest
angles.

The three layers of scintillator are arranged in an axial-$u$-$v$
geometry, with a $u$ stereo angle of $23.774^\circ$ and a $v$ stereo angle of 
$24.016^\circ$.  Each layer contains 
1280 strips.  The WLS fibers are split at $z=0$ and read out from each end 
resulting in 2560 readout channels/layer.  
The geometry of the CPS axial layer matches that of the CFT for Level~1
(Section~\ref{sec:l1}) 
triggering purposes.  Each group of sixteen WLS fibers from the CPS axial layer
corresponds to one of the eighty CFT sectors in $\phi$.  As with the CFT, the
stereo layers are not used in the Level~1 trigger.  However, unlike the CFT, 
the stereo layers of the CPS are used in the Level~2 trigger.

\begin{figure}
\centerline{\includegraphics[width=3.in]{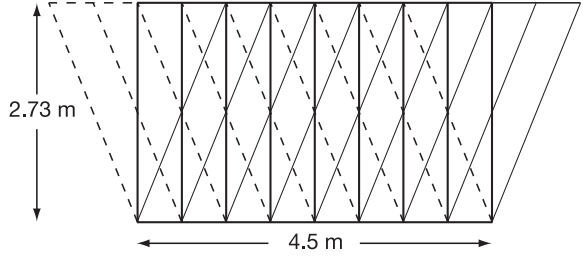}}
\caption{The CPS unwrapped in a plane.  Each rectangle (trapezoid) is one 
octant module of the axial (stereo) layer.  Note that the stereo octant edges 
are precisely aligned with the axial octant edges.}
\label{fig:cps-unwrapped}
\end{figure}

Each layer is made from eight octant modules.  The modules consist of two 
1/32"
stainless steel skins with the scintillator strips sandwiched in between.  The
ends of the stereo modules align to the ends of alternating axial modules, as
shown in Figure~\ref{fig:cps-unwrapped}.  The modules are attached directly to
the solenoid by bolts at each corner.  Eight $\quart$" pins at each end of the 
solenoid provide additional registration.  Connector blocks are spot welded
between the stainless steel skins.  The blocks provide structural integrity to
the end region of the modules and mounting locations for the WLS connector and
the cover for the light-guide connector.

\subsection{Forward preshower detector}
\label{sec:fps}

\begin{figure}
\centerline{\includegraphics[width=3.in]{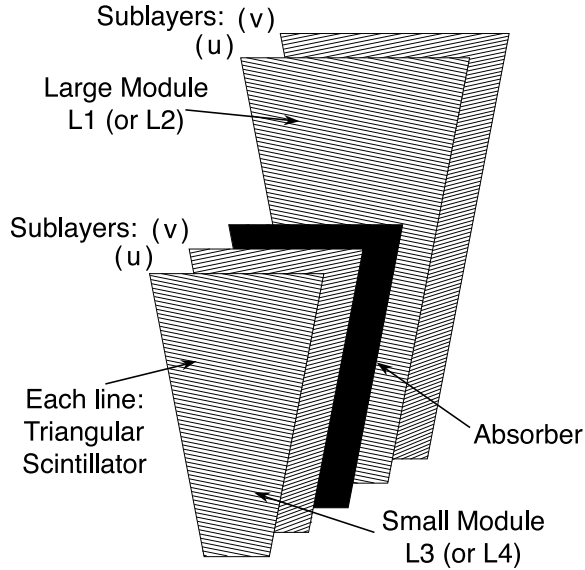}}
\caption{Complete $\phi$-segment of a FPS module showing the overlapping
$u-v$ MIP and shower layers, separated by a lead and stainless steel absorber.}
\label{fig:fps-module}
\end{figure}

The two FPS detectors (north and south) are mounted on the spherical heads of 
the end calorimeter cryostats, occupying the region between the luminosity 
monitor (Section~\ref{sec:lum-monitor}) at the inner edge and the 
intercryostat detectors (Section~\ref{sec:icd}) at the outer edge.  Each 
detector is made from two layers, at different $z$, of two planes of 
scintillator strips.  A $2X_0$-thick lead-stainless-steel absorber separates 
the two layers, as shown in Figure~\ref{fig:fps-module}.  The upstream layers 
(those nearest the interaction region) are known as the minimum 
ionizing particle, or MIP, layers while the downstream layers behind the 
absorber are called the shower layers.  Charged particles passing through 
the detector will register minimum ionizing signals in the MIP layer, allowing 
measurement of the location (in $\eta$, $\phi$, and $z$) of the track.  
Electrons will readily shower in the absorber, leading to a cluster of energy, 
typically on the order of three strips wide, in the shower layer that is 
spatially matched with the
MIP-layer signal.  Heavier charged particles are less likely to shower,
typically
producing a second MIP signal in the shower layer.  Photons will not generally
interact in the MIP layer, but will produce a shower signal in the shower 
layer.  The shower layers cover the region $1.5 < |\eta| < 2.5$ and the 
MIP layers and the absorber cover the region 
$1.65 < |\eta| < 2.5$.  The outer region of the FPS, $1.5 < |\eta| < 1.65$, 
lies in the shadow of the solenoidal magnet coil, which provides up to $3X_0$ 
of material in front of the FPS.  This amount of material readily induces 
showers that can be identified in the shower layers of the detector.  


\begin{figure}
\centerline{\includegraphics[width=3.in]{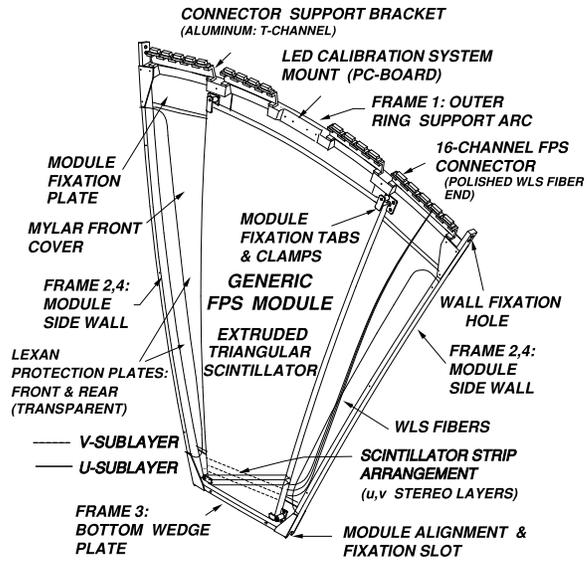}}
\caption{Three dimensional view of an FPS module mounted within its supporting
frame.}
\label{fig:fps-module-3d}
\end{figure}

\begin{figure}
\centerline{\includegraphics[width=5.in]{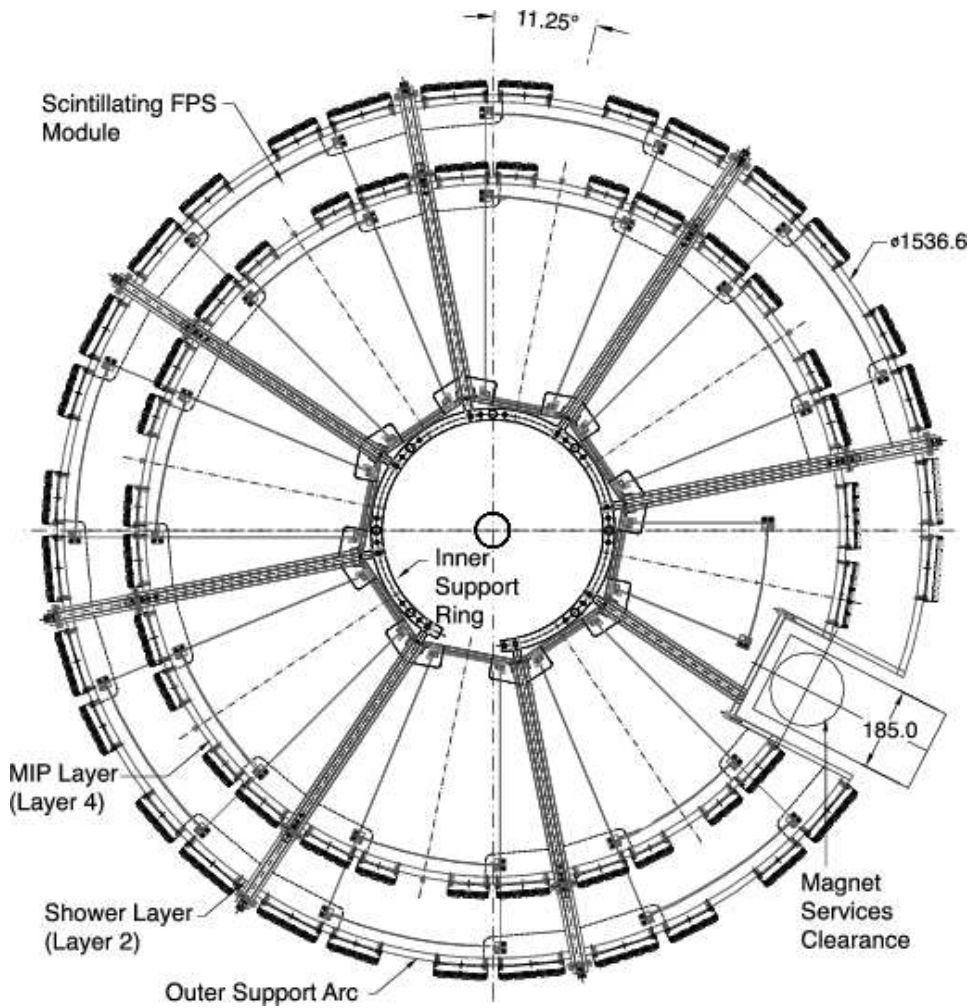}}
\caption{$r-\phi$ view of the north FPS detector.  For clarity, only layers 2
(shower) and 4 (MIP) are shown; layers 1 (shower) and 3 (MIP) are rotated by 
$22.5^\circ$ in $\phi$ with respect to these layers so their supports do not
overlap.  The south FPS detector 
is a mirror image of the north detector.  The cutback near four o'clock
accommodates the solenoidal magnet cryogenic services on the south side.}
\label{fig:fps-layer-2-4}
\end{figure}

\begin{figure}
\centerline{\includegraphics[width=6.in]{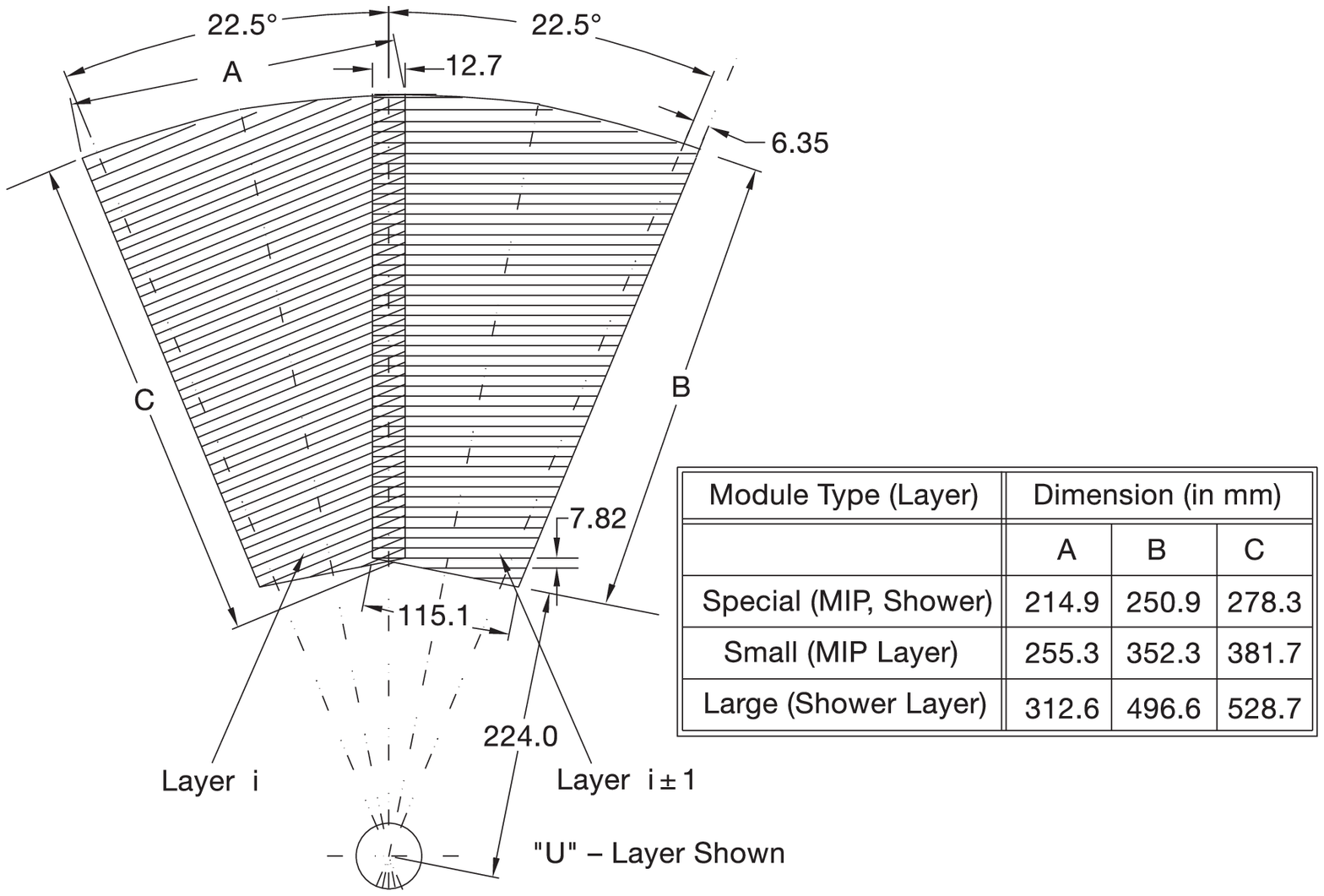}}
\caption{Relative orientation of two modules in successive FPS layers. 
The 12.7~mm overlap region is shown, along with the dimensions of the three 
types of modules.}
\label{fig:fps-module-overlap}
\end{figure}

Each FPS detector has four measuring planes: MIP $u$ and $v$
and shower $u$ and $v$.  Each measuring plane is constructed from two layers,
each containing eight 22.5$^\circ$ wedges (Figure~\ref{fig:fps-module-3d}) of 
active material separated by eight 
wedges of inactive material.  The active material consists of two sublayers of 
nested scintillator strips with a stereo angle of 22.5$^\circ$ with 
respect to one another.  The two layers are separated in $z$; the WLS
fibers are brought to the periphery of the detector in the space between the 
layers where connection blocks facilitate the transition to clear waveguides
take the light to the VLPC cassettes.  Each MIP plane has 206 
scintillator strips
which are perpendicular to an edge at constant $\phi$.  Each shower layer has
288 strips, also perpendicular to a constant $\phi$ edge.  Four special
wedges in the vicinity of the solenoid cryogenics service pipes are notched to
allow these to enter and have 142 strips per wedge in both the MIP and shower 
layers.  The presence of these special wedges reduces the coverage to 
$1.8 < |\eta| < 2.5$ in this area.  Successive FPS layers are offset to prevent 
projective cracks in $\phi$.  An $r-\phi$ view of two layers of the north 
FPS detector is shown in Figure~\ref{fig:fps-layer-2-4} and 
the three distinct types of
wedges and their dimensions within the successive FPS $\phi$, $z$ layers are
described in Figure~\ref{fig:fps-module-overlap}.


The absorber is divided into forty-eight wedge-shaped segments for 
easier handling, each subtending $7.5^\circ$ in $\phi$ and weighing 
approximately 5 lbs.  Each segment consists of two lead absorber
elements epoxied to each side of a 1/8"-thick stainless-steel 
plate.  Each steel plate is approximately 0.5" longer
radially than the lead plates, to allow connection to inner and outer support
rings that are nested within the overall FPS layers.  Similarly to the modules
occupying the MIP and shower layers, the absorber segments are
individually bent in three dimensions to conform to the spherical geometry of
the end calorimeters.  The $\phi$ edges of each segment
are beveled to allow nesting of adjacent absorber segments with minimal 
non-projective gaps between them.  The total thickness of a
lead-stainless-steel-lead segment is 11 mm, or $2X_0$.

\FloatBarrier

\section{Calorimetry}
\label{sec:calorimeters}

The D\O\ calorimeter system consists of three sampling calorimeters
(primarily uranium/liquid-argon) and an intercryostat detector.

\subsection{Calorimeters}

\begin{figure}
\centerline{\includegraphics[width=6.in]{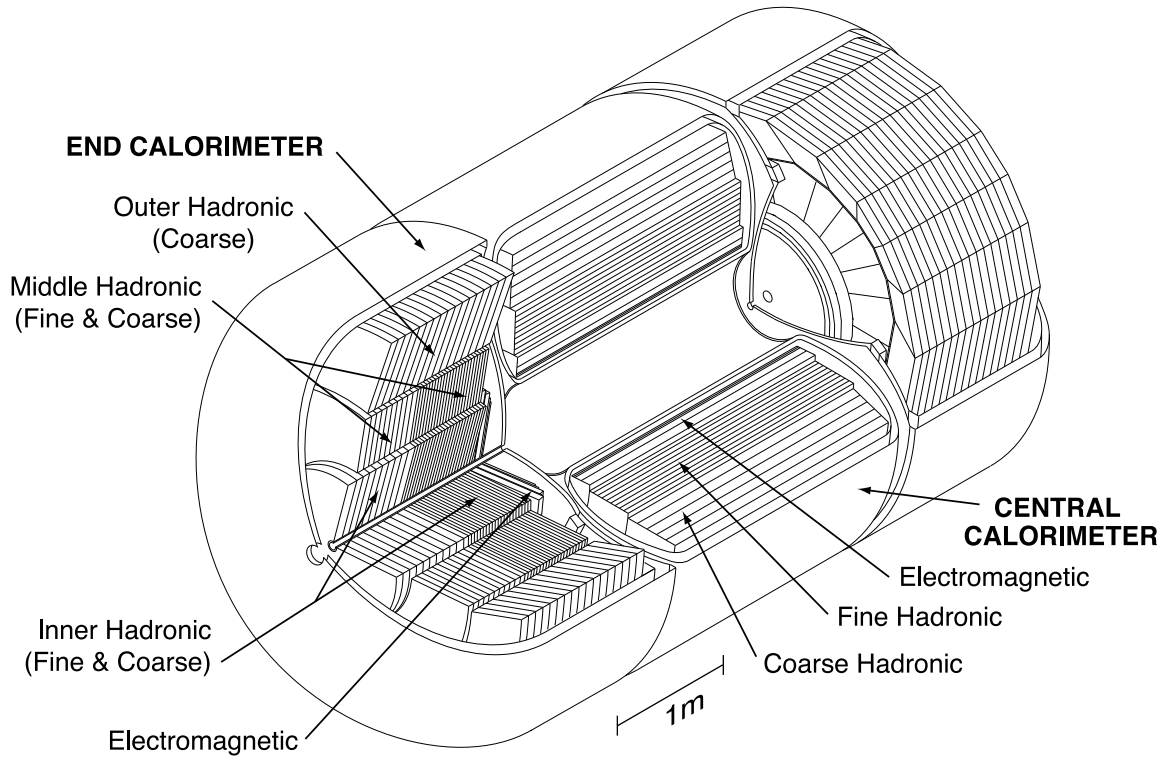}}
\caption{Isometric view of the central and two end calorimeters.}
\label{fig:cal-isometric}
\end{figure}

The calorimeters were designed to provide energy measurements for electrons, 
photons, and jets in the absence of a central magnetic field (as was the case 
during Run I of the Tevatron), as well as assist in identification of 
electrons, photons, jets, and muons and measure the transverse energy balance 
in events.  The calorimeters themselves are unchanged from Run I and are
described in detail in Ref.~\cite{d0_nim}.  They are illustrated in 
Figure~\ref{fig:cal-isometric}.

\begin{figure}
\centerline{\includegraphics[width=6.in]{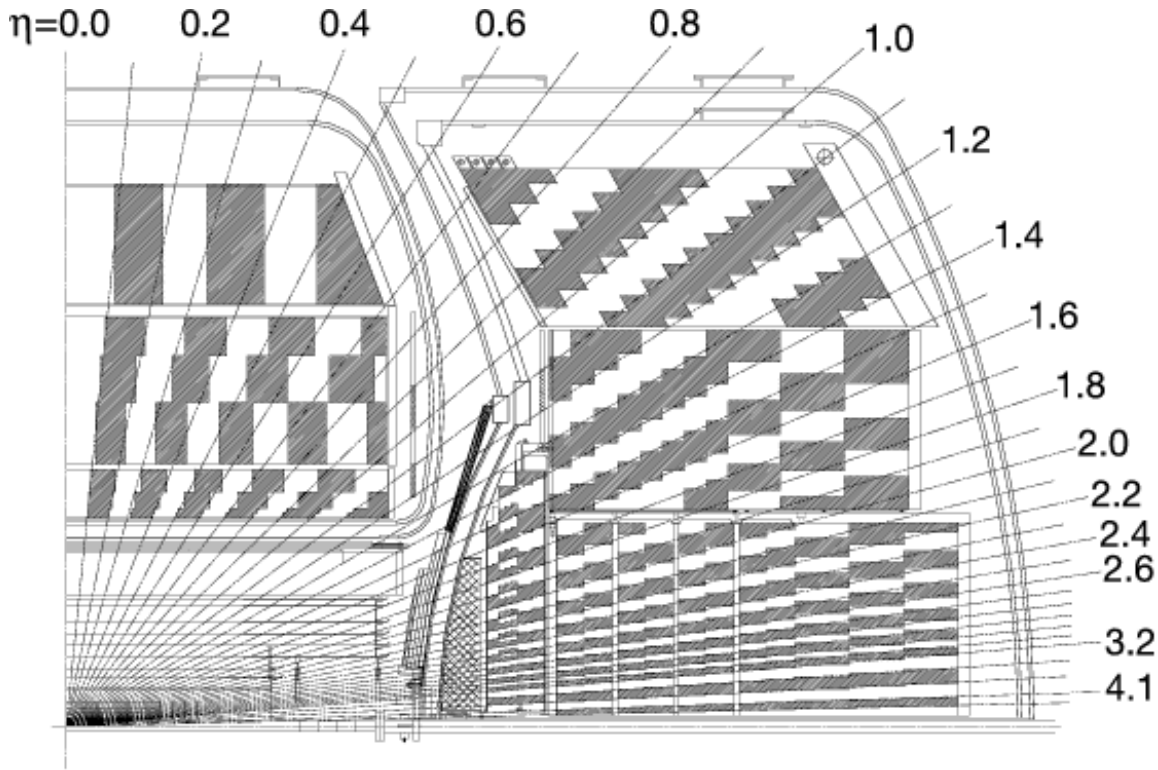}}
\caption{Schematic view of a portion of the D\O\ calorimeters showing the
transverse and longitudinal segmentation pattern.  The shading pattern 
indicates groups of cells ganged together for signal readout.  The rays 
indicate pseudorapidity intervals from the center of the detector.}
\label{fig:cal-side-eta}
\end{figure}

As shown in Figure~\ref{fig:cal-side-eta}, the central calorimeter 
(CC) covers $|\eta| \lsim 1$ and the two end calorimeters, ECN (north) and ECS
(south), extend coverage to $|\eta| \approx 4$.  Each calorimeter contains an
electromagnetic section closest to the interaction region followed by fine and
coarse hadronic sections. 
The active medium for the calorimeters is liquid argon and each of the
three calorimeters (CC, ECN, and ECS) is located within its own cryostat that
maintains the detector temperature at approximately 90~K.  Different absorber 
plates are 
used in different locations.  The electromagnetic sections (EM) use thin plates
(3 or 4~mm in the CC and EC, respectively), made from nearly pure depleted 
uranium.  The fine hadronic 
sections are made from 6-mm-thick uranium-niobium (2\%) alloy.  The coarse 
hadronic modules contain relatively thick (46.5 mm) plates of copper 
(in the CC) or stainless steel (EC).

\begin{figure}
\centerline{\includegraphics[width=3.in]{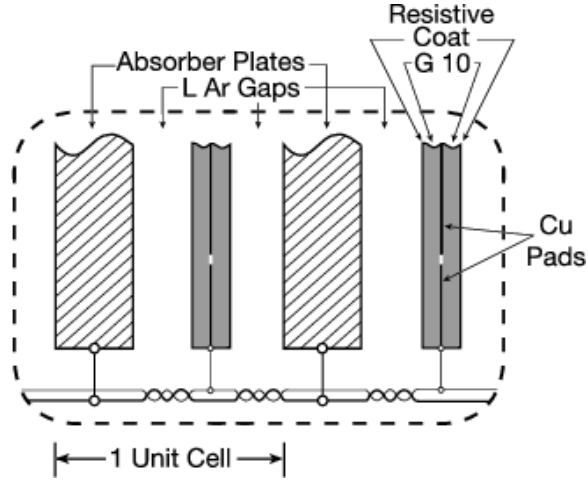}}
\caption{Schematic view of the liquid argon gap and signal board unit cell for
the calorimeter.}
\label{fig:cal-cell}
\end{figure}

A typical calorimeter cell is shown in Figure~\ref{fig:cal-cell}.  The electric
field is established by grounding the metal absorber plates and connecting the
resistive surfaces of the signal boards to positive high voltage (typically
2.0 kV).  The electron drift time across the 2.3~mm liquid argon gap is 
approximately 450~ns.  Signal boards for all but the 
EM and small-angle hadronic modules in the EC are made from two 0.5~mm G-10 
sheets.  The surfaces of the sheets facing the liquid argon gap are coated with 
carbon-loaded epoxy \cite{cal-epoxy} with a typical sheet resistivity of 
40~M$\Omega/\Box$; these surfaces serve as the high voltage electrodes for
the gap.  For one sheet, the other surface is bare G-10; the facing inner
surface of the second sheet, 
originally copper-coated, is milled into the pattern necessary for segmented
readout.  Several such pads at approximately the same $\eta$ and $\phi$ are
ganged together in depth to form a readout cell. 

The two smallest-angle modules (EM and hadronic) in each EC have the added 
problem that even small gaps between neighboring azimuthal sectors would give 
undesirable dead regions.  Thus, the signal boards for these monolithic 
modules are made from multilayer printed circuit boards.  The outer surfaces 
are coated with the same resistive epoxy as for the other signal boards.  
Etched pads on an interior surface give the desired segmentation.  Signal 
traces on another interior surface bring the signals to the outer periphery.  
The pad and trace layers are connected by plated-through holes.  The signals 
from these multilayer boards in the EM and small-angle hadronic modules are 
ganged together along the depth of the modules.
 
Calorimeter readout cells form pseudo-projective towers as shown in
Figure~\ref{fig:cal-side-eta}, with each tower subdivided in depth.  
We use the term ``pseudo-projective'' because the centers of cells of
increasing shower depth lie on rays projecting from the center of the
interaction region, but the cell boundaries are aligned perpendicular to the
absorber plates. 

There are four separate depth layers for the EM modules in the
CC and EC.  In the CC, the layers are approximately 1.4, 2.0, 6.8 and 9.8$X_0$
thick.  In the EC, they are approximately 1.6, 2.6, 7.9 and 9.3$X_0$ thick.  
The values given for the first layers include all material in the calorimeters
themselves from the outer warm walls to the first active liquid argon gap.  
The detector components between the interaction region and the first active gap
in the CC at $\eta = 0$ provide about 4.0$X_0$ of material; those between the 
interaction region and the first active gaps of the ECs at 
$\eta = 2$ are 4.4$X_0$ thick.

In the CC, the fine hadronic modules have three longitudinal gangings of
approximately 1.3, 1.0, and 0.76$\lambda_A$.  The single coarse hadronic module
has a thickness of about 3.2$\lambda_A$.  The two EC inner hadronic modules
(Figure~\ref{fig:cal-isometric}) are cylindrical, with inner and outer radii of
3.92 and 86.4~cm.  The fine hadronic portion consists of four readout cells,
each $1.1\lambda_A$ thick.  The coarse hadronic portion has a single readout
cell $4.1\lambda_A$ thick.  Each of the EC middle hadronic modules has four
fine hadronic readout cells of about $0.9\lambda_A$ each and a single coarse
hadronic section of $4.4\lambda_A$.  The outer hadronic modules of the ECs are
made from stainless steel plates inclined at an angle of about $60^\circ$ with
respect to the beam axis (see Figure~\ref{fig:cal-side-eta}).  The maximum
thickness is $6.0\lambda_A$.  Each layer is offset from the others 
to provide hermeticity.

\begin{figure}
\centerline{\includegraphics[width=4.in]{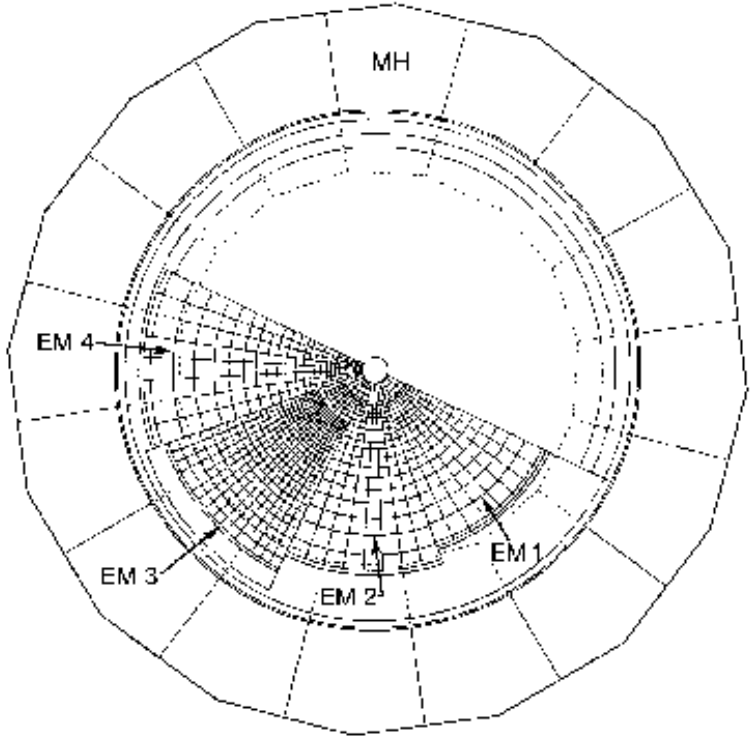}}
\caption{Layout of the EC electromagnetic readout cells for the four
longitudinal EM layers.  EM1 is closest to the interaction region. 
MH indicates the EC middle hadronic cells.}
\label{fig:ecem-segmentation}
\end{figure}

The transverse sizes of the readout cells are comparable to the transverse 
sizes of showers: 1--2~cm for EM showers and about 10~cm for hadronic showers. 
Towers in both EM and hadronic modules are $\Delta\eta = 0.1$ and 
$\Delta\phi = 2\pi/64 \approx 0.1$.  The third layer of the EM modules, located
at the EM shower maximum, is segmented twice as finely in both $\eta$ and 
$\phi$ to allow more precise location of EM shower centroids. 
As can been seen in Figure~\ref{fig:ecem-segmentation}, cell sizes increase
in $\eta$ and $\phi$ at larger $\eta$ to avoid very small cells.

\subsubsection{Calorimeter electronics}

\begin{figure}
\centerline{\includegraphics[width=6.in]{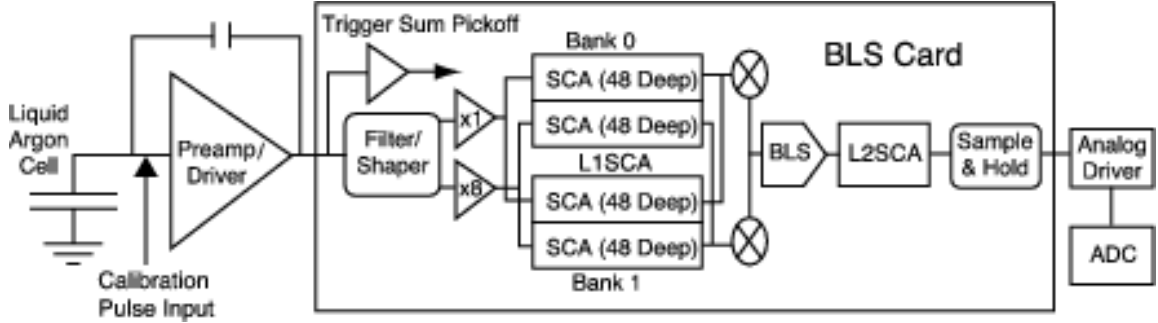}}
\caption{Readout chain of the calorimeter in Run~II indicating the
three major components: preamplifiers, baseline subtractor and storage
circuitry (BLS), and the ADCs.}
\label{fig:chain}
\end{figure}

Figure~\ref{fig:chain} illustrates the main components
in the calorimeter readout chain. 
There are 55,296 calorimeter electronics channels to be read out; 47,032 
correspond to channels connected to physical readout modules in the cryostats. 
The remaining electronics channels are not connected to the detector.  (The
ADC cards are identical and contain enough channels to read out the most
populated regions of the detector.)
The readout is accomplished in three principal stages. In
the first stage, signals from the detector
are transported to charge preamplifiers located on the cryostats via
low impedance coaxial cable. In the second stage, signals from the
preamplifiers are transported on twisted-pair cables to the analog signal
shaping and storage circuits on baseline subtractor (BLS) boards.
The precision signals from the BLSs are transmitted on an analog bus and 
driven by analog drivers over 130~m of twisted-pair cable to ADCs.  These 
signals then enter the data acquisition system for the Level~3
trigger decision (Section~\ref{sec:l3trigger}) and storage to tape.  The 
preamplifiers and BLSs are completely new for Run~II, and were necessary to 
accomodate the significant reduction in the Tevatron's bunch spacing.

\paragraph{Front-end electronics}
Signals from the calorimeter cells are transported
on 30~$\Omega$ coaxial cables (with a typical length of 10~m) 
to a feedthrough port (the interface
between the cold region and the warm region) on the cryostat and then
on to the preamplifiers which are mounted as close as possible to the
feedthroughs at the top of the cryostat.  The cables from the feedthrough port 
to the preamplifier were replaced for Run~II to provide better impedance 
matching to the
preamplifier input ($\approx 30~\Omega$) and to equalize the lengths to provide
better timing characteristics by minimizing the spread of signal arrival times.
Electron drift time across the 2.3~mm liquid-argon gap remains approximately 
450~ns at 2.0~kV for Run~II, which provides a challenge for 
signal charge integration with beam crossings occurring every 396~ns.  The
calorimeter electronics were designed to maintain the good signal-to-noise
ratio obtained during Run~I in the expected high instantaneous luminosity 
environment of the original Tevatron Run~II design, with a minimum bunch 
crossing time of 132~ns.

The preamplifiers are individually packaged transimpedance
(i.e.\ charge to voltage) hybrid amplifiers on ceramic substrates. Forty-eight
individual preamplifier hybrids are mounted on a motherboard, with ninety-six
motherboards housed in a single preamplifier box. Twelve such boxes,
mounted on top of the calorimeter cryostats, make up the full calorimeter 
preamplifier system.  Since access to these preamplifier boxes
is very limited, the power supplies for the preamplifiers are
redundant with each preamplifier box having two supplies, a primary and a
backup that can be switched in to replace the primary supply.  Low noise
commercial switching power supplies are used to provide the necessary
power.

The preamplifier shaping networks are modified for cells of different
capacitance to give approximately the same output waveform for all cells.
Given the large range of detector capacitances at the input to the
preamplifiers, there are fourteen species of preamplifier (plus one for
readout of the intercryostat detector (ICD), Section~\ref{sec:icd}) which 
provide similarly shaped output signals to the BLS boards.  This is important 
to maintain good timing for the peak-sampling circuit. 
The characteristics of the preamplifier species are
shown in Table~\ref{tab:cal_preamp}.
The ICD feedback capacitor is 22~pF to reduce the gain of the output
signals and thus preserve their dynamic range.  

Dual parallel input JFETs help maintain low noise performance. Two
output driver stages provide the capability to drive a terminated
115~$\Omega$ twisted-pair cable. The power requirement of a single
preamplifier is 280~mW. The noise performance can be evaluated from
the width of the pedestal distribution, which is a function of the
detector input capacitance.

\begin{table}
\begin{center}
\caption{Characteristics of the different preamplifier species used for the
calorimeter and the ICD readout.  Since the cell capacitances depend on the 
size of the cell, with larger cells having higher capacitance, preamplifiers 
A -- D are used in various locations in both the central and end calorimeters.
FH$n$ stands for fine hadronic, layer $n$; 
CH$n$ is similar for the coarse hadronic layers; 
and MG for the CC and EC massless gaps (Section~\ref{sec:icd}).}
\label{tab:cal_preamp}
\vskip 0.1 in
\begin{tabular}{ p{60pt} p{60pt} p{95pt} p{50pt} p{33pt} p{42pt} }
\hline
Preamplifier \par species & $<$Cell~cap.$>$ \par (nF) & Layer \par read out & 
Feedback \par cap.\ (pF) & RC \par (ns) & Total \par preamp. \par channels  \\
\hline
A  & 0.26--0.56 & EM1,2, MG \par EC CH1,2 & 5 & 0 & 13376 \tabularnewline
B  & 1.1--1.5   & MG \par EC FH1,2,4, CH1 & 5  & 26 &  2240 \tabularnewline
C  & 1.8--2.6   & CC CH1 \par EC FH1--4, CH1--3 & 5 & 53 & 11008 \tabularnewline
D  & 3.4--4.6   & CC FH1--3, CH1 \par EC FH1--4, CH1--3 & 5  & 109  & 8912 
\tabularnewline

E      & 0.36--0.44 & CC EM3               & 10 & 0       & 9920 \tabularnewline
F      & 0.72--1.04 & EC EM3,4             & 10 & 14      & 7712 \tabularnewline
G      & 1.3--1.7   & CC EM4 \par EC EM3,4 & 10 & 32      & 3232 \tabularnewline
Ha--Hg & 2--4       & EC EM3,4             & 10 & 47--110 & 896 \tabularnewline
I      & --         & ICD                  & 22 & 0       & 384  \tabularnewline
\hline
\end{tabular}
\end{center}
\end{table}

The preamplifier motherboard is an eight-layer printed circuit board, with
ground or power planes of solid copper separating planes containing
signal traces to minimize noise pickup and cross-talk.  The
motherboard houses the precision resistors (0.1\% 10~k$\Omega$ or 
20~k$\Omega$, depending on the preamplifier feedback capacitor) for the 
calibration voltage pulse.  A single input line pulses six
preamplifiers at once using a low capacitance trace. 

All of the electronics is located in an area where there can be a
residual magnetic field of a few hundred gauss, and so the switching 
supplies are magnetically shielded.  All other devices (including
cooling fans) have been verified to function in the residual field.
New heat exchange systems were installed in the existing preamplifier
boxes to deal with increased power dissipation.

\paragraph{Signal shaping and trigger}
The single-ended preamplifier signals are routed from the calorimeter
over approximately 25~m of twisted-pair cable
to the BLSs located below the cryostats, where access is significantly easier.
The BLSs use switched capacitor
arrays (SCAs) as analog storage devices to hold the signal for about
4~$\mu$s until the Level~1 trigger decision is made, and for an additional
2~ms (on average, with an allowed maximum of 25~ms) until the Level~2 trigger 
decision is made.  They also provide baseline
subtraction to remove any low frequency noise or pileup from the
signal. In addition, faster shaped analog sums of the signals are
picked off to provide prompt inputs to the calorimeter trigger system
for both the Level~1 and Level~2 calorimeter trigger decisions 
(Sections~\ref{sec:l1cal} and \ref{sec:l2cal}). 

To minimize the effects of pile-up (more than one event in the detector 
due to multiple interactions in a single beam crossing or to interactions
in multiple beam crossings), only two-thirds of the charge
collected by the preamplifier circuit, corresponding to the first
$\approx 260$~ns of signal collection from the gap, is used in
the shaper circuit.  The preamplifier output is an integral of the
detector signal characterized by a rise time of about 450~ns
and a recovery time of 15~$\mu$s.  Shaped signals are sampled
every 132~ns.
The shaper circuit produces a unipolar signal with a peak
at about 320~ns and a return to zero after $\approx 1.2~\mu$s.
The shaped signal is sampled at $t_0 = 320$~ns, close to the
peak.  To subtract the baseline, the signal three samples earlier
($t_0 - 396$~ns) is subtracted by the BLS circuitry.

The calorimeter Level~1 and Level~2 triggers are based on the energy
measured in trigger towers of size $0.2
\times 0.2$ in $\Delta\eta \times \Delta\phi$, which is obtained by
making appropriate sums (via resistor packs on the BLS boards) of the fast
pickoffs at the shaper inputs.  The resistor packs have been tuned
based on a sampling weight optimization study which sought to
maximize electron resolution first, and jet resolution second, for the channels
included in the trigger (the coarse hadronic sections do not contribute to the
trigger).

\paragraph{Digitization}
A BLS board processes signals from the forty-eight channels from four 
pseudo-projective calorimeter towers. Each tower corresponds to up to 
twelve preamplifier signals.
There are signal shapers for each channel on the BLS motherboard, and
trigger pick-off and summation circuits tap the preamplifier signal
prior to the shaper circuitry.  The shaped preamplifier signals are
fed to daughterboards, one per tower, each of which holds five SCA
chips.  The SCAs contain an array of 48 capacitors to pipeline the
calorimeter signals.  The first and last buffers are not used in the
readout to avoid edge effects in the chips.  The SCA is not designed
for simultaneous read/write operations, therefore two SCA banks are
alternately employed to provide the capability to write and read 
the integrated charges.  This scheme provides the
4.2~$\mu$s buffering necessary prior to the arrival of the Level~1
(Section~\ref{sec:l1}) 
trigger decision.  There are also two gain paths ($\times 1$ and
$\times 8$) to extend the ADC readout dynamic range, thus four of the SCAs are
used to store the signals for the twelve channels on a daughtercard
until the Level 1 trigger decision has been made.  Once a positive
Level 1 decision is received, baseline subtractor circuitry on each
daughterboard decides channel-by-channel which gain path to use and
subtracts the stored baseline from
the peak signal.  It then stores the result in the Level~2 SCA that
buffers the data until a Level~2 (Section~\ref{sec:l2}) trigger decision has 
been made.  Once a 
Level~2 trigger accept is issued, the data are transferred from the Level 2
SCA to a sample-and-hold circuit on the daughterboard and an analog
switch on the BLS motherboard buses the data on the BLS backplane to
analog drivers which transfer the signal up to the ADC across 130~m of
twisted-pair cable.  The gain
information is sent simultaneously on separate digital control cables.

The ADC successive approximation digitizers, reused from Run~I, have
a 12-bit dynamic range, but the low and high gain paths for each
readout channel effectively provide a 15-bit dynamic range.  This matches the
measured accuracy of the SCA.  The readout system effectively introduces no
deadtime up to a Level~1 trigger accept rate of 10~kHz, assuming
no more than one crossing per superbunch\footnote{A superbunch is a group of 
twelve $p$ or \pbar\ bunches, where each bunch is separated by 396~ns; the 
Tevatron contains three evenly-spaced superbunches of each particle.} is read 
out.  

A master control board synchronizes twelve independent controllers
(one for each readout quadrant of the calorimeter) in a shared VME crate.  
These controllers
provide the timing and control signals that handle the SCA requirements and
interface to the Level~1 and Level~2 trigger systems.  Each control
board houses three Altera FPGA (field programmable gate array) 
chips~\cite{FLEX10K} (10K series, 208-pin packages).
Three FIFO pipelines buffer up to about forty events awaiting readout. 
Timers on these event buffers ensure that
``stale'' data are appropriately flagged to the data acquisition
system.  Other readout errors are also flagged in the readout.  The
control boards permit the readout VME crate to be run in numerous
diagnostic and calibration modes.

\paragraph{Calibration system}
\label{sec:cal-calibration}
The calorimeter calibration system consists of twelve identical units
for the calorimeter and one slightly modified unit for the ICD 
(Section~\ref{sec:icd}). Each unit is composed of one pulser board, its power
supply located in the BLS racks on the platform underneath the detector, and 
six active fanout boards housed inside the preamplifier boxes mounted on the
cryostats.  The pulser boards are controlled via a VME
I/O register to set the amplitude and the timing of the
calibration signal and enable the channels to be pulsed.
A DAC sets the current delivered by the pulser boards using 17 bits, with
an avarage current/DAC of 0.825~$\mu$A/DAC, and the timings are
controlled through a 8-bit delay line in steps of 2~ns.

\begin{figure}
\centerline{\includegraphics[width=3.in]{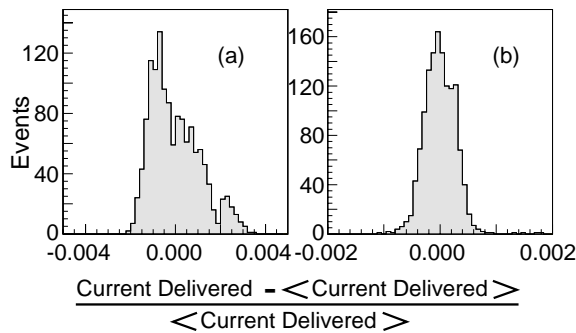}}
\caption{The mean current/DAC delivered by the pulser boards is 
825~mA/DAC with a spread of (a) $\pm0.2$\% for all 15 pulser boards
produced and (b) $\pm0.1$\% within one board.}
\label{fig:uni}
\end{figure}

Both the pulser board and the active fanout have been shown
to provide a pulser signal with a linearity at the per mil level, and
all the currents delivered are uniform within 0.2\% between all boards and 
0.1\% within a board (Figure~\ref{fig:uni}).
All of the pulse shapes have been measured to 
estimate systematic effects on the signal amplitude, the timing, and the
charge injected.

\paragraph{Gain determination}
For all calorimeter channels, the gain calibration factors are
determined from deviations of the slope ADC/DAC from its ideal
value of 0.25 ADC counts per DAC unit.  The
dispersion of the coefficients for electromagnetic (hadronic) channels
is about 5\% (10\%).  Systematic shifts of the slope values can be
observed for different preamplifier types.

Part of these differences is due to the injection of the calibration
signal at the preamplifier input, outside of the cryostat. A fraction
of the pulser signal travels down the signal cable and is reflected
from the calorimeter cell, depending on the capacitance of the cell,
and therefore has a different measured shape.  The effects are largest
for hadronic cells with high capacitance, where differences in the
timing can produce large tails in the distribution of the calibration
coefficients.  Corrections for these effects have been derived
for the calibration coefficients, and the delay settings of the pulsers
have been tuned to maximize the response in the electromagnetic layers.

\begin{figure}
\centerline{\includegraphics[width=3.in]{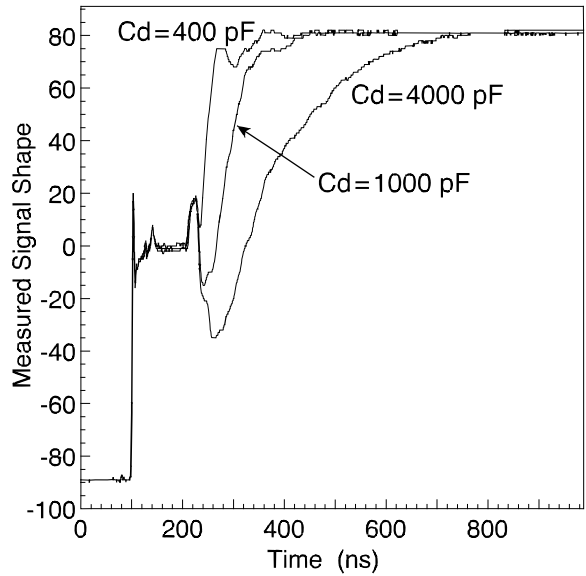}}
\caption{Response to reflection measurement for three different channels.
The vertical voltage scale is in arbitrary units.  Cd is the detector cell
capacitance.}
\label{fig:reflection}
\end{figure}

Models of the electronics chain have been made to evaluate the
differences between the electronics response to a calibration signal
and a detector signal. To render these models realistic, all stable
parameters of the signal path from the detector to the preamplifier
input have been determined from reflection measurements. The reflected
response (in arbitrary voltage units) to a step function is shown in
Figure~\ref{fig:reflection} for three different channel
types. Quantitatively the values for the cable resistivity outside and
inside the cryostat, the inductance of the feedthrough and the
signal strips, as well as the capacitance of each cell, have been
determined and used in the simulation model.  Correction factors to the 
calibration coefficients indicate these effects are 
below the percent level for electromagnetic channels, when sensed
close to the signal maximum.

\subsubsection{Liquid argon monitoring}

The purity of the liquid argon is critical to the detector performance
as electronegative contaminants (particularly oxygen) can combine
with electrons traversing the gap and can severely impact the energy
measurement.  The liquid argon was recovered from Run~I and, before
refilling the cryostats for Run~II, the purity was remeasured \cite{ar-purity} 
with an external argon test cell equipped with $^{241}$Am and $^{106}$Ru
radioactive sources.  The $^{241}$Am is an alpha source (5~MeV 
$\alpha$-particles, 
432~yr half-life) that was used for purity measurements during Run~I; 
the $^{106}$Ru is a new beta source (maximum 3.5~MeV $\beta$-particles, 
$\approx$1~yr half-life) with activity of 30--40~kBq.  
Calibration of the test cell was
performed by injecting controlled amounts of $\rm{O}_2$ into pure argon.
The contamination of the liquid argon was measured to be less than
$0.30 \pm 0.12$~ppm for all three calorimeter cryostats.  A 5\% signal loss is
expected for a contamination level of 1~ppm.

Radioactive sources are used to monitor the contamination levels {\it in situ}.
Each of the three cryostats is equipped with four
$^{241}$Am sources with an activity of 0.1~$\mu$Ci and four $^{106}\rm{Ru}$ 
sources.  Three of
the beta sources in each cryostat now have very low levels of activity
($\lsim 1$~Bq), fifteen years after the initial detector construction.  One
stronger source has an activity of about 4~Bq.  The charge
liberated in the liquid argon gap by the alpha sources is about 4~fC
(about 25,000 electrons) and about half as much for the beta source.  The
trigger rate is about 500~Hz for the alpha source and 0.3~Hz
for the strong beta source.  The argon purity of the three calorimeters has 
been extremely stable over time.

A new readout board with a preamplifier, dual operational-amplifier, and
differential driver sends the amplified signals via a shielded
twisted-pair cable to a differential receiver board to be digitized by
a 12-bit ADC using a Xilinx FPGA (Spartan XCS40XL)~\cite{Xilinx}.  Histograms 
can be accumulated very rapidly (few kHz) on the receiver board which are
read out via CAN-bus to a PC running National Instruments LabView
\cite{labview}.
This design is a modification of the liquid argon monitoring system
\cite{atlas-lar} for ATLAS and sensitive to less than 1~ppm of oxygen
equivalent contamination.  

The signal response is also a function of the liquid argon
temperature so this is monitored as well.  The temperature of
the liquid argon is maintained at $90.7 \pm 0.1$~K.

\subsection{Intercryostat detector and massless gaps}
\label{sec:icd}

Since the calorimeter system is contained in three separate cryostats, 
it provides incomplete coverage in the pseudorapidity region
$0.8<\mid\eta\mid<1.4$, as can be seen in Figure~\ref{fig:cal-side-eta}.
In addition, there is substantial unsampled material in this
region, degrading the energy resolution.  To address this
problem, additional layers of sampling have been added.  
Within the central and end cryostats are single-cell structures called 
massless gaps --- calorimeter readout cells in front of the first layer of 
uranium.  The ICD provides scintillator sampling that is attached to the 
exterior surfaces of the end cryostats.  It covers the region 
$1.1<\mid\eta\mid<1.4$; its position with respect to the rest of the detector 
can be seen in Figure~\ref{fig:tracker}.  The massless gaps are those
used in Run~I, while the ICD is a replacement for a similar detector in the 
same location (the Run~I detector had approximately twice the coverage in
$\eta$; the space was needed for SMT and CFT cabling in Run~II).

\begin{figure}
\centerline{\includegraphics[width=3.5in]{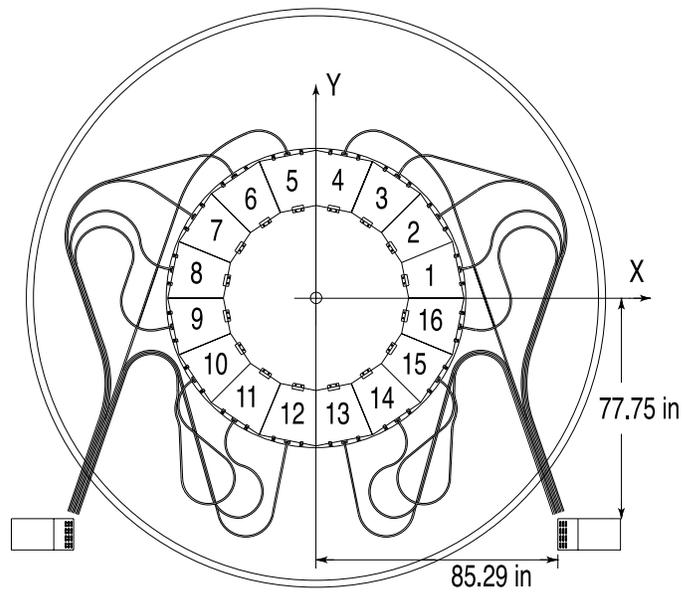}}
\caption{The arrangement of the ICD tiles on the endcap cryostats.  
The rectangles represent the iron block and fiber backplane assemblies in 
which the ICD electronics and PMTs are installed.  The beamline is 
perpendicular to the page.}
\label{fig:icd-tile-array}
\end{figure}

\begin{figure}
\centerline{\includegraphics[angle=0,width=4.0in]{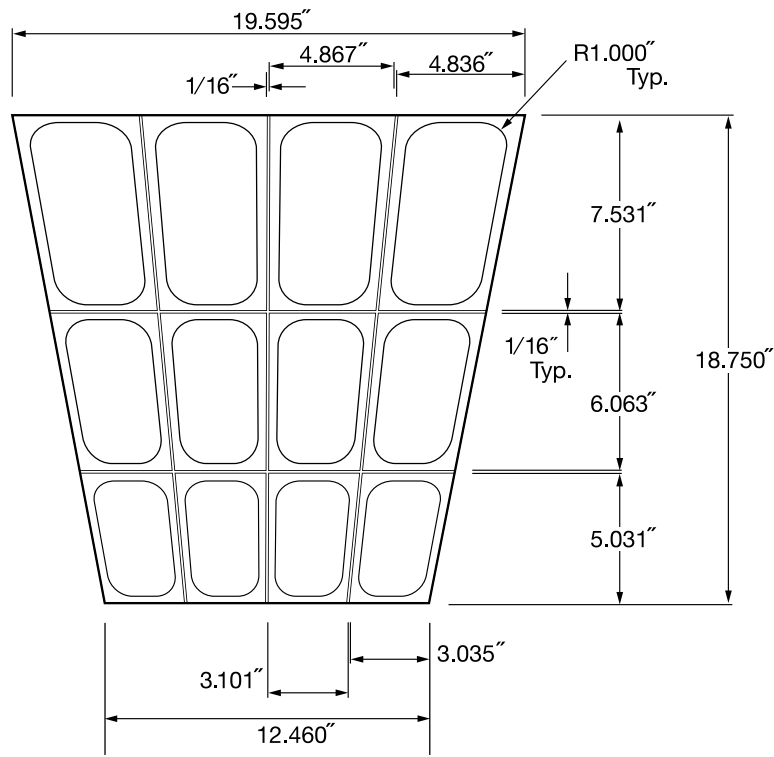}}
\caption{Each ICD tile is subdivided into twelve straight-edged trapezoidal 
subtiles.  WLS fibers for readout are embedded into each subtile along the
curved trapezoids.}
\label{fig:icd-tile-detail}
\end{figure}

The ICD, shown in Figure~\ref{fig:icd-tile-array}, is a series of 0.5"-thick 
scintillating tiles (Bicron BC-400 \cite{bicron}) enclosed in light-tight 
aluminum boxes.  Each tile covers an area of 
$\Delta\eta \times \Delta\phi \approx 0.3 \times 0.4$ and is
divided into twelve subtiles, as shown in Figure~\ref{fig:icd-tile-detail},
each covering $\Delta\eta \times \Delta\phi \approx 0.1 \times 0.1$.
Because of the cryogenic services for the solenoid, one half of a tile is 
missing at the south end of the detector, giving a total of 378 channels.

The subtiles are optically isolated from one another by grooves filled with 
epoxy mixed with highly reflective pigment.  Each subtile is read out via 
two 0.9-mm-diameter wavelength shifting fibers (Bicron BCF-91A) embedded in a 
groove cut to a depth of 3.5~mm near the outside edge of each subtile.  A third
fiber, used for calibration, is also located in the groove. 
The wavelength-shifting fibers are mated to 1.0~mm clear optical fibers 
(Bicron BCF-98) at the outer radius of the ICD tile enclosure.
The clear fibers terminate at a Hamamatsu R647 \cite{hamamatsu}
photomultiplier tube (PMT).  

The ICD electronics are located in a low-magnetic-field region 
away from the tiles and are contained in a drawer system.
An ICD drawer houses PMTs, PMT electronics, and preamplifiers for
six channels of readout.
Clear fibers from the ICD tiles are mated to clear fibers
contained in a light-tight enclosure called the ``fiber
backplane.'' Between the fiber backplane and the crate containing
the drawers is a block of iron, with holes drilled to accept the
PMTs.  Fibers from the fiber backplane terminate in ``cookies'' that
hold the ends of the fibers in the correct location with respect to
the photocathodes of the PMTs.  When a drawer is closed, six PMTs are
aligned with six fiber bundles (two per channel plus one for a
calibration signal in each bundle) from the backplane. The PMTs are 
spring-loaded to ensure good contact with the fibers. The iron block, in
combination with tubular magnetic shields, protects the PMTs from
fringe magnetic fields.  O-rings at the base of the PMTs and the drawers
themselves provide a dark environment for the PMTs.

The signal electronics are designed to be compatible with the calorimeter 
BLS/ADC system and electronics calibration. The ICD uses a
modified version of the calorimeter preamplifiers,
designed to stretch the PMT signal into a signal similar to that
of the liquid argon readout. A modified form of the calorimeter electronic
pulser system is used for electronics commissioning. The responses of the 
PMTs are monitored
using an LED calibration system, similar in design to that 
used for the muon trigger scintillation counters
(Section~\ref{sec:muon-led-monitoring}).  An LED pulser is housed in 
each of the fiber backplanes, and the amount of light is software-controlled.

\begin{figure}
\centerline{\includegraphics[width=3.0in]{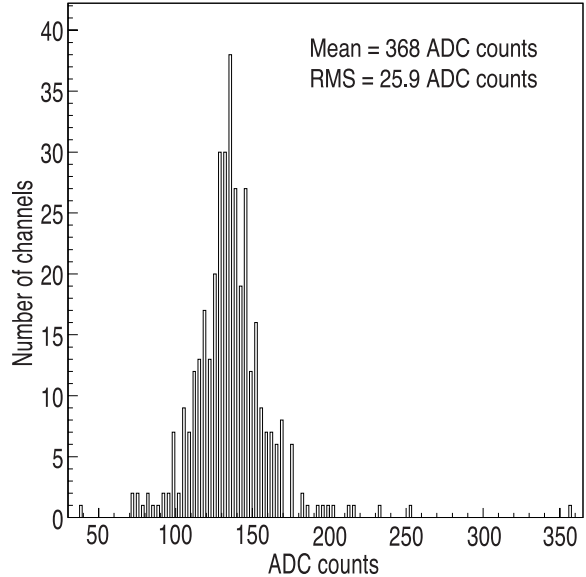}}
\caption{Cosmic ray signals in all channels of the ICD.}
\label{fig:ICDmips}
\end{figure}

Individual channel responses were measured on a test stand using
cosmic rays. This test stand used the same combination of tiles,
fibers, and electronics that was mounted in the 
experiment after testing.
Figure~\ref{fig:ICDmips} shows the distribution of cosmic ray peaks
from the test stand. ICD sampling weights are taken from the full
Monte Carlo simulation of the D\O\ detector and are tuned using
dijet and photon-jet $E_T$\ balance.

\FloatBarrier

\section{Muon system}
\label{sec:muon}

For muon triggering and measurement, the upgraded detector uses the original
central muon system proportional drift tubes (PDTs) and toroidal magnets 
\cite{d0_nim}, central scintillation counters (some new and some installed 
during Run~I), and a completely new forward muon system. 
The central muon system provides coverage for $|\eta| \lsim 1.0$.  The new 
forward muon system extends muon detection to $|\eta| \approx 2.0$, uses
mini drift tubes (MDTs) rather than PDTs, and includes trigger scintillation
counters and beam pipe shielding.  
The small angle muon system \cite{d0_nim} of the original detector, including 
its associated magnets, has been removed.

During Run I, a set of scintillation counters, the cosmic cap 
\cite{cosmic_cap_nim}, was installed on the top and upper sides of the outer 
layer of central muon PDTs.  This coverage has been extended to the lower sides 
and bottom of the detector, to form the cosmic bottom.  These trigger 
scintillation counters are fast enough to allow us to associate a muon in a 
PDT with the appropriate bunch crossing and to reduce the cosmic ray 
background.  
Additional scintillation counters, the $A\phi$ counters, have been installed on
the PDTs mounted between the calorimeter and the toroidal magnet.  The
$A\phi$ counters provide a fast detector for triggering and identifying muons
and for rejecting out-of-time background events.

The scintillation counters are used for triggering; the wire chambers are used
for precise coordinate measurements as well as for triggering.  Both types of
detectors contribute to background rejection:  the scintillator with timing
information and the wire chambers with track segments.

\begin{figure}
\centerline{\includegraphics[width=6.in]{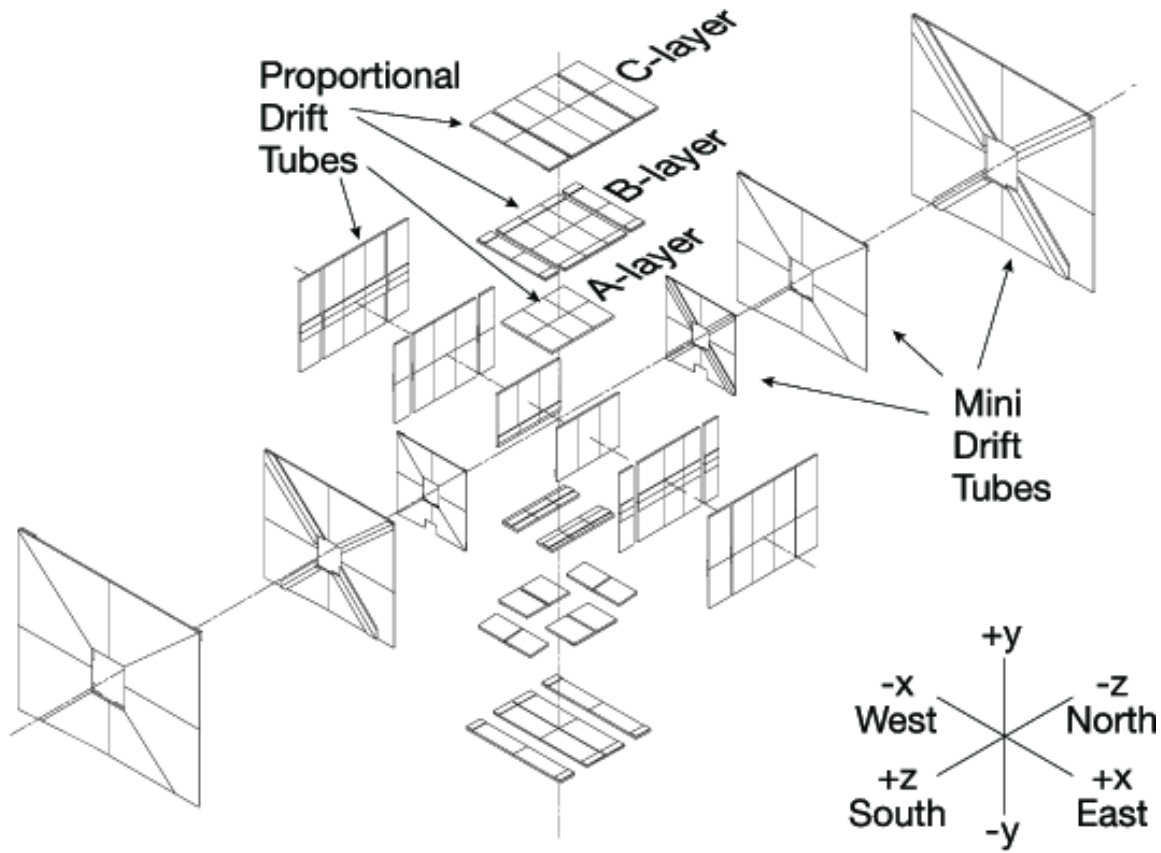}}
\caption{Exploded view of the muon wire chambers.}
\label{fig:mudrift}
\end{figure}

\begin{figure}
\centerline{\includegraphics[width=6.in]{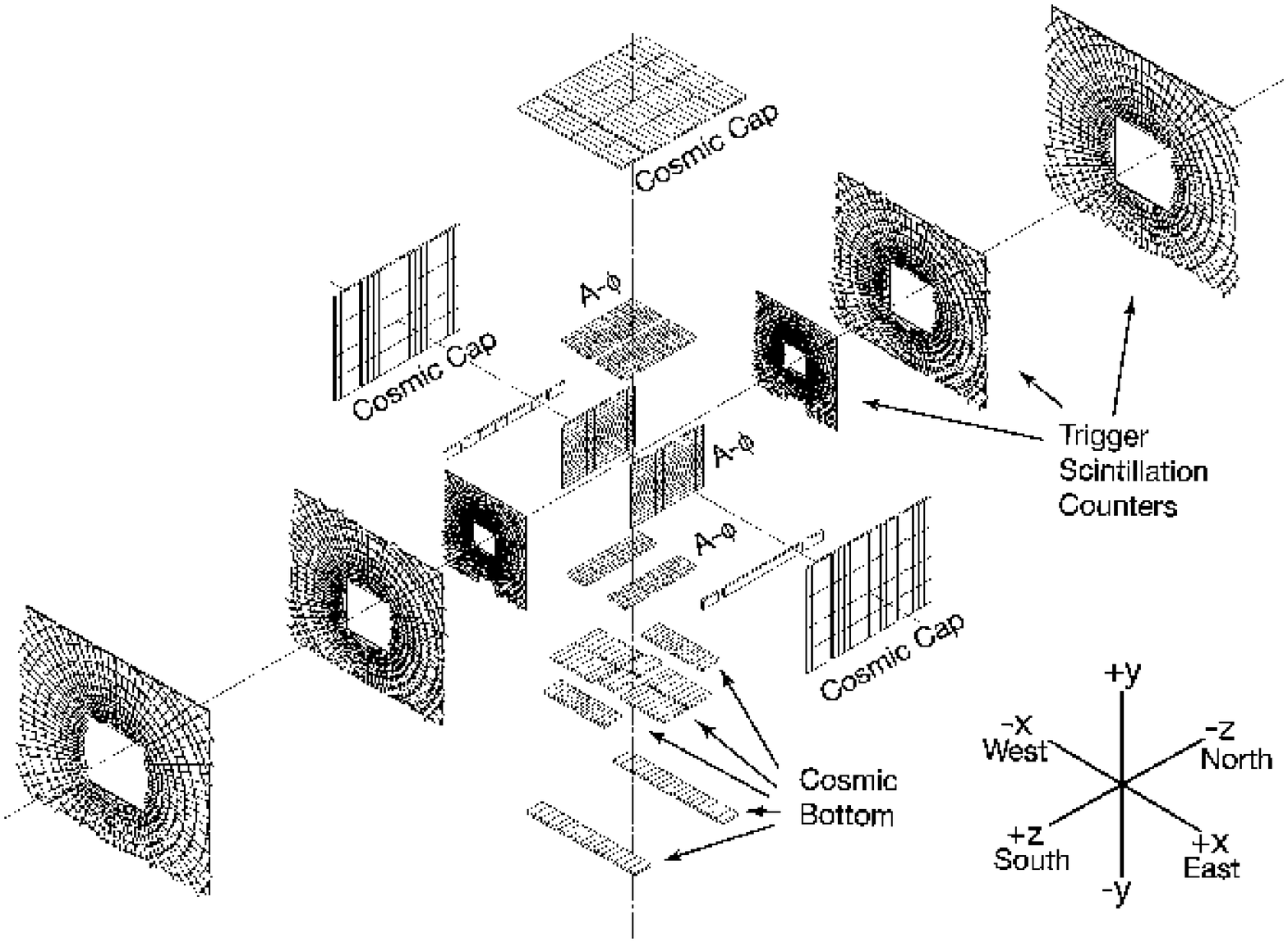}}
\caption{Exploded view of the muon scintillation detectors.}
\label{fig:muscint}
\end{figure}

New detectors and the modifications made to the original system are discussed
in detail in the following sections; original components are described briefly.
Exploded views of the muon system are shown in Figures~\ref{fig:mudrift} and
\ref{fig:muscint}.

\subsection{Toroidal magnets}
\label{sec:toroids}

The toroidal magnets are described in detail in Ref.~\cite{d0_nim} and visible 
in Figures~\ref{fig:detector}, \ref{fig:solenoid_perspective}, and
\ref{fig:solenoid-map}.  Having a stand-alone muon-system momentum measurement
{\it i}) enables a low-$p_T$ cutoff in the Level~1 muon trigger, 
{\it ii}) allows for cleaner matching with central detector tracks,  
{\it iii}) rejects $\pi/K$ decays, and
{\it iv}) improves the momentum resolution for high momentum muons.

The central toroid is a square annulus 109~cm thick whose inner surface is
about 318~cm from the Tevatron beamline; it covers the region 
$|\eta| \lsim 1$.  To allow access to the 
inner parts of the detector, it was constructed in three sections.  The 
center-bottom section is a 150-cm-wide beam, fixed to the detector platform, 
which provides a base for the calorimeters and central tracking detectors.  Two 
C-shaped 
sections, which can be moved perpendicularly to the center beam, complete the 
central toroid.  The magnet is wound using twenty coils of ten turns each.
The two end toroids are located at $454 \le |z| \le 610$~cm.  In the center of 
each end toroid is a 183~cm
square hole centered on the beamline; in $x$ and $y$ the magnets extend 426~cm
from the beamline.  The end toroid windings are eight coils of 
eight turns each. 

During Run~I, the central and end toroid coils were operated in series at a 
current of 2500~A;  the internal fields in the central toroid were 
approximately 1.9~T and those in the end toroids were approximately 2.0~T.  
During Run II, the 
magnets are again being operated in series, but at a current of 1500~A.  The 
magnetic field is about 6\% lower than that of Run I.  Now, however, the 
primary measurement of the muon momentum is performed using the new central 
tracking system and the reduced current provides a substantial cost savings.  
As in Run~I, the polarity of the magnets during data collection is regularly 
reversed.

\subsection{Central muon detector}

The central muon system consists of a toroidal magnet
(Section~\ref{sec:toroids}), drift chambers, the 
cosmic cap and bottom scintillation counters, and the $A\phi$ scintillation 
counters.  

\subsubsection{Central muon drift chambers}

The drift chambers are described in detail in Ref.~\cite{d0_nim}.
The three layers of drift chambers are located inside (A layer) and
outside (B and C layers) of the central toroidal magnet and cover 
$|\eta| \lsim 1$.  Approximately 55\% of
the central region is covered by three layers of PDTs; close to 90\% is covered
by at least two layers.  The drift chambers are large, typically $2.8 \times
5.6$~m$^2$, and made of rectangular extruded aluminum tubes.  The PDTs outside
of the magnet have three decks of drift cells; the A layer has four decks with
the exception of the bottom A-layer PDTs which have three decks.  The cells are
10.1~cm across; typical chambers are 24 cells wide and contain 72 or 96
cells.  Along with an
anode wire at the center of each cell, vernier cathode pads are located above
and below the wires to provide information on the hit position along the wire.
The wires are ganged together in pairs within a deck and then read out by
electronics located at one end of each chamber.

For each PDT hit, the following information is recorded:  the electron drift 
time, the difference $\Delta T$ in the arrival time of the signal pulse at the
end of the hit cell's wire and at the end of its readout partner's 
wire, and the 
charge deposition on the inner and outer vernier pads.  Both $\Delta T$ and the 
charge deposition are used to determine the hit position along the wire.  The 
drift distance resolution is $\sigma \approx 1$~mm.  The resolution of the 
$\Delta T$ measurement varies depending on whether the muon passes through the 
cell close to or far from the electronics.  If the hit occurs far from the 
electronics, the
resolution is approximately 10~cm.  If it is close, the signal propagates two 
wire lengths and the dispersion in the signal causes the resolution to degrade 
to about 50~cm.  Using charge division, the pad signal resolution is about
5~mm.  However, only the A-layer pads are fully instrumented with electronics; 
about 10\% of the B- and C-layer pads are instrumented.  There are several
reasons for this:  {\it i}) for tracks traversing all three layers, the pad
coordinate does not improve the pattern recognition or resolution significantly,
{\it ii}) for tracks that only reach the A-layer, the additional information
could help with track matching and background rejection, {\it iii}) the pad
signals are used to monitor the gain to track aging in the PDTs --- the 
instrumented B- and C-layer pads serve this purpose, and {\it iv}) fully
instrumenting the B- and C-layer pads was too expensive. 

To reduce the charge collection time,
we are using a faster gas mixture than we used during Run I.  The new mixture
is 84\% argon, 8\% methane, and 8\% CF$_4$.  The operating high voltage is
2.3~kV for the pads and 4.7~kV for the wires.  The drift velocity
is approximately 10~cm/$\mu$s, for a maximum drift time of about 500~ns. 
The contribution to the hit position uncertainty due to diffusion is about
0.4~mm, worse than the 0.3~mm achieved using a slower gas during Run I. 
The poorer resolution is offset by the reduced occupancy and benefits to
triggering due to decreasing the number of crossings in the drift time interval 
to two for 396~ns bunch spacing.  The gas flow rate is 200 liters per minute,
providing an exchange rate of three volumes per day.  The gas is recirculated
and filtered to remove contaminants.

Vapors from the copper-clad Glasteel (polyester and epoxy copolymer sheets with
chopped glass fibers) \cite{Glasteel} cathode pads are deposited 
on the wires in a sheath whose thickness is proportional to the accumulated 
integrated charge.  This effect was observed after PDT chamber construction
was completed.  As the coating 
thickens, the gain of the chamber drops and the chamber becomes inefficient.  
This aging during Run~II has been significantly reduced by the increased gas 
flow, improved shielding to reduce background particles, and removal of the 
Main Ring.  The chamber wires could be cleaned if necessary, 
but no aging has been observed up to an integrated luminosity of 0.8~fb$^{-1}$.
Since access to the
A-layer chambers and the four central B-layer chambers directly under the
detector is difficult, the cathode pads in these chambers have been replaced by
copper-clad G10 and no aging is anticipated.

\subsubsection{Cosmic cap and bottom counters}

The cosmic cap and bottom counters are installed on the top, sides and bottom
of the outer layer of the central muon PDTs.  They provide a fast 
timing signal to associate a muon in a PDT with the appropriate bunch 
crossing and discriminate against the cosmic ray background.

The cosmic cap counters are described in detail in Ref.~\cite{cosmic_cap_nim}. 
They are made from grooved 0.5" Bicron 404A scintillator with Bicron BCF 91A 
and Kuraray Y11 wave-shifting fibers glued into the grooves using Bicron 600 
optical epoxy.  There are 240 counters, 25" wide, and 81.5"--113" long. 
The counters are positioned with their width along $z$ and length 
along $\phi$.  The grooves are 1.75~mm deep and 4~mm wide; they run along the 
length of the counter, from the end to just past the center.  The grooves on 
each half of the counter
are offset so they do not overlap at the center of the counter.  They are
spaced 8~mm apart so that half of the counter surface is covered with fibers.  
Each groove contains four fibers.  The fibers are glued with five-minute epoxy 
at the ends of the counters and polished using a diamond cutter.  To 
increase 
the light yield, a 1/32" anodized aluminum sheet is attached to the ends with 
aluminized mylar tape.  The sides of the counters are milled.  The fibers are 
gathered at the center of the counters, divided into two bundles with two fibers
from each groove in one bundle and two in the other, and epoxied 
into two acrylic plastic disks with holes in the center of each.  The ends of 
the disks are polished using the diamond cutter.

The scintillator is wrapped in a layer of Tyvek~\cite{Tyvek} 
with a hole for the fibers and
disks; around the hole is an aluminum lip.  A layer of 1/8" thick Styrofoam 
is placed over the fibers on the counter top; aluminum sheets 0.020" thick 
cover the bottom and top surfaces.  An outer frame of Unistrut provides 
support for the counter.  A piece of black molded plastic fits over the outside
of the aluminum lip and covers the phototubes, cookies and fibers.

The fibers are read out using two 1.5" 10-stage EMI~\cite{EMI} 
9902KA phototubes mounted on the counters.  The 
light yield varies depending on the distance from the phototube and the 
proximity to the counter edge.  It is typically 30 photoelectrons 
per PMT for hits near the PMTs and 18 for hits near the distant corners.  

The cosmic bottom counters complete the scintillator coverage of the central 
toroidal magnet.  There are 132 counters, of two different designs.  The 
forty-eight counters
located on the outside of the center bottom B layer of PDTs (where there is no 
C layer) are nearly identical to the cosmic cap counters described above.
Some minor improvements were made in the placement of the edge fibers to
increase the light yield and the counter frames are made from 1/8" steel bent
into U-shaped channels.  The counters are suspended from the B-layer PDTs. 

The sixty-eight counters located on the undersides of the remaining 
B and C layers of the PDTs are similar to the cosmic 
cap counters except that the bottom counters have fewer fibers and they are 
placed in vertical, rather
than horizontal, grooves.  The grooves are approximately 6~mm deep and 6--10~cm
apart.  This distribution of fibers results in the same light yield as the
horizontal arrangement used for the cap counters.  These counters use 
1"-diameter MELZ~\cite{melz} FEU-115M photomultiplier tubes.  
They are 12-stage PMTs with a 2~ns risetime and good quantum efficiency and 
uniformity.  The PMTs are placed within 42-mm-diameter magnetic shields.
The B-layer counters are suspended from the
strong edges of the PDTs; the C-layer counters roll underneath the C-layer PDTs
with one set of wheels in a track to maintain the counter position. 

An important difference between the cosmic cap and cosmic bottom counters is
that the bottom counters are positioned with their narrow dimension along
$\phi$ and their long dimension along $\eta$.  This orientation has better 
matching in $\phi$ with the central fiber tracker trigger.  The widths of the 
counters are approximately $4.5^\circ$ in $\phi$ and they
are as long or slightly longer than their respective PDTs are wide. 
Table~\ref{tab:cosmic_counters} lists the location, number and size of the
cosmic cap and bottom counters.

\begin{table}
\begin{center}
\caption{Location, number, and size of cosmic cap and bottom scintillation
counters.  All of the scintillator is 0.5" thick.}
\label{tab:cosmic_counters}
\begin{tabular}{lcccc}
\hline 
Location        & Number & Width (inches) & Length (inches) & PMT \\
\hline
Cosmic cap top	           & 80 & 25      & 113 & EMI \\  
Cosmic cap upper sides     & 80 & 25      & 108 & EMI \\
Cosmic cap lower sides     & 80 & 25      & 81.5 & EMI \\
Central side B layer       & 12 & 35      & 49 & EMI \\
Central side B layer       &  4 & 35      & 67 & EMI \\
Central bottom B layer     & 20 & 22.375  & 98.125 & EMI \\
Central bottom B layer     & 20 & 15.75   & 98.125 & EMI \\
Central bottom B layer lap &  8 & 18.5    & 99.5 & EMI \\
End bottom B layer         & 20 & 13.375  & 91.062 & MELZ \\
End bottom B layer         & 12 & 19.25   & 91.062 & MELZ \\
Central bottom C layer     & 20 & 22.062  & 68.062 & MELZ \\
Central bottom C layer     & 16 & 29.3    & 68.062  & MELZ \\
\hline
\end{tabular}
\end{center}
\end{table}

\subsubsection{$A\phi$ scintillation counters}

The $A\phi$ scintillation counters cover the A-layer PDTs, those between 
the calorimeter and the toroid.  They provide a fast detector for triggering on 
and identifying muons and for rejecting out-of-time backscatter from the 
forward direction.  In-time scintillation counter hits are matched with tracks 
in the CFT in the Level~1 trigger (Section~\ref{sec:trigger}) for 
high-$p_T$ single muon and low-$p_T$ dimuon triggers.  The counters 
also provide the time stamp for low-$p_T$ muons which do not penetrate the 
toroid and thus do not reach the cosmic cap or bottom counters.

\begin{figure}
\centerline{\includegraphics[width=6.in]{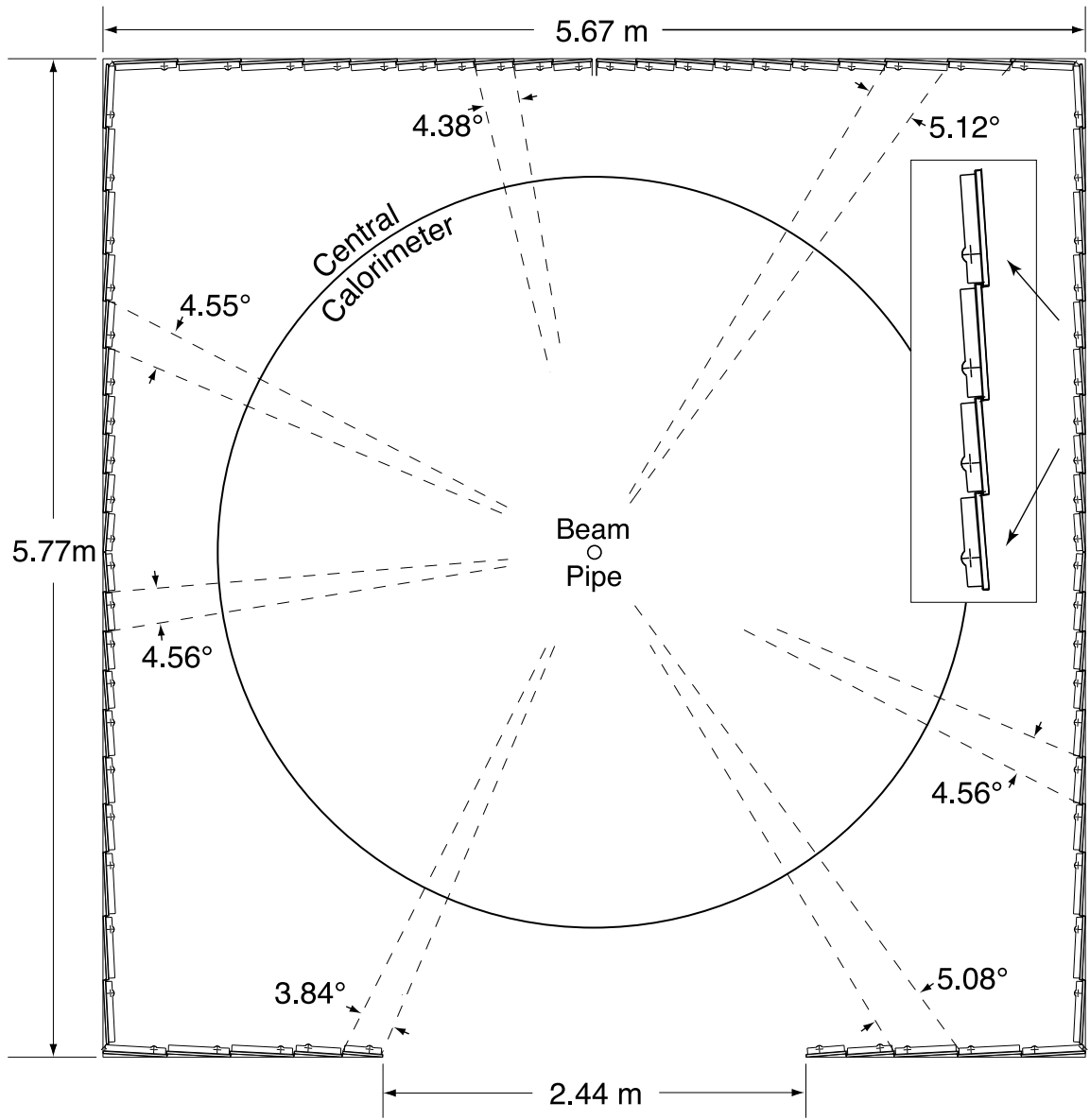}}
\caption{End view of the layout of the $A\phi$ scintillation counters.  The
inset box shows an enlarged view of four counters.  Azimuthal coverage is shown
for seven of the counters.  The bump on each counter represents the
photomultiplier tube attached to the counter case.}
\label{fig:A_phi_layout}
\end{figure}

An end view of the $A\phi$ counter layout is shown in
Figure~\ref{fig:A_phi_layout}.  The $\phi$ segmentation is approximately
$4.5^\circ$ which matches the central fiber tracker trigger
(Section~\ref{sec:l1ctt}) segmentation.  The longitudinal segmentation is 
33.25" which provides the necessary
time resolution and a match to the size of the PDTs; nine counters are 
required along the detector in the $z$ direction.  The nearly constant 
segmentation in $\phi$ is accomplished through the use of three sizes of
counter:  14.46", 10.84", and 9.09" wide.   The widest counters are located at
the corners of the detector, the narrowest at the center of each side.  There is
a gap at the bottom of the detector where the calorimeter support is located. 
The counters overlap an average of about 3\% in $\phi$ to reduce the 
possibility of muons passing through cracks.  Along the
length of the detector, the counters are butted end-to-end with a small gap
between each.  There are 630 $A\phi$ counters.

The counters are made from Bicron 404A scintillator with G2 fiber embedded in
vertical grooves \cite{evdokimov} in a manner similar to the cosmic bottom 
counters described above.  The grooves are about 1.75" apart and run along the 
length of each counter from the edge to the middle.  Six fibers are spot glued 
into each groove and taper out of the groove at the middle of the counter.  
The fibers are routed to a MELZ FEU-115M photomultiplier tube which is secured 
to the counter case. 
The counter and fibers are wrapped in a layer of Tyvek, followed by a layer of
black TEDLAR~\cite{tedlar}.  The counter case is an aluminum box with welded 
corners.  It
provides mechanical protection, support of the PMT and counter mounts, and
protection against light leaks.  The counters are mounted on aluminum
cross-members attached to steel brackets which are fastened to the edges of the
A-layer PDTs.  

The $A\phi$ counters operate in a magnetic field of 200--350~G due to the 
residual magnetic field of the toroidal and solenoidal magnets.  Magnetic 
shields made of 1.2-mm-thick $\mu$-metal and 6-mm-thick soft iron with a 
diameter of 48~mm provide 
shielding from the fringe field (Section~\ref{sec:mu-trigger-counters}).  
The effect of the magnetic field on the PMT gain is less than 10\% for any 
field direction.

The performance of an $A\phi$ prototype counter was studied using cosmic rays. 
The average muon signal corresponds to 50--60 photoelectrons.
Comparison of the amplitudes of signals from the ends of the counters and from 
the center of the counter shows that the counters are uniform to $\pm7$\%.  The
timing resolution of the counters is about 2~ns, due to photoelectron 
statistics, amplitude variation along the length of the counter, variation in 
the $z$-position of the vertex, variation in the $z$-position of the hit over 
the length of the counter, and variation of the time of collision.  Different 
times-of-flight 
for particles at different polar angles are compensated for by varying cable 
lengths since the front-end electronics do not allow such timing adjustments.

\subsection{Forward muon system}
\label{sec:famus}

The layout of the forward muon system is shown in Figure~\ref{fig:detector}.  
It covers $1.0 \lsim |\eta| \lsim 2.0$ and consists of four
major parts:  the end toroidal magnets (Section~\ref{sec:toroids}, 
three layers of MDTs for muon track reconstruction, three layers
of scintillation counters for triggering on events with muons, and shielding
around the beam pipe.  

\subsubsection{Mini drift tubes}

Mini drift tubes were chosen for their short electron drift time
(below 132~ns), good coordinate resolution (less
than 1~mm), radiation hardness, high segmentation, and low occupancy.
The MDTs are arranged in three layers (A, B, and C, with A closest to the
interaction region inside the toroidal magnet and C furthest away), each of 
which is divided into eight octants, as illustrated in 
Figure~\ref{fig:mudrift}.  A layer consists of three (layers B and C) 
or four (layer A) planes of tubes  
mounted along magnetic field lines (the field shape 
in the forward toroids is more ``square'' than ``circular'').  The entire 
MDT system contains 48,640 wires; the maximum 
tube length is 5830~mm in layer C.  Since the flux of particles
drops with increasing distance from the beam line, the occupancy of individual
tubes is the same within a factor of two over an entire layer.

\begin{figure}
\centerline{\includegraphics[width=3.in]{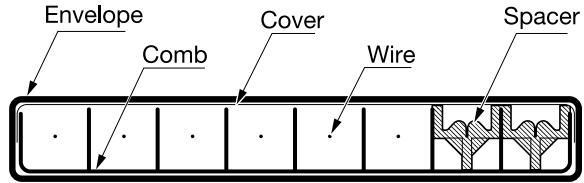}}
\caption{Cross-sectional view of a mini drift tube.}
\label{fig:mdt-xsec}
\end{figure}

An MDT tube consists of eight cells, each with a $9.4 \times 9.4$~mm$^2$ 
internal cross section and a 50~$\mu$m W-Au anode wire in the center, see
Figure~\ref{fig:mdt-xsec}.  The 
tubes are made from commercially available aluminum extrusion combs (0.6~mm 
thick) with a stainless steel foil cover (0.15~mm thick) and are inserted 
into PVC sleeves.  They are closed by endcaps that provide accurate 
positioning of the anode wires, wire tension, gas tightness, and electrical 
and gas connections.  The anode wires are supported by spacers; the  
unsupported wire length never exceeds 1~m.

\begin{figure}
\centerline{\includegraphics[width=3.in]{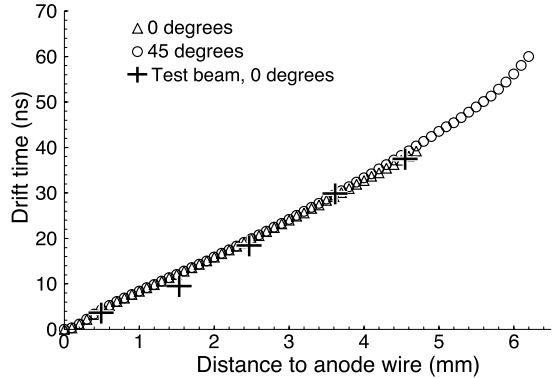}}
\caption{Time-to-distance relationship for a mini drift tube.  The points 
labeled 0~degrees and 45~degrees are calculated using GARFIELD.  The crosses
indicate measurements done at $0^\circ$.}
\label{fig:mdt-t-to-d}
\end{figure}

The MDT system uses a CF$_4$-CH$_4$ (90\%-10\%) gas mixture.  It is 
non-flammable, fast, exhibits no radiation aging, and has 
a wide operational plateau.  The maximum drift time for tracks that are
perpendicular to the detector plane is 40~ns; for tracks inclined
at $45^\circ$, the maximum drift time is 60~ns. Figure~\ref{fig:mdt-t-to-d}
shows the time-to-distance relationship for inclinations of $0^\circ$ and
$45^\circ$ calculated using GARFIELD \cite{garfield} and for test beam data.

Negative high voltage is applied to the cathode ($-3200$~V); the anode wire is 
grounded at the amplifier.  Each anode wire is connected to an amplifier and a
discriminator located as close as possible to the detector.  Each
amplifier discriminator board (ADB) contains 32 channels and detects signals
with a 2.0~$\mu$A threshold.  Output logical differential signals
from the ADB are sent to digitizing electronics which measure the signal 
arrival time with respect to the beam crossing with an accuracy of 18.8~ns
(1.9~mm).  This large time bin (equal to the Tevatron RF bucket size since the
main accelerator clock is 53~MHz) limits the 
coordinate resolution of the MDTs and was selected to reduce the cost of almost 
50,000 TDCs (time-to-digital converters).  Hit information is sent to the
trigger and data acquisition systems; the data acquisition system also receives 
drift times.

The efficiency of the MDTs is 100\% in the active area of the
cells for tracks that are perpendicular to the MDT plane.  The overall plane
efficiency is less, due to the wall thickness and PVC sleeves, and is 
approximately 95\%.  The MDT efficiency is reduced near the 
wire-support spacers as well.  Although each spacer is only 5~mm wide along 
the wire, the degradation in the electric field causes the efficiency to drop 
to approximately 20\% for about 10~mm along the wire.  An additional 
inefficiency of about 5\% is caused by the tube 
endcaps and gaps between octants for mounting, gas connections, and high 
voltage and signal cables. 

The momentum resolution of the forward muon spectrometer is limited by multiple
scattering in the toroid and the coordinate resolution of the tracking 
detector.  Although the MDT coordinate resolution measured in a test beam is 
about 350~$\mu$m, the 18.8~ns time bin of the digitizing electronics leads to 
a resolution of about 0.7~mm per hit.  The standalone 
momentum resolution of the forward muon system is approximately 20\% for muon 
momentum below 40~GeV/$c$.  The overall muon
momentum resolution is defined by the central tracking system for muons 
with momentum up to approximately 100~GeV/$c$; the forward muon system 
improves the resolution for higher momentum muons and is particularly 
important for tracks with $1.6 \lsim \eta \lsim 2.0$, i.e. those which 
do not go through all layers of the CFT.

\subsubsection{Trigger scintillation counters}
\label{sec:mu-trigger-counters}

\begin{figure}
\centerline{\includegraphics[width=6.in]{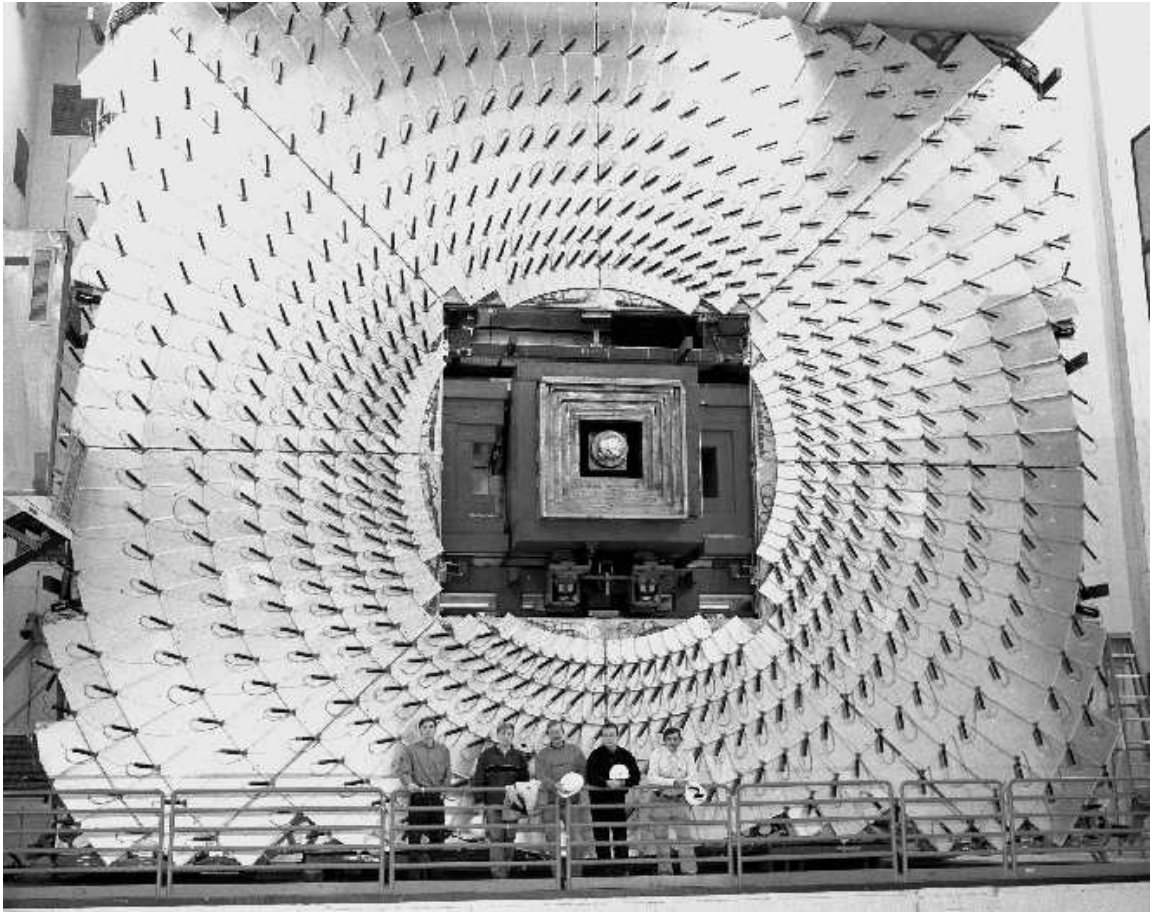}}
\caption{Photograph of the C layer of muon trigger scintillation counters 
of the forward muon system.}
\label{fig:c-layer-photo}
\end{figure}

The muon trigger scintillation counters are mounted inside (layer A)
and outside (layers B and C) of the toroidal magnet 
(Figure~\ref{fig:detector}).  The C layer of 
scintillation counters is shown in Figure~\ref{fig:c-layer-photo}.  Each layer 
is divided into octants containing about ninety-six counters.  The $\phi$ 
segmentation is $4.5^\circ$ 
and matches the CFT trigger sectors.  The $\eta$ segmentation is 0.12 (0.07) 
for the first nine inner (last three) rows of counters.  The largest counters, 
outer counters in the C layer, are $60 \times 110$~cm$^2$.  The B and C layers 
have geometries similar to that of the A layer, but limited in places by the
collision hall ceiling and floor.

\begin{figure}
\centerline{\includegraphics[width=6.in]{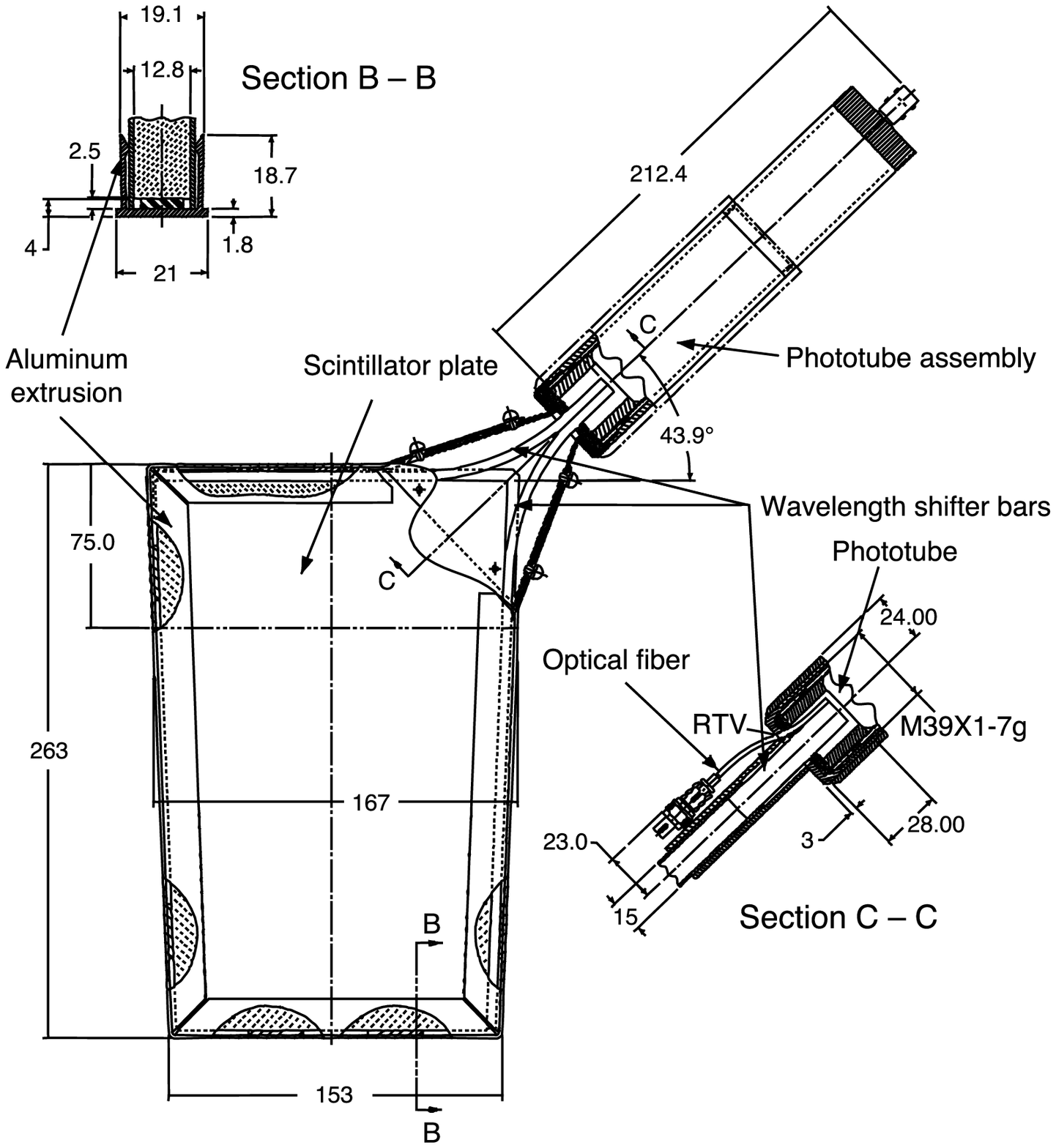}}
\caption{Design of a scintillation counter for the forward muon trigger 
system.  Dimensions are in mm.}
\label{fig:mu-trig-counter}
\end{figure}

The counter design was optimized to provide good time resolution and amplitude 
uniformity for background rejection, high muon detection efficiency, and 
reasonable cost for the production of nearly five thousand counters.  A 
typical counter is shown in Figure~\ref{fig:mu-trig-counter}.  The counters 
are made of 0.5" Bicron 404A scintillator plate cut in a trapezoidal shape.  
Kumarin 30 \cite{kumarin} WLS bars are
attached to two sides of the plate for light collection.  The bars are 4.2~mm 
thick and 0.5" wide and bent at $45^\circ$ to transmit light to a 1" phototube. 
The phototubes are fast green-extended phototubes, FEU-115M, from MELZ. 
They have a quantum efficiency of 15\% at 500~nm and a gain of about $10^6$.

The sides of scintillator plates where the WLS bars are attached were left 
unpolished after milling.  The unpolished sides result in a larger number of
photoelectrons and better uniformity than polished sides do and cost less
\cite{evdokimov}.
The scintillator and WLS bars are wrapped in Tyvek for better light
collection and black paper for light tightness.  The wrapped counters are placed
in 1-mm-thick aluminum containers with stainless steel transition pieces for 
connection of the phototube assembly.  

The fringe magnetic field due to the solenoidal and the toroidal magnets 
reaches 300~G in the region where the A-layer phototubes are located
(Figure~\ref{fig:solenoid-map}).  To 
reduce the field in the area of the phototubes to approximately 1~G, they are 
placed in 48-mm-diameter magnetic shields made of 1.2-mm-thick mu-metal and 
3- or 6-mm-thick soft 
iron.  Since the field near the A layer is larger, the shields there use 6~mm 
of iron.  Those for the B and C layers use 3~mm of soft iron.  The effect of 
the magnetic field on the phototube signal is less than 10\% for fields up to 
350~G for the 6-mm-thick shields for any field orientation with respect to the
phototube. 

\begin{figure}
\centerline{\includegraphics[width=3.in]{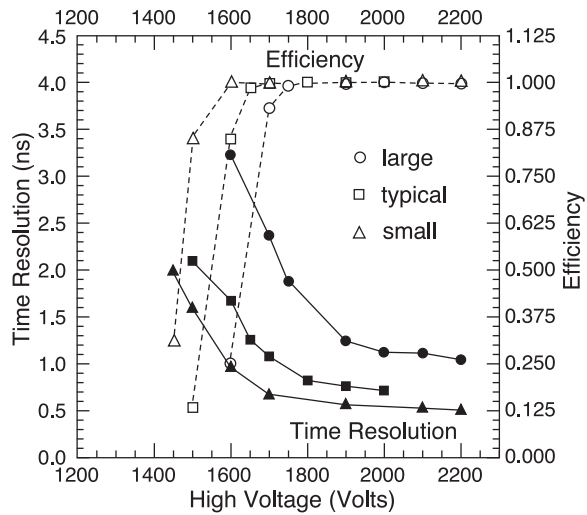}}
\caption{Time resolution and detection efficiency of muon trigger
scintillation counters.  Large counters are $60 \times 106$~cm$^2$; typical
counters are $24 \times 34$~cm$^2$; small counters are $17 \times 24$~cm$^2$.}
\label{fig:mu-trig-res-eff}
\end{figure}

The performance of three counters of different sizes was studied in a 
125~GeV/$c$ muon test beam.  Figure~\ref{fig:mu-trig-res-eff} shows the 
dependence of the counter efficiency and time resolution as a function of high 
voltage for three counters: ``large'' $60 \times 106$ cm$^2$,  ``typical'' 
$24 \times 34$ cm$^2$, and ``small'' $17 \times 24$ cm$^2$.  A single
discrimination threshold of 10~mV was used for these measurements. 
Time resolution 
of better than 1~ns and detection efficiency above 99.9\% can be achieved at
appropriate high voltage settings for all counter sizes.  The non-uniformity 
of the counter response was measured by irradiating counters at different 
points using a $^{90}$Sr source and checked via cosmic ray studies.  For all 
counter sizes, the rms non-uniformity is less than 10\%.  Cosmic ray muons 
were used to determine the average number of photoelectrons detected in the
counters.  The largest counters have an average of 61 
detected photoelectrons; the smallest counters give about three times as many 
photoelectrons.

Phototube signals are sent to 48-channel VME-based scintillator front-end 
(SFE) cards.  Input 1:1 transformers at the SFE card isolate the SFE from DC
noise picked up between the phototubes and electronics modules which are 
located about 10~m apart.  After amplification, the signals are sent to a 
10-bit ADC and to a discriminator.  Discriminated and gated signals are passed 
to the Level~1 trigger system (Section~\ref{sec:l1}) and to the SFE TDC with a 
1.03~ns time bin.  
After digitization, amplitude and time information is sent to the Level~2 
trigger system (Section~\ref{sec:l2}) and to the data acquisition system
(Section~\ref{sec:daq}).  The amplitude is measured 
in one out of sixteen channels per event for counter response monitoring.

\subsubsection{Shielding}

Three sources contribute to non-muon background in the central and forward muon
systems: {\it i}) scattered proton and antiproton fragments that interact with 
the end of the calorimeter or with the beampipe produce background in the 
central and forward A layer; {\it ii}) proton and antiproton fragments 
interacting with the Tevatron low-beta quadrupole magnets produce hits in 
the B and C layers of the forward muon system; and {\it iii}) beam halo 
interactions from the tunnel.  Shielding installed in the accelerator tunnel 
during Run~I \cite{run1-shielding} significantly reduced the background from 
beam halo.  New shielding has been installed for Run~II to reduce the 
background due to proton and antiproton remnants.  Reduction in backgrounds 
along with the use of radiation-hard detectors helps ensure the long-term, 
reliable operation of the muon system.

The shielding consists of layers of iron, polyethylene, and lead in a steel
structure surrounding the beam pipe and low-beta quadrupole magnets.  Iron is 
used as the hadronic and electromagnetic absorber due to its relatively short 
interaction (16.8~cm) and radiation (1.76~cm) lengths and low cost.  
Polyethylene is a good absorber of neutrons due to its high hydrogen 
content.  Lead is used to absorb gamma rays.
 
The position of the shielding is shown in Figure~\ref{fig:detector}.  It 
extends from the end calorimeter cryostat, through the end toroid magnet, to 
the wall of the collision  hall.  It consists of three 
rectangular cross section pieces that are 84", 85" and 60" long, starting at 
the calorimeter and moving away from the detector center.  The piece closest 
to the toroid has a 20" square hole at the center
followed by 16" of iron, 6" of polyethylene, and 2" of lead.  The two outer 
pieces are identical except that the hole is 25" square followed by 20" of 
iron.  The most-forward section is split vertically down the center so that it 
can be moved out of the way when the end toroid magnet is repositioned for
access to the central parts of the detector. 
Figure~\ref{fig:c-layer-photo} shows the shielding in the ``open'' position.

Monte Carlo studies based on MARS code \cite{mars1,mars2} 
show that the shielding provides a factor of 50--100 reduction in the energy 
deposition in muon detector elements.  Reduction of backgrounds reduces the
occupancy and detector aging, and provides almost background-free muon
triggering and reconstruction.  MARS Monte Carlo predictions of the number of
hits in the muon detectors agree with the observed occupancies within 50\%.

\subsection{Muon scintillation counter monitoring}
\label{sec:muon-led-monitoring}

All muon scintillation counters are calibrated and 
monitored using an LED-based pulser system~\cite{mu-led-monitoring}. 
Given the large number of counters
involved, it is difficult to use cosmic ray and beam muons to check the 
timing and do the PMT gain calibration quickly.  The 
LED system allows us to find dead PMTs, isolate the behavior of the PMTs
from the front-end electronics, adjust the relative timing between channels,
set initial PMT gains, monitor PMT gains, and track timing changes.  

A pulser triggers an LED driver board that drives four LEDs embedded in an
acrylic block.  The light is further mixed in two acrylic mixing
blocks.  The upstream side of each block is frosted by sand-blasting; the sides
of the blocks are polished to maximize total internal reflection.  The light is
then distributed to an array of clear optical fibers embedded in a
fourth acrylic block.  These light distribution fibers are connected to
fibers which are butted up to each PMT.  To match the peak wavelengths and
emission spectra of the wavelength-shifting fibers used in the counters,
different LEDs are used for the central and forward counters.  In the central
region, blue-green NSPE510S LEDs from Nichia America Corp. are used; in the
forward region, blue NSPB320BS LEDs, also from Nichia America Corp., have been
installed.  The clear fibers are Hewlett Packard HFBR-500 optical fibers.

The light intensity of the LEDs is monitored using a PIN diode (Hamamatsu S6775)
mounted at the downstream end of the first mixing block; this works well
because the mixing blocks
evenly distribute the light to the fiber array.  As long as the PIN diode is
stable over time, variation in the gain of the PMTs can be measured
independently of variation in the light output of the LEDs.

\begin{figure}
\centerline{\includegraphics[width=3.in]{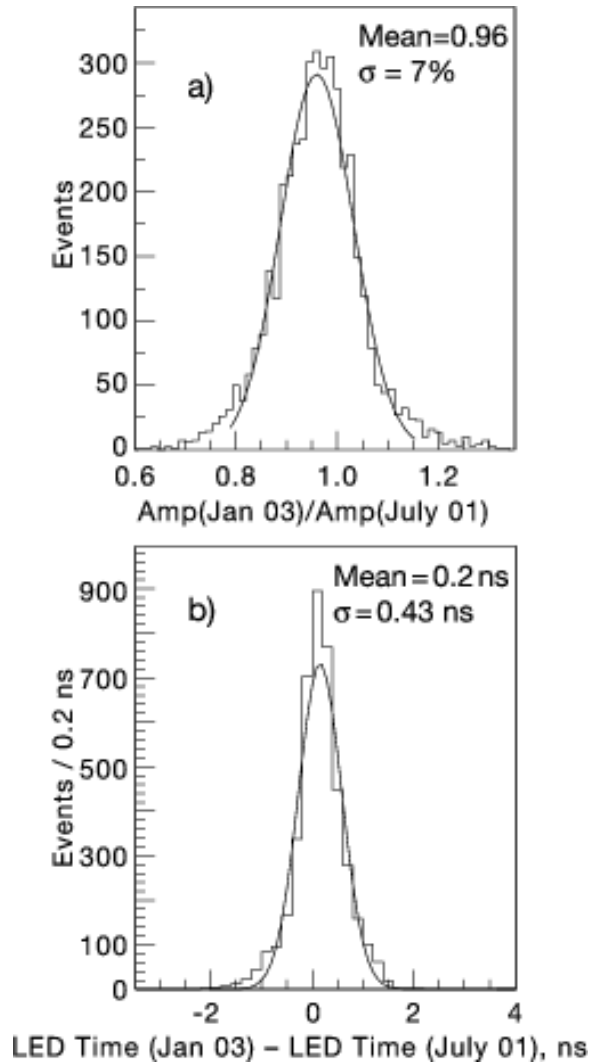}}
\caption{Results from the forward muon system monitoring by the LED calibration 
system over one and one-half years; the smooth curves are Gaussian fits to the 
data. a) shows the LED amplitude ratio, b) shows the absolute difference in the
LED timing.}
\label{fig:amptime}
\end{figure}

Figure \ref{fig:amptime} shows results from the forward muon system monitoring 
between July 2001 and January 2003.  The ratio of the LED amplitudes 
is stable within 4\% with a standard 
deviation of about 7\%.  The timing stability over this interval shows a peak
stability within 0.2~ns with a standard deviation of about 0.43~ns.  The 
typical time resolution of counters during data taking is 2~ns, so stability 
is not a major factor in the overall time resolution of the system; 
photoelectron statistics, beam spot size, and bunch timing are all more 
important.

\FloatBarrier

\section{Forward proton detector}
\label{sec:fpd}

The forward proton detector (FPD)~\cite{FPD} measures protons and antiprotons 
scattered at small angles (on the order of 1~mrad) that do not impinge upon 
the main D\O\ detector.  During Run I, such diffractive events were tagged 
using a rapidity gap (the absence of particles in a region of the detector), 
however a forward particle detector is necessary for access to the full 
kinematics of the scattered particle.

\subsection{The detector}

The FPD consists of a series of momentum spectrometers that make use of 
accelerator magnets in conjunction with position detectors along the beam line. 
The position detectors operate a few millimeters away from the beam and have to 
be moved out of the beamline during injection of protons or antiprotons 
into the accelerator.  Special 
stainless steel containers, called Roman pots~\cite{ROMAN}, house the position
detectors, allowing them to function outside of the ultra-high vacuum of the 
accelerator, but close to the beam.  The scattered $p$ or $\bar{p}$ traverses 
a thin steel window at the entrance and exit of each pot. The pots are remotely 
controlled and can be moved close to the beam during stable conditions.

\begin{figure}
\centerline{\includegraphics[width=6.in]{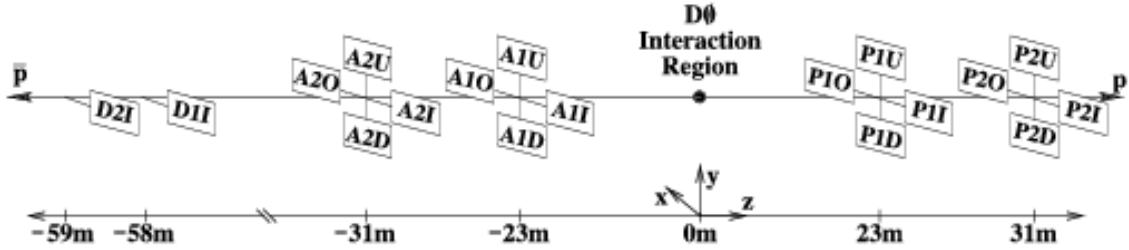}}
\caption{FPD layout. Quadrupole castles are designated with a leading 
P or A when placed on the 
$p$ side or the $\bar{p}$ side, respectively; the number designates the station
location; while the final letter indicates pot potition (U for up, D down, 
I in, O out).  D1I and D2I are dipole castles.} 
\protect\label{fig:fpdlayout}
\end{figure}

The Roman pots are housed in stainless steel chambers called castles. The FPD 
consists of eighteen Roman pots arranged in six castles. The castles
are located at various distances from the D\O\ interaction point
and in locations that do not interfere with the accelerator.
The arrangement of the FPD is shown in Figure~\ref{fig:fpdlayout}. 
Four castles are located 
downstream of the low beta quadrupole magnets on each side of the collision
point: two on the $p$ side (P1 and P2) and two on the $\bar{p}$ side (A1 and 
A2). Each of these quadrupole castles contains four Roman pots arranged to 
cover most of the area around the beam. Two castles (D1 and D2) are located on 
the outgoing $\bar{p}$ side after the dipole magnet.  Each of these dipole 
castles contains only one Roman pot.  
There are nine spectrometers: the two dipole castles form one, and on each side
of the interaction region the two up, two down, two in, and two out pots are
paired to form the other eight (Figure~\ref{fig:fpdlayout}).

\subsubsection{The position detector}

\begin{figure}
\centerline{\includegraphics[width=3.in]{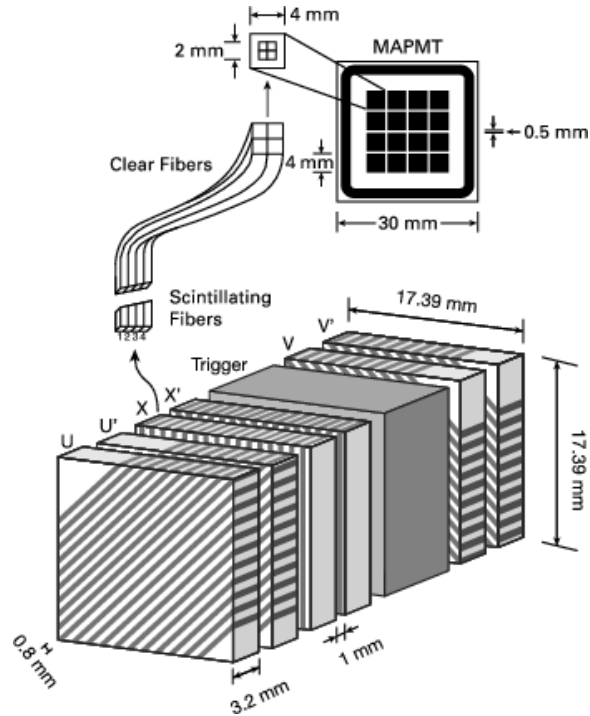}}
\caption{Detector fibers and the MAPMT.} 
\label{fig:fpd-cartoon}
\end{figure}

The construction of the position detectors is illustrated in 
Figure~\ref{fig:fpd-cartoon}.  Each detector is made of 0.8-mm-thick 
double-clad square scintillating fibers (Bicron BCF10 \cite{bicron}) bundled 
in groups of four parallel fibers, forming a 
scintillating structure measuring 0.8~mm $\times$ 3.2~mm. One end of the 
detector element is aluminized (about a 3-$\mu$m-thick layer) to increase 
the light yield and the other end of each scintillating fiber is spliced to a 
double-clad clear fiber of square cross section (Bicron BCF98) with the 
same dimensions.  The use of square fibers gives an increase of 
about 20\% in light output compared to round fibers. 
The scattered $p$ or $\bar{p}$ goes through 3.2~mm of scintillating 
material.  The four clear fibers take the light of one element to 
a single channel of the Hamamatsu H6568 16-channel multi-anode 
photomultiplier (MAPMT), yielding approximately ten photoelectrons.

As shown in Figure~\ref{fig:fpd-cartoon}, each detector consists of six planes 
in three views ($u$, $x$ and $v$) 
to minimize ghost hit problems and to reduce reconstruction ambiguities. Each
view is made of two planes ($u-u'$, $x-x'$, and $v-v'$), the primed layers 
being offset by two-thirds of a fiber with respect to the unprimed layers. 
The $u$ and $v$ planes are oriented at $\pm45^\circ$ with respect to the 
horizontal bottom of the detector, while the $x$ plane is at $90^{\circ}$. 
There are twenty channels in each layer of the $u$ and $v$ planes and sixteen 
channels in each of the $x$ layers.  There are 112 channels (each with four 
fibers) per detector giving a total of 2016 channels in the eighteen Roman 
pots.  The readout of each detector requires seven MAPMTs and includes a 
trigger scintillator read out by a 
fast photomultiplier tube (Phillips XP2282 \cite{phillips}).  The FPD uses CFT 
electronics (Section~\ref{sec:AFE}) for the 
scintillating fiber detector read out, luminosity monitor electronics 
(Section~\ref{sec:lum-monitor}) 
for the trigger read out, and the muon LED system for calibration
(Section~\ref{sec:muon}).  The trigger manager designed for the muon system 
incorporates this information in making a Level~1 trigger decision on FPD 
tracks (Section~\ref{sec:l1fpd}).

\subsubsection{The castle}

\begin{figure}
\centerline{\includegraphics[width=4.5in]{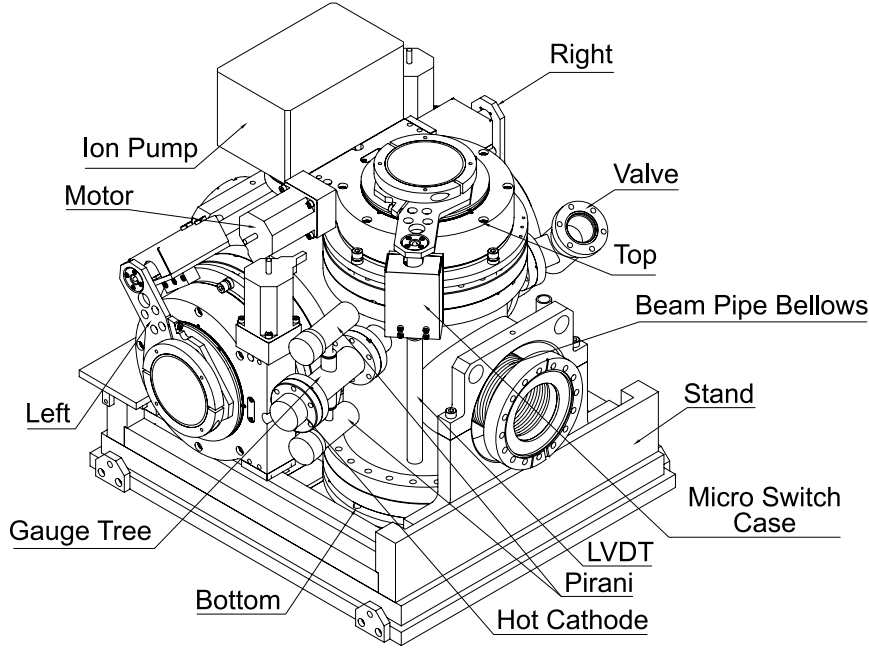}}
\caption{FPD quadrupole castle.} 
\label{fig:castle}
\end{figure}

Figure~\ref{fig:castle} shows an FPD quadrupole castle. It has four arms 
(dipole castles are similar, but have just one arm), each containing 
a Roman pot housing a detector. The castles are made of 316L stainless steel 
and, due to the ultra-high vacuum necessary in the interior of the castles, 
all parts were cleaned with demineralized water and alkaline detergent in an
ultrasound bath and dried with a hot air jet before being TIG welded. 
The quadrupole castles (dipole castles are not in ultra-high vacuum) are baked 
at 150$^\circ$ C when the vacuum has been broken.  
A set of hot cathode and convection Pirani-style sensors monitors
the vacuum in the chamber. Each castle has an associated ion pump 
to provide the ultra-high vacuum.  

The castle sits on a stand that allows adjustment of its position in all
directions over a range of 15 mm with an accuracy of 0.1 mm.
The pot is connected to a driving system that makes it possible to move it 
perpendicularly to the beam. A 200-$\mu$m-thick window separates the detector 
(inside the pot) from the castle ultra-high vacuum.  The system is operated by 
a step motor and a set of reduction gears allows pot motion with a precision of 
approximately $5~\mu$m. A system of cylindrical and conical bearings allows 
adjustment of the pot alignment and a linear variable differential 
transducer (LVDT) monitors the pot position.

\subsection{Acceptance}

The FPD acceptance is maximized by minimizing the distance 
between the detectors and the beam axis. This distance is limited primarily by 
interaction with the beam halo which increases as the pots are inserted closer 
to the beam.  FPD acceptance is determined as a function of $t$, the 
four-momentum transfer squared of the scattered proton or antiproton, and 
$\xi = 1 - x_p$ where $x_p$ is the fractional longitudinal momentum of the 
scattered particle.  For the dipole spectrometer, the acceptance is highest for
$|t| \lsim 2$~GeV$^2$/$c^4$, $0.04 \lsim \xi \lsim 0.08$ and extends to 
$|t| \lsim 4.3$~GeV$^2$/$c^4$, 
$0.018 \lsim \xi \lsim 0.085$ (coverage is incomplete).  The acceptance in the 
quadrupole spectrometers covers most of the region 
$0.6 \lsim |t| \lsim 4.5$~GeV$^2$/$c^4$, $\xi \lsim 0.1$.

For elastic events, both particles must be detected by diagonally opposite
spectrometers with no activity detected in any other D\O\ subdetector.  A 
sample of elastic events collected during special runs was used to measure the 
position resolution of the FPD by comparing the $x$ coordinate determined by 
combining information from the $u$ and $v$ planes to the $x$ coordinate from 
the $x$ plane.  This process gives a resolution of 130~$\mu$m.

\FloatBarrier

\section{Luminosity monitor}
\label{sec:lum-monitor}

The primary purpose of the luminosity monitor (LM) is to determine 
the Tevatron luminosity at the D\O\ interaction 
region. This is accomplished by detecting inelastic \pbarp\ 
collisions with a dedicated detector. The LM also serves to measure beam halo 
rates and to make a fast measurement of the $z$ coordinate of the interaction 
vertex.  

\subsection{The detector}

\begin{figure}
\centerline{\includegraphics[width=3.in]{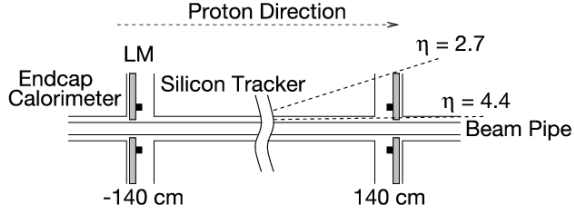}}
\caption{Schematic drawing showing the location of the LM detectors.}
\label{fig:lum-fig1}
\end{figure}

\begin{figure}
\centerline{\includegraphics[width=3.in]{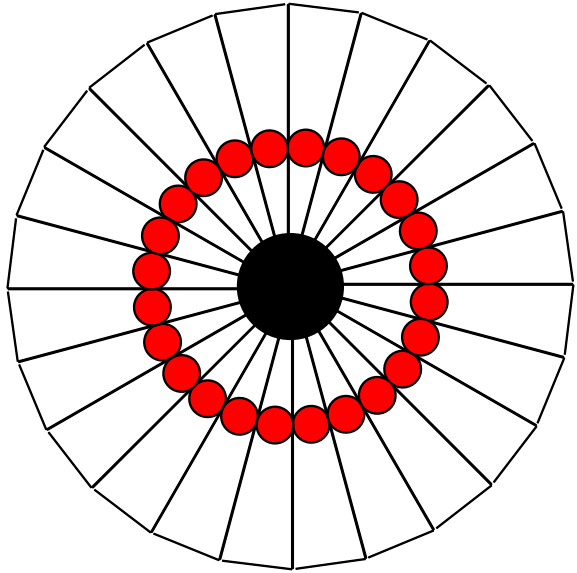}}
\caption{Schematic drawing showing the geometry of the LM counters
and the locations of the PMTs (solid dots).}
\label{fig:lum-fig2}
\end{figure}

The LM detector consists of two arrays of twenty-four plastic
scintillation counters with PMT readout located at $z = \pm 140$~cm
(Figure~\ref{fig:lum-fig1}). 
A schematic
drawing of an array is shown in Figure~\ref{fig:lum-fig2}.
The arrays are located in front of the end calorimeters and occupy the 
radial region between the beam pipe and the forward preshower detector.
The counters are 15~cm long and cover the pseudorapidity range
$2.7 < |\eta | < 4.4$. 

Scintillation light produced in the Bicron BC-408 scintillator is
detected by Hamamatsu \cite{hamamatsu} R5505Q fine mesh PMTs.  
Due to space constraints and the characteristics of the PMTs,
they are mounted on the faces of the scintillators
with the axes of the PMTs parallel to the $z$ axis. They
have no magnetic shielding, and their gain is reduced by a 
factor of about 30 when the solenoidal magnet is turned on due to the 
approximately 1~T magnetic field in this region~\cite{LM-Ref1}.
The time-of-flight resolution for the counters is about 0.3~ns,
with the dominant contribution to the resolution being the variation in light
path length for particles striking different locations on the scintillator. 

Radiation damage is a concern for detectors located this close to the
beams. Much of the radiation dose seen by these detectors comes from
the \pbarp\ collision products and is thus unavoidable.
The PMTs are exposed to a radiation flux of
about 25~krad/fb$^{-1}$, which is sufficient to cause darkening of
the borosilicate glass window typically used for PMTs.
The R5505Q PMTs have fused silica (quartz) windows which are
largely immune to radiation damage~\cite{hamamatsu}.
The radiation flux increases rapidly with decreasing radius, reaching a level
of approximately 300~krad/fb$^{-1}$ at the innermost scintillator edge.
Based on the radiation damage study in Ref.~\cite{LM-Ref3}, modest
($\approx 10$\%) light loss is expected for the innermost scintillator
edge after 3~fb$^{-1}$.

The scintillation counters are enclosed in light-tight enclosures, with
each enclosure holding twelve counters.
Preamplifiers inside the enclosures amplify the PMT signals
by a factor of five. The fused silica PMT windows are much
more permeable to helium gas than borosilicate glass \cite{LM-Ref2}.
To avoid damage from the widely fluctuating helium concentration
present in the collision hall, the enclosures are purged
with dry nitrogen.

For accurate timing of the PMT signals, low-loss 
cables~\cite{LM-Ref4} are used to bring the signals from the detector to 
the digitization and readout electronics.
The signals are equalized in time and split into two paths. On one
path, currently in use for luminosity measurements, analog sums are formed 
from the PMT signals for
each of the two arrays, which are then timed using a digital TDC to 
identify \pbarp\ collisions \cite{LM-Ref5}. On the other path, which is an
upgrade currently being installed, two types of
custom VME boards provide the required signal processing. Six LM-TDC
boards are used to digitize the time and charge for each PMT
and apply charge-slewing corrections to generate fully calibrated
time-of-flight measurements. A single LM-VTX board utilizes the
measurements made on the LM-TDC boards to calculate the average time
for each counter array and the $z$ coordinate of the interaction vertex.

\subsection{Luminosity determination}

The luminosity {$\mathcal{L}$} is determined from the average number of 
inelastic collisions per beam crossing $\bar N_{LM}$ measured by the LM:  
${\mathcal{L}} = {f \bar N_{LM} \over \sigma_{LM}}$\ 
where $f$ is the beam crossing frequency and $\sigma_{LM}$ is the effective
cross section for the LM that takes into account the
acceptance and efficiency of the LM detector \cite{LM-note}.
Since $\bar N_{LM}$ is typically greater than one, it is important to 
account for multiple \pbarp\ collisions in a single beam crossing. This
is done by counting the fraction of beam crossings with no collisions and
using Poisson statistics to determine $\bar N_{LM}$.

To accurately measure the luminosity, it is necessary to distinguish
\pbarp\ interactions from the beam halo backgrounds.  We separate these
processes by making precise time-of-flight measurements of particles
traveling at small angles with respect to the beams.
We first assume that particles hitting the LM detector
originate from a \pbarp\ interaction and estimate the $z$ coordinate
of the interaction vertex $z_{v}$ from the difference in time-of-flight: 
$z_{v} = {c\over 2}(t_- - t_+)$ where $t_+$ and $t_-$ are the times-of-flight 
measured for particles hitting the LM detectors placed at $\pm 140$~cm.  
Beam-beam collisions are selected by requiring
$|z_v| < 100$~cm, which encompasses nearly all \pbarp\ collisions
produced by the Tevatron ($\sigma_{z} \approx 30$~cm).
Beam halo particles traveling in the $\pm \hat z$ direction will
have $z_v \approx \mp 140$~cm, and are eliminated
by the $|z_v| < 100$~cm requirement. 

Level~1 triggers (Section~\ref{sec:l1}) are grouped together so that they 
have common deadtime, i.e., common sources of enable, disable, and readout.  
This allows the readout to be partitioned so that different triggers may
read out independent portions of the detector.  The luminosity
associated with each trigger takes into account the instantaneous
luminosity, the deadtime, and losses in the data acquisition system.

The luminosity block is the fundamental unit of time for the
luminosity measurement.  Each block is indexed by the luminosity block
number (LBN), which monotonically increases throughout Run~II.  The LBN
is incremented upon run or store transitions, TFW or SCL initialization, by
request, or after 60~seconds have elapsed.  The time period is short
enough so that the instantaneous luminosity is effectively constant
during each luminosity block, introducing negligible uncertainty into
the measurement of the luminosity due to the width of the time slice.
Raw data files are opened and closed on LBN boundaries.  Luminosity
calculations are made independently for each LBN and 
averaged over the luminosity block.

\subsection{Luminosity data acquisition system}

The luminosity data acquisition system (LDAQ) is a stand-alone
data acquisition system running on the online cluster.  The LDAQ was designed 
to collect sufficient data to measure, verify, and monitor the luminosity
delivered to and used by D\O.  The LDAQ connects to the
following systems: L1~trigger framework (Section~\ref{sec:tfw}), 
accelerator controls system, 
D\O\ controls system (Section~\ref{sec:controls}), Level~3
(both ScriptRunner and DAQ, Sections~\ref{sec:l3trigger} and \ref{sec:daq}), 
COOR (Section~\ref{sec:coor}), and the datalogger (Section~\ref{sec:daq}).  
Data from different sources are correlated, loaded into a database, and 
used for luminosity calculations.

\FloatBarrier

\section{Triggering}
\label{sec:trigger}

\begin{figure}
\centerline{\includegraphics[width=6.in]{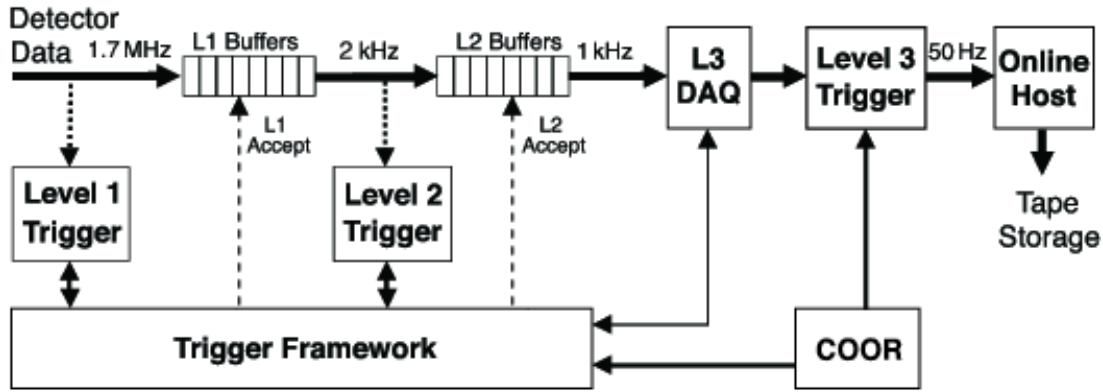}}
\caption{Overview of the D\O\ trigger and data acquisition systems.}
\label{fig:trigger-overview}
\end{figure}

\begin{figure}
\centerline{\includegraphics[width=3.in]{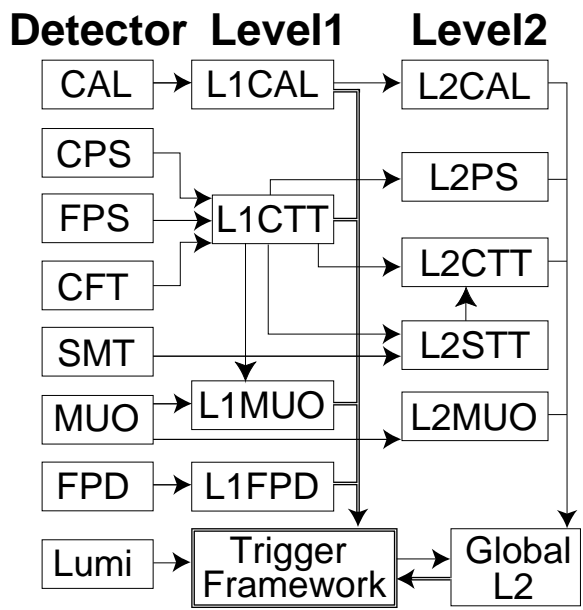}}
\caption{Block diagram of the D\O\ L1 and L2 trigger systems.  
The arrows show the flow of trigger-related data.}
\label{fig:l1-l2-layout}
\end{figure}

With the increased luminosity and higher interaction rate delivered by the 
upgraded Tevatron, a significantly enhanced trigger is necessary to select the 
interesting physics events to be recorded.  Three distinct levels form this 
new trigger system with each succeeding level examining fewer events but in 
greater detail and with more complexity.  The first stage (Level~1 or L1) 
comprises a collection of hardware trigger elements that provide a trigger 
accept rate of about 2~kHz.  In the second stage (Level~2 or L2), hardware 
engines and embedded microprocessors associated with specific
subdetectors provide information to a global processor 
to construct a trigger decision based on individual objects as well as object 
correlations.  The L2 system reduces the trigger rate by a factor of about two 
and has an accept rate of approximately 1~kHz.
Candidates passed by L1 and L2 are sent to a farm of Level~3 (L3)
microprocessors; sophisticated algorithms reduce the rate to about 50~Hz and
these events are recorded for offline reconstruction.  
An overview of the D\O\ trigger and data acquisition system is shown in 
Figure~\ref{fig:trigger-overview}.  A block diagram of the L1 and L2 trigger
systems is shown in Figure~\ref{fig:l1-l2-layout}.

The trigger system is closely integrated with the 
read out of data, as illustrated in Figure~\ref{fig:trigger-overview}.
Each event that satisfies the successive L1 and L2 triggers is 
fully digitized, and all of the data blocks for the event are
transferred to a single commodity processor in the L3 farm.  
The L1 and L2 buffers play an important role in minimizing the experiment's 
deadtime by providing FIFO storage to hold event data awaiting a Level~2
decision or awaiting transfer to Level~3.  

The overall coordination and control of D\O\ triggering is handled by the 
COOR package (Section~\ref{sec:coor}) 
running on the online host.  COOR interacts directly with the trigger framework
(for L1 and L2 triggers) and with the DAQ supervising systems (for the L3
triggers).  
The data acquisition system responsible for the data flow of the fully 
digitized event into L3 is described in Section~\ref{sec:daq}.

\subsection{The Level 1 trigger}
\label{sec:l1}

L1 is implemented in specialized hardware and examines every event for 
interesting features.  The calorimeter trigger (L1Cal) looks for energy 
deposition patterns exceeding programmed limits on transverse energy 
deposits; the central track trigger (L1CTT) and the muon system trigger 
(L1Muon) compare tracks, separately and together, to see if they exceed preset 
thresholds in transverse momentum.  The L1 forward proton detector trigger 
(L1FPD) is used to select diffractively-produced events by triggering on 
protons or antiprotons scattered at very small angles. 

All events awaiting L1 trigger decisions are pipelined and thus make minimal 
contributions to the deadtime.  In order to participate in the trigger 
decision, the L1 trigger decision must arrive at the trigger framework in 
3.5~$\mu$s or less. The rate of L1 trigger accepts is limited by the maximum 
readout rates of the participating subsystems and by a desire to minimize 
the deadtime associated with the readout.

\subsubsection{Trigger framework}
\label{sec:tfw}

The trigger framework (TFW) gathers digital information from each of the
specific L1 trigger devices and chooses whether a particular event is to be
accepted 
for further examination.  In addition, it coordinates various vetoes that can
inhibit triggers, provides the prescaling of triggers too copious to pass on
without rate reduction, correlates the trigger and readout functions, manages
the communication tasks with the front-end electronics and the trigger
control computer (TCC), and provides a large number of scalers that allow
accounting of trigger rates and deadtimes.

\begin{figure}
\centerline{\includegraphics[width=5.in]{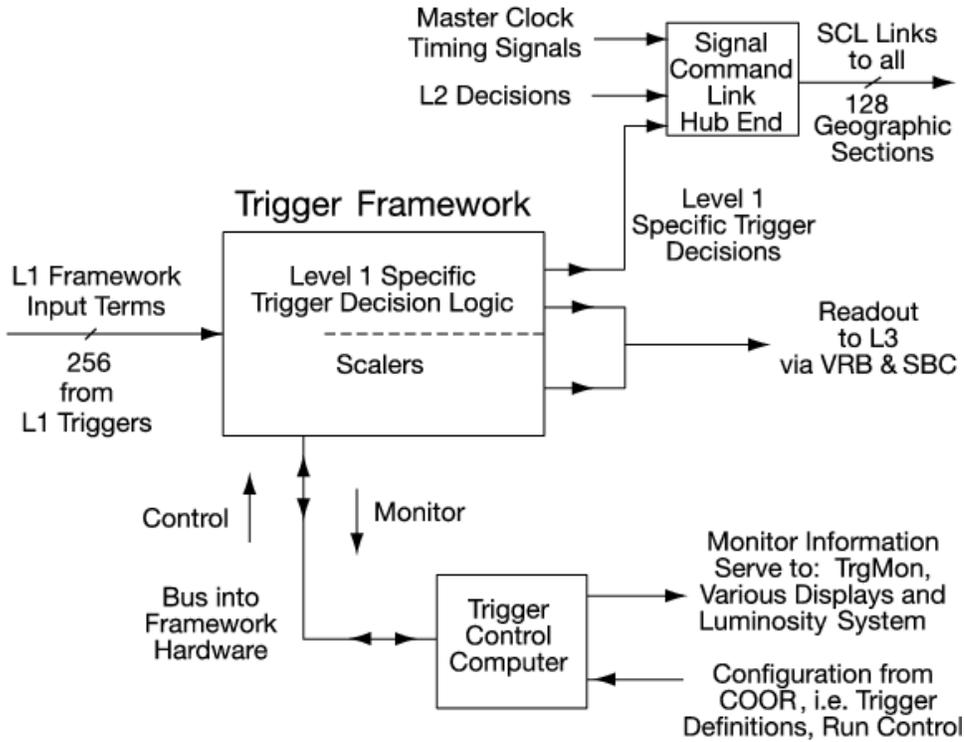}}
\caption{Block diagram of the trigger framework.}
\label{fig:framework}
\end{figure}

The TFW for Run~II is built out of 9U 400~mm cards housed in 
customized VME crates. All of the cards in the TFW use the same general 
circuit board, the same front and rear panel layout, and the same 
connectors and make extensive use of field programmable gate array (FPGA) 
technology to implement different functions.  A 
block diagram of the principal functions of the TFW is shown
in Figure~\ref{fig:framework}.

The functions of the TFW are summarized below:
\begin{itemize}

\item It receives up to 256 ``AND-OR'' terms (bits) from various 
parts of the experiment, which are used to form specific triggers.

\item
Up to 128 specific triggers can be programmed using the AND-OR terms.  
The ``or'' of all of these triggers determines 
whether or not a given crossing had a valid trigger.  These are called 
``physics'' triggers.  Each of the 128 specific triggers has a separate 
programmable ``beam condition'' trigger associated with it.  For a trigger to 
occur, both the ``physics'' trigger and ``beam condition'' conditions must be 
satisfied.

\item 
All the resources, including the triggers, are programmed from COOR
(Section~\ref{sec:coor}) via 
text-based commands interpreted in the TCC which then configures the TFW.

\item
Any trigger can be prescaled under program control. The TFW 
manages the prescale ratios.

\item
At the level of data acquisition, the 
detector is divided into subsections or groups of subsections served 
by a single serial command link (SCL) called geographic sectors. 
Typical geographic sectors are detector front-ends but other systems such as
the TFW readout are also geographic sectors. The TFW supports up to 128 
geographic sectors and any specific trigger can be programmed to 
request digitization and readout of any subset of the 128 geographic sectors. 
This permits a partitioning of the data 
acquisition system into several separate systems, which is particularly 
useful during setup, installation, and calibration. Each geographic sector 
receives an L1Accept signal from the TFW over the SCL, followed by an 
L2Accept or L2Reject from the L2 global processor. 

\item
The TFW can be programmed so that only particular bunch crossings 
contribute to a given trigger.
Beam crossing numbers and L1Accept numbers provide a unique identifier for 
events.

\item
The TFW provides all of the scalers needed to count the triggers as well as 
scalers to monitor the live time of combinations of these triggers
(known as exposure groups).  

\item
The TFW offers an extensive menu of lower level and specialized commands for 
diagnosis of the trigger hardware.

\item
TFW scalers and registers are accessed through the TCC to perform trigger
system programming, diagnostics, and monitoring.

\item
During Run~I, elements of the L1 trigger could only be ANDed to give a
specific trigger.  For Run~II, the capability to allow both ANDs and ORs of
trigger terms has been added to the TFW, using resident firmware.
The resulting combinations are called pseudo-terms, since they are not
formed in hardware.  The pseudo-terms have the ability
either to sharpen trigger turn-on curves, or to reduce combinatorial
backgrounds and thus reduce the rate of low-$p_T$ triggers.
                                                                                
\end{itemize}

\subsubsection{Level~1 calorimeter trigger}
\label{sec:l1cal}

The underlying architecture of the L1 calorimeter trigger for Run~II is 
the same as it was for Run~I \cite{l1cal-run1}.  There are, however, a number 
of improvements in the way the data are received by the trigger and the way the 
results are extracted. 

\paragraph{Summary of operational principles}
The trigger inputs consist of electromagnetic (EM) and hadronic (H) trigger 
tower energies made up from sums in depth and transverse coordinates 
($\Delta\eta \times \Delta\phi = 0.2 \times 0.2$) of fast analog pickoffs from 
the BLS circuits.  There are 12 EM towers and 1280 H 
towers: forty slices in $\eta$, covering the region $|\eta| < 4$, and 
thirty-two slices in $\phi$, covering the full $2\pi$ of the azimuth.  
The tower energies are converted to $E_T$ on input, and have pedestals 
subtracted and energy scales adjusted if necessary.  The variables used in 
trigger calculations are actually the EM transverse energies and the total 
transverse energies (EM+H) formed by adding the corresponding EM and H towers.
The triggers available for use in the experiment include:

\begin{itemize}
\item Global variables ~\\
There are two global variables: $\Sigma E_T$, the sum of all tower $E_T$s, with
four thresholds; and \met, again with four thresholds. 

\item Local variables ~\\
Each EM tower and each EM+H tower is compared to four programmable $E_T$ 
threshold sets.  A given threshold set could contain different values of $E_T$ 
for each tower although in practice they are all set to the same value.  A bit 
is set if a tower exceeds its reference value.  (A feature vetoing the EM 
tower if its corresponding EM+H tower exceeds a programmed reference value is 
available but has never been used.)  The number of EM towers 
and EM+H towers exceeding their thresholds in a given reference set is counted 
and trigger bits are generated if this count exceeds any of several programmed 
count limits.  We chose to use two such count limits for EM towers 
and four for EM+H towers.  

\item Large tiles  ~\\
Since the trigger towers are small relative to the size of jets, we use 
some of the partial $E_T$ sums needed for the \met\ calculation to trigger on
jets.  Conveniently available are sums covering $4 \times 8$ trigger towers in 
$\eta \times \phi$, corresponding to an area of 1.28 in 
this space.  For generating trigger bits, four reference sets are 
available with two count thresholds for each set but only for EM+H and not 
for EM-only large tiles.

\end{itemize}

\paragraph{Modifications for Run~II}
A number of changes were made to the calorimeter trigger in anticipation of 
the higher crossing rates and luminosities expected during Run~II.  

\begin{itemize}
\item{Signal receiver electronics} ~\\
In Run I, the conversion of the BLS energy signals to properly-scaled $E_T$ 
values was done on the calorimeter trigger front-end (CTFE) cards by a network
of hard-wired precision resistors, making gain adjustments difficult.  For 
Run~II, we bypassed this problem by building new front-end 
receiver cards capable of gain adjustment up to a factor of two for 
each channel under computer control.  This change also allowed us to add 
circuitry to decrease the trigger signal rise times to be compatible with the 
132~ns accelerator bunch spacing originally anticipated for Run~II.  Each card 
processes the signals from two 
trigger towers and two such cards were needed to upgrade each Run~I CTFE card.

\item{Readout electronics} ~\\
To speed up the readout of L1 calorimeter trigger data and 
therefore improve the L1 trigger accept rate, the old readout system of 
two eight-bit buses was replaced with ten sixteen-bit buses.  With this 
improvement, the L1Cal accept rate can exceed 10~kHz.

\item{Trigger coverage} ~\\
Trigger tower energy data are read out for the full calorimeter.  
Because of signal to noise considerations, only the trigger towers for 
$|\eta| < 3.2$ are used for triggering.  In addition to the 
calorimeter trigger towers, ICD towers are embedded in the data as well.  

\end{itemize}

\subsubsection{Level~1 central track trigger}
\label{sec:l1ctt}

The L1CTT~\cite{l1ctt-1} reconstructs the trajectories of 
charged particles using fast discriminator data provided by three  
scintillator-based detectors: the central fiber tracker 
(Section~\ref{sec:CFT}) and the central and forward preshower detectors 
(Section~\ref{sec:preshower}). 
Data processed by the L1CTT are used to make L1 
trigger decisions, otherwise known as trigger terms.  While the L1CTT is 
optimized for making fast L1 trigger decisions, the electronics also store 
more-detailed event data (e.g. sorted lists of tracks and preshower clusters) 
for later L2/L3 readout, or for use as seeds by other D\O\ trigger systems.

\paragraph{Overall system design}
The input to the L1CTT system consists of the discriminator bits generated on 
the AFE boards every 132~ns.  These discriminator bits are sent from the AFE 
boards over point-to-point low voltage differential signal (LVDS) links to 
chains of digital front-end  (DFE) boards.  Each of the DFE boards in the 
L1CTT system is built 
using a common 6~U $\times$ 320~mm motherboard that supports one or 
two daughterboards.  Currently there are two different daughterboard layouts 
with different sizes and numbers of FPGAs.  A transition board allows a 
DFE motherboard to drive LVDS, fiber 
optic, and coaxial copper links.  L1CTT-specific protocols define the data 
format for communication between all DFE boards and consumers. This hardware 
modularity, when coupled with the flexibility of FPGAs, enables 
application-specific functionality to be located in firmware, 
minimizing the number of unique DFE boards in the system. 
   
The L1CTT system comprises three subsystems: CFT/CPS axial, CPS stereo, and 
FPS.  Of these, the CFT/CPS axial and FPS subsystems provide L1 trigger terms 
to the trigger framework.  All three subsystems participate 
in L2/L3 readout by sending track and cluster lists to various preprocessor 
and readout crates. 

\paragraph{L1CTT subsystems}

\begin{itemize}
\item{CFT/CPS axial} ~\\ 
The CFT/CPS axial subsystem (Figure~\ref{fig:cft-cps-axial}) is designed to 
provide triggers for charged particles with $p_T > 1.5$~GeV/$c$. 
In addition to finding tracks, it 
must also find CPS clusters, match tracks to clusters, and 
report the overall occupancy of the CFT axial layers.  Significant resources 
are allocated for triggering on isolated tracks. 
The CFT/CPS axial system also supplies 
the L1Muon and L2STT systems with lists of seed 
tracks, and sends track and cluster information to the L2CTT and L2PS 
preprocessors (Figure~\ref{fig:l1-l2-layout}).

\begin{figure}
\centerline{\includegraphics[width=4.in]{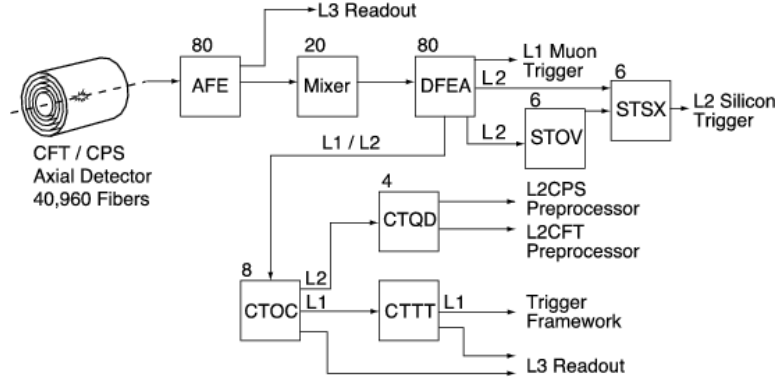}}
\caption{The CFT/CPS axial components of the L1CTT system.  The numbers in the
top left corners of the boxes indicates the number of board of each type.}
\label{fig:cft-cps-axial}
\end{figure}

For mechanical reasons, the CFT and CPS axial fibers are grouped by cylinder 
layer before routing to the AFE boards.  However, the track finder algorithms 
require the fiber information to be arranged in 4.5$^\circ$ sectors in the 
transverse plane (Figure~\ref{fig:trig-sector}). Furthermore, each
track-finder daughterboard must receive information from each of its 
neighboring sectors to find tracks that cross sector boundaries.  
A set of twenty mixer boards \cite{ctt-mixer} handles this data reorganization 
and duplication. After data duplication, the mixer output data rate is a 
constant 475~Gbits/s.  Total latency through the mixer system is 
200~ns.

\begin{figure}
\centerline{\includegraphics[width=3.in]{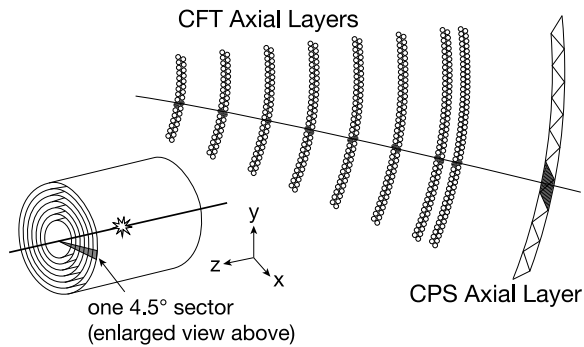}}
\caption{Transverse schematic view of a single 4.5$^\circ$ sector.  A 
hypothetical track is overlaid on the eight CFT axial doublet layers and CPS 
axial layer.  The track equations require a fiber hit on all eight CFT axial 
layers.  Note that the CPS is relatively farther from the CFT than shown.}
\label{fig:trig-sector}
\end{figure}
   
From the mixer system, the properly organized discriminator bits are sent to 
the first tier of DFE boards.  This first tier consists of 
forty motherboards, with each motherboard having two daughterboards 
(DFEAs).  Each DFEA daughterboard unpacks the CFT data and compares the fiber 
hits against approximately 20,000 predefined track equations.  To 
minimize latency, this operation is performed in parallel using combinatorial 
logic in FPGAs.  Device resources are balanced by grouping track equations into 
four $p_T$ bins, with each $p_T$ bin corresponding to a single FPGA, as shown 
in Table~\ref{tab:track-equation}.
   
\begin{table}
\begin{center}
\caption{Track equation distribution in each track-finder FPGA}
\label{tab:track-equation}
\begin{tabular}{lccc}
\hline 
           & $p_T$ range &                 & FPGA Resources \\
 $p_T$ bin & (GeV/$c$)   & Track equations & (system gates) \\
\hline
Maximum    & $> 10$     & 3000             & 200k \\
High       & 5 -- 10    & 3000             & 200k \\
Medium     & 3 -- 5     & 4000             & 300k \\
Low        & 1.5 -- 3   & 10000            & 500k \\
\hline
\end{tabular}
\end{center}
\end{table}
   
Each track-finder FPGA outputs the six highest-$p_T$ tracks it finds; these 
tracks are passed to a fifth FPGA for additional processing which includes 
sorting, matching tracks and CPS clusters, counting tracks, and calculating 
sector occupancy and total $p_T$.  Tracks from each sector are sent over 
gigabit coaxial cables to L1Muon, where the tracks are matched 
to hits in the muon detector.  Counts of tracks and occupancy data are passed 
downstream to the octant boards over LVDS links.

The next tier of DFE boards (CTOC) collects and sorts data within an octant 
(ten sectors).  The CTOC boards receive track counts and sector 
occupancy data from the DFEAs and then sum up the number of 
tracks, determine which sector had the most fibers hit, and check for isolated 
tracks.  This information is passed on to a single DFE board (CTTT) 
where the individual trigger term bits are generated and sent to the trigger 
framework within 2.5~$\mu$s of the beam crossing.  The CTTT can provide up to 
96 trigger terms of which 55 are currently defined.  Two examples of CTTT 
trigger terms are {\it i}) at least one track above a 
$p_T$ threshold of 1.5, 3, 5, or 10~GeV/$c$ and {\it ii}) at least one 
isolated track above 5 or 10~GeV/$c$ and at least one track with a 
confirmation in the CPS for triggering on electrons.

The TFW issues an L1Accept control bit by
considering trigger term bits from many different 
subsystems and makes a global L1 decision.  Upon receiving an L1Accept, the 
AFE boards digitize the fiber data from the appropriate analog buffers in the
analog pipeline with 
8-bit resolution.  When the L1Accept control bit embedded in the data 
stream reaches the DFEA boards, they send a sorted list of tracks 
and CPS clusters to the CTOC boards.  At this processing stage, finer $p_T$ 
information is available than that used at L1.  The lists are sorted by 
the CTOC and CTQD boards and passed downstream to preprocessor crates for 
more detailed analysis.  These track lists, remapped onto the geometry of 
the silicon tracker, are also used as seeds for the L2STT.  Additionally 
the CTOC and CTTT boards send copies of their input data to the L3 readout 
system for monitoring and debugging.
   
\item{CPS stereo} ~\\
The CPS stereo subsystem (Figure~\ref{fig:cps-stereo}) provides 
information on the clusters in the two stereo layers of the CPS.  Unlike the 
other L1CTT subsystems, CPS stereo does not generate L1 trigger terms.  Rather, 
the DFE stereo (DFES) boards store discriminator bits and begin processing 
only after an L1 trigger decision has been made.

\begin{figure}
\centerline{\includegraphics[width=3.in]{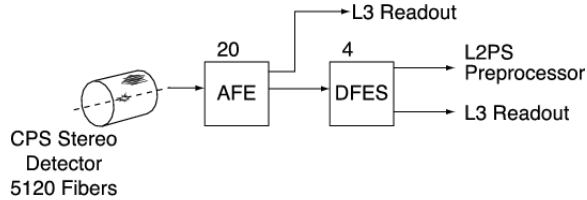}}
\caption{CPS stereo subsystem hardware.}
\label{fig:cps-stereo}
\end{figure}
   
Event buffers in the DFES FPGAs store discriminator bits sent from the AFE 
boards. Upon receipt of the L1Accept control bit, the DFES boards extract 
the discriminator bits and search for clusters of hits in the stereo layers.  
Sorted lists of CPS clusters are sent on to the L2PS preprocessor for 
additional analysis.  The DFES boards also send a copy of either their
inputs (the discriminator bits) or their L2 outputs (the hit clusters) 
to the L3 readout system.
   
\item{FPS} ~\\
The FPS subsystem (Figure~\ref{fig:ctt-fps}) produces its own set of L1 
trigger terms, which are passed to the TFW.  The overall 
structure of the FPS subsystem is similar to the CFT/CPS axial subsystem in 
that the FPS has three tiers of DFE boards: a finder (DFEF), a concentrator 
(FPSS), and a trigger term generator (FPTT).
   
\begin{figure}
\centerline{\includegraphics[width=3.in]{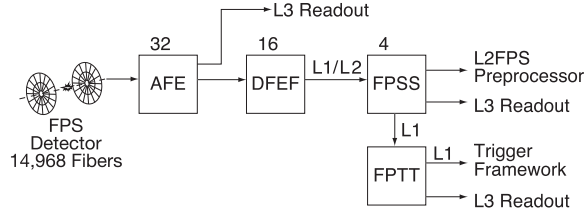}}
\caption{FPS subsystem hardware.}
\label{fig:ctt-fps}
\end{figure}

The DFEF boards receive discriminator bits from the AFE boards and search for 
clusters of hits in the FPS fiber layers.  A list of clusters is stored in the 
DFEF for later L2 readout while counts of clusters are passed downstream to 
the FPSS boards.  The FPSS boards sum these cluster counts and pass this 
information to the FPTT, where the L1 trigger terms are produced.

When the TFW issues an L1Accept, the FPS subsystem switches 
into readout mode. The AFE boards begin digitizing and reading out fiber data, 
while the DFEF boards extract cluster lists from their buffers.  These 
lists of clusters are concatenated and sorted by the FPSS boards before being 
sent to the L2PS preprocessor.  The FPSS and FPTT boards send their L1 input 
data to the L3 readout system.  

\item{STT} ~\\
The STT subsystem, described in detail in Section~\ref{sec:l2stt}, 
is used to map the DFEA outputs onto the sixfold STT/SMT geometry.
After an L1Accept has been issued, lists of L1CTT seed tracks generated by
the DFEAs are concatenated and reformatted for the L2STT
by an L1CTT subsystem consisting of two types of boards:
six STOV (``overlay'' or ``overlap'') and six STSX (``sextant'') boards 
(Figure~\ref{fig:cft-cps-axial}).  Two types of board are necessary 
because the maximum number of input/output links on each
board does not allow the remapping to be done in one set of boards.
The DFEAs have duplicate outputs: one set feeds the CTOCs,
the other the STOVs (DFEA outputs shared between L2STT sextants) and STSXs
(outputs unique to a sextant). The STOV outputs are fed into the STSXs; the 
STSXs send the data (lists of L1CTT tracks) via six optical fibers to the 
L2STT system.

\end{itemize}

\subsubsection{Level 1 muon trigger}
\label{sec:L1Muon}

L1Muon looks for patterns consistent with
muons using hits from muon wire chambers, muon scintillation counters,
and tracks from the L1CTT.  Field
programmable gate arrays are used to perform combinatorial
logic on roughly 60,000 muon channels and up to 480 tracks from L1CTT
for every bunch crossing.  Data from the detector front-ends are
transmitted on custom Gbit/s serial links over coaxial cable.
The serial link receivers and FPGAs are located on VME cards that
reside in four custom VME crates on the detector platform.

\begin{figure} 
\centerline{\includegraphics[width=5.in]{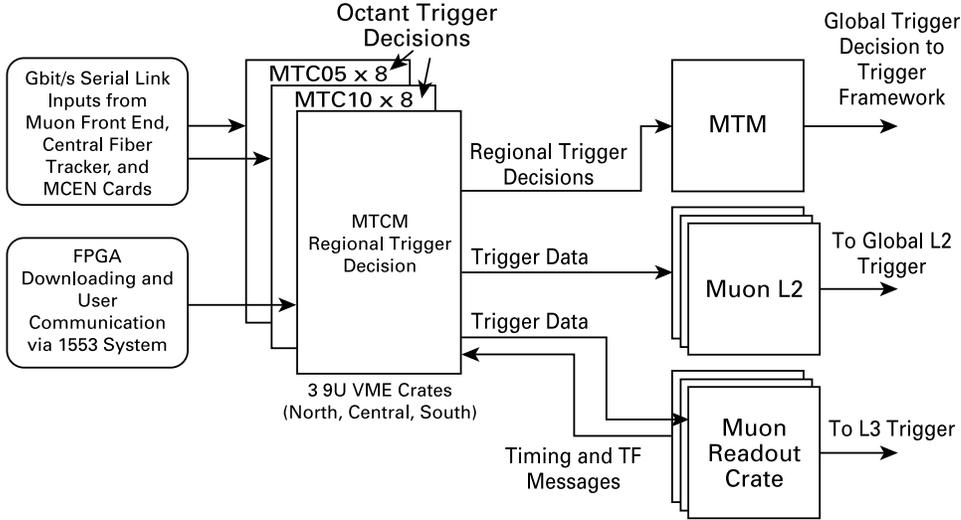}} 
\caption{Level~1 muon trigger system overview.  Each octant has
two trigger cards that process muon detector hit information and
L1CTT tracks for that octant.  The octant triggers for a given
region are summed on the MTCM and sent to the MTM, which
combines the three regions and sends up to 32 triggers to the TFW.}
\label{L1Muon_overview}
\end{figure}

The muon system (and L1Muon) is divided into central, north, and south
regions.  Each region is further divided into octants.  Front-end data
from each octant are processed by two L1Muon trigger cards (Figure
\ref{L1Muon_overview}).  The scintillator trigger cards (MTC05) match
central tracks to muon scintillator hits while the wire trigger cards
(MTC10) match scintillator-confirmed track stubs in wire chambers
between the two or three layers of the muon system.  The octant decisions
from each MTC05/MTC10 pair in a region are summed in the muon trigger
crate managers (MTCMs) and sent to the muon trigger manager (MTM).
The MTM forms 256 global L1Muon triggers and sends up to 32 of these to
the TFW.
The download of the specific triggers is handled via EPICS software
(Section~\ref{sec:controls}).
The total latency of the L1Muon trigger is about 3.20~$\mu$s,
driven by the central wire chambers (PDTs) and tracks from L1CTT.

\paragraph{L1Muon hardware}
All detector inputs to L1Muon use Gbit/s serial links.
The transmitters and receivers are 1.5'' $\times$ 2.2''
daughter cards that are mounted on the muon front-end boards and on the L1Muon
trigger cards.  Each serial link can transmit up to $16 \times 7 = 112$ bits
every 132~ns crossing.  All MTC05, MTC10, and MTM trigger cards
use a common motherboard with sixteen serial links and different
flavor daughter cards that perform the MTC05, MTC10, and MTM logic.

The MTC05 cards match tracks from L1CTT to hits in the muon
scintillator system.  Each octant trigger card receives tracks from
the L1CTT for the ten 4.5$^\circ$ sectors in that octant plus one
sector of overlap on either side.  Each sector sends the six highest-$p_T$ 
tracks to L1Muon, and each track contains the CFT fiber position in the
outer layer, $p_T$ value, and sign of the track curvature in the central
magnetic field.  The triggers formed by the MTC05 cards include loose (track
matched to A-layer scintillator) and tight (track matched to a
scintillator road using the A and B layers) for four $p_T$ thresholds
(roughly 1.5, 3, 5, and 10~GeV/$c$).  Loose and tight scintillator-only
triggers are also formed.  

The MTC10 cards form triggers based on wire hits.  In the central
region, the hits from the wire chambers (PDTs) are sent directly to
the trigger cards.  The hits for each layer are used to form track
stubs, or centroids, which are then used to confirm scintillator hits
in each layer.  Triggers are formed by matching centroid-confirmed
scintillator hits between layers.  In the forward region, the centroid
finding is done by separate centroid-finding cards (MCENs), which
subsequently send the centroids to the MTC10 cards.  The MTC10 cards
then use the centroid-confirmed scintillator hits to form loose
(A-layer) and tight (A- and B-layer) triggers.

The data from the various front-end systems arrive asynchronously at
L1Muon and must be synchronized before triggers can be formed for a
given event.  To accomplish this, all received data are written
directly into FIFOs which are initially empty.  When all FIFOs are
not empty (i.e., they have all received data for the first bunch
crossing), the data are read from the FIFOs and sent to the MTC05,
MTC10, or MTM cards for trigger formation.  In addition to
synchronizing the data for a given event, the trigger cards also
buffer the input data and trigger decisions pending global L1 and
L2 trigger decisions.  The input data and trigger decisions are
stored in dual port memories and a pointer to the data is
written into a FIFO.  When an L1 or L2 accept is received,
the pointer is used to read the data for a particular event.  The L1Muon
trigger can also send all of the received input data from the detector
front-ends to aid debugging.

\subsubsection{Level 1 forward proton detector trigger}
\label{sec:l1fpd}

\begin{figure}
\centerline{\includegraphics[width=6.in]{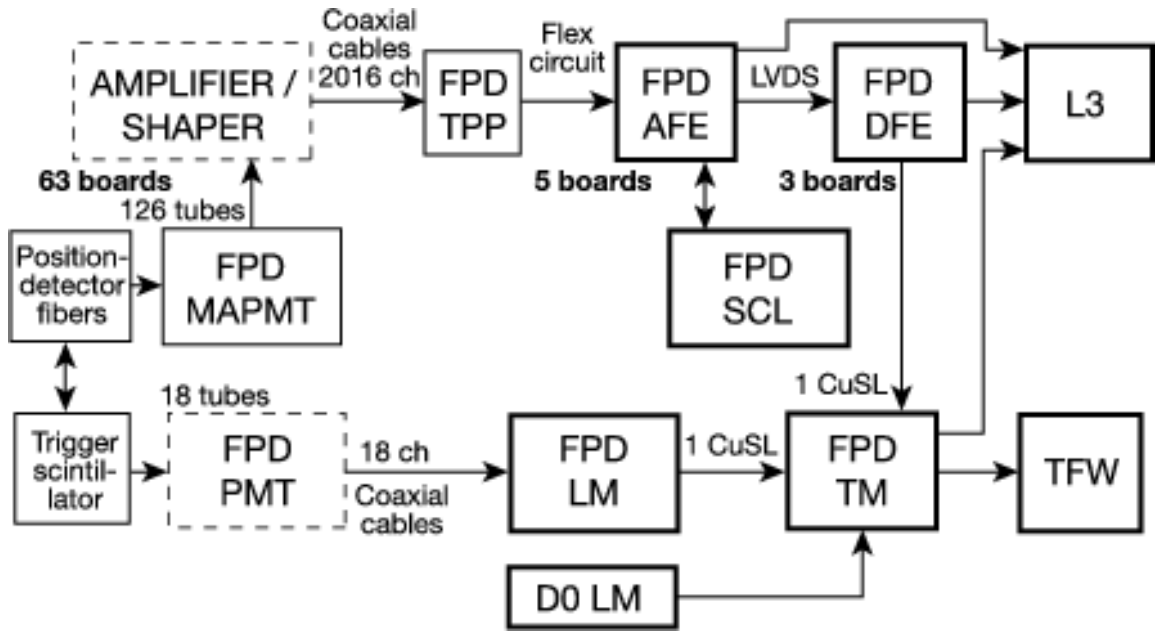}}
\caption{A schematic view of the L1FPD trigger.  The dashed boxes indicate 
Run~I surplus boards; the bold boxes indicate Run~II standard; and the thin
boxes indicate custom FPD electronics.  CuSL is a copper serial link.}
\label{fig:l1fpd-flow}
\end{figure}

L1FPD selects events in which the outgoing beam particles pass through one 
or a combination of the nine FPD spectrometers (Section~\ref{sec:fpd}).  
A schematic view of the 
L1FPD design is shown in Figure~\ref{fig:l1fpd-flow}. 

The signals produced in the FPD scintillating fibers are read out via 
16-channel MAPMTs, and are then amplified and shaped before being passed to 
the AFE boards. In addition to providing analog fiber charge integration 
information to the L3 data acquisition system, the AFE boards pass 
discriminated signals via LVDS connections to 
three DFE boards every 132~ns.  At the DFE boards, pixel hit pattern 
recognition (a pixel is defined to be a $u$, $x$, and $v$ fiber crossing)
is used to compare the fiber hit information with detector 
hit patterns stored in FPGAs. Three 
FPGAs containing between 600,000 and 1.5 million gates store 
generated L1 hit patterns for the nine spectrometers.  The DFE boards also 
send their processed data to the L3 trigger system.
The relative positions of the fibers in the different planes of each 
detector are used to define a finely segmented grid in which to search for
hit patterns with optimal spatial resolution.  Hit trigger conditions are set 
with different thresholds corresponding to different numbers of fiber layer 
hit coincidences. 

The FPD trigger manager (TM) searches for 
coincidences between the discriminated hit signals of both position 
detectors of any FPD spectrometer.  We use a set of nine FPD spectrometer  
(single diffractive) triggers with differing hit thresholds to 
select events in which at least
one of the outgoing beam particles leaves the interaction region intact.
Coincidences between the spectrometers on different sides of the 
interaction region are required to trigger on events in which both 
outgoing beam particles leave the interaction region intact in 
back-to-back configurations (elastic diffractive 
triggers) and in the larger set of other two-spectrometer configurations
(double pomeron triggers). 
To reduce the contribution from beam halo spray particles that mimic a trigger
signal, events with large hit multiplicities are rejected.
The FPD information can be combined with timing information from 
the FPD trigger scintillator photomultiplier tubes (FPD LM), veto
counters, and the luminosity monitor (D\O\ LM).
The decision is then passed to the TFW.

\subsection{The Level~2 trigger}
\label{sec:l2}

The L2 trigger provides detector-specific preprocessing engines 
and a global stage (L2Global) to test for correlations in physics signatures
across detector subsystems.
The L2 trigger system was designed to handle input rates of 
up to 10~kHz with a maximum accept rate of 1~kHz.   
L2 preprocessors collect data from the front-ends and L1 
trigger system and analyze these data to form physics 
objects.  L2 can also combine data across detectors to form higher 
quality physics objects and examine event-wide correlations in all L2 physics
objects.  The 
L2Global processor selects events based on the set of 128 selections 
applied at L1 and additional script-controlled criteria.
Events passing L2 are tagged for full readout and further 
analysis in the L3 trigger.

\subsubsection{Architecture and components of the L2 trigger}

\begin{figure}
\centerline{\includegraphics[width=6.in]{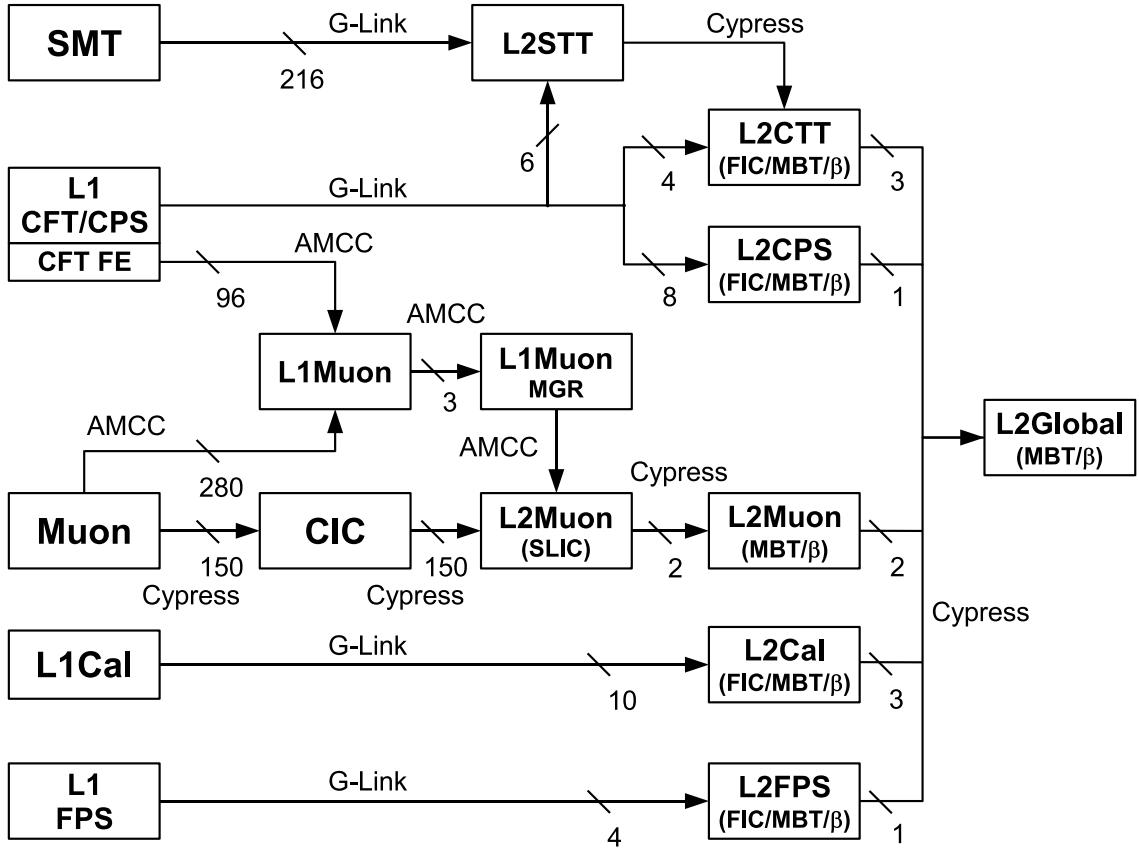}}
\caption{L2 data paths and connections.  Data paths are labeled according
to transmission protocol, either 1.3~Gb/s fiber G-Links, 160~Mb/s copper
Cypress Hotlinks, or 1.4 Gb/s copper AMCC links.  FIC indicates a fiber
converter card, MBT an Mbus transceiver card, and $\beta$ an L2$\beta$eta card.
The number of transmission cables is given for each pathway.}
\label{fig:l2_connections}
\end{figure}

L2 includes preprocessors for each detector subsystem and 
a global processor for integration of the data.  Preprocessor subsystems 
include tracking, calorimeter, preshower, and muon systems.  The subsystems  
work in parallel and trigger decisions are made in the L2Global stage 
based on physics objects reconstructed in the preprocessors.  
Preprocessing is performed either with serial CPU-based cards or with 
CPU cards plus highly parallelized DSP or programmable logic-based cards.
The preprocessor and global stages function as 2- or 3-stage 
stochastic pipelines as illustrated in Figure~\ref{fig:l2_connections}.  

Data arrive at the L2 system via three transmission protocols.
Calorimeter and tracker data and signals from the TFW are
transmitted by 1.3~Gbit/s serial G-Links~\cite{l2-glink} on optical fibers.
The muon system uses 160~Mbit/s Cypress Hotlink~\cite{l2-cypress-hotlink}
transmitters on coaxial cables or standard CAT/6 cables, 
unshielded twisted pair (UTP) Hotlinks.

\paragraph{L2 crates}
The L2 system occupies 9U VME crates built to the VME64/VIPA \cite{VIPA} 
standard.  In addition to a 64-bit VME bus, these crates are instrumented with 
a custom high-bandwidth 128-bit bus (Magic Bus~\cite{l2-mbus} or Mbus) 
for fast intra-crate data flow and communication.  The Mbus 
supports data rates of up to 320~Mbit/s.  
The VME backplane is used to read out events accepted by L2 into L3 and for 
control and monitoring operations.

A typical L2 preprocessor/global crate contains the following devices: 
\begin{itemize}
\item VME controller and dual-port-memory (DPM) card used for 
downloading runtime parameters and for reporting monitoring data to the TCC
\item a VME single-board computer (SBC) used to send data to L3
\item one or more Mbus transceiver (MBT) cards (Hotlink-to-Mbus 
interface, Section~\ref{sec:mbt})
\item one or more sets of fiber converter (FIC) and VME transition (VTM) 
cards (fiber-to-hotlink interface, Section~\ref{sec:mbt})
\item one or more L2$\beta$eta processor cards (VME and Mbus hosted
SBCs, Section~\ref{sec:beta}) for data processing.
\end{itemize}

\begin{figure}
\centerline{\includegraphics[width=3.5in]{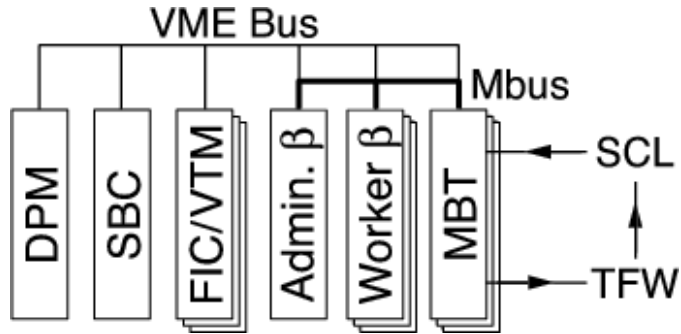}}
\caption{Typical layout of cards in an L2 preprocessor crate.
Cards that could be installed in parallel to increase input channels 
or processing power are designated by stacked boxes.}
\label{fig:l2-crate}
\end{figure}

The layout of the preprocessor/global crates is shown in 
Figure~\ref{fig:l2-crate}.  L2STT (Section \ref{sec:l2stt}) and 
L2Muon (Section \ref{sec:l2muon}) use additional specialized cards for 
processing their data before sending track 
information to the L2CTT and L2Muon preprocessors respectively.

\paragraph{Magic bus transceiver and data conversion cards}
\label{sec:mbt}
Data arriving at L2 must first be converted into the correct
physical format before processing can begin.  The coaxial Hotlink 
data from the muon system is received by analog cable input 
converter (CIC) cards.  These cards retransmit the data to 
the first stage of L2Muon as differential UTP Hotlink signals.  A relative
of the CIC card, the serial fanout (SFO), selectively fans out 1 to 12 
UTP Hotlinks where signal duplication is necessary.
Similarly, FIC cards receive G-Link data
and buffer up to sixteen events while converting the inputs to differential
UTP Hotlink signals.  The incoming fibers are received by a VME transition 
module (VTM) located at the rear of the crate.  

Data are passed to L2$\beta$eta processors via the Mbus.  
An MBT card transfers data to and from the Mbus.
The MBT accepts up to seven Hotlink inputs, assembles the data into events, 
and broadcasts them on the Mbus backplane to the L2$\beta$eta cards.    
The high speed, 128-bit input path is equipped with a buffer 
for up to sixteen events.  There are two Hotlink
output paths that are targets of the 128-bit Mbus programmed 
I/O on the L2$\beta$eta cards.  These
are used by preprocessors to send their output to L2Global.  

One MBT, designated the ``pilot,'' coordinates the event broadcast across all
MBTs and receives SCL information for the crate from the TFW, including the
L1Accept information, L2 decision information, and SCL initialize messages.  The
pilot MBT in the L2 global crate transmits the L2 decision to the TFW over
sixteen PECL output channels.  A demultiplexing card receives eight 16-bit words
and transmits the 128-bit decision to the TFW.

\paragraph{L2$\beta$eta processors}
\label{sec:beta}
L2$\beta$eta processor cards are used for analysis and control of 
data flow in each preprocessor subsystem crate.  The L2$\beta$eta processors 
replaced the original L2Alpha boards\footnote{L2Alpha processors were similar 
to those constructed by the CDF 
experiment~\cite{l2-alpha-1,l2-alpha-2}.  They were based on the PC164 
motherboard manufactured by Digital Semiconductor (now Compaq).  Each card 
contained a 500~MHz 21164 64-bit Digital Alpha CPU and 21164 PCI
interface integrated with the necessary hardware interfaces. 
The L2Alphas have been supplanted by the higher performance and 
upgradable L2$\beta$eta cards.} early in Run~II and 
are composed of commercially produced (Compact PCI 
standard~\cite{l2-compact-pci}) SBCs mounted on 6U-to-9U VME adapter cards.
Each SBC provides dual 1~GHz Pentium processors.
The SBC connects to the adapter via a 64-bit, 33/66 MHz PCI bridge.  
The 9U adapter, controlled by the SBC, implements 
all D\O -specific protocols for Mbus and
TFW connections.  Custom I/O functions on this card are
implemented in a single FPGA (Xilinx XCV405E) plus assorted logic converters
and drivers.  The FPGA is used to implement 
internal data FIFOs and address translation tables for
broadcasting data from the Mbus to CPU memory, reducing the complexity
of the adapter card.  Mbus programmed I/O and various other trigger system
interfaces, via front panel connections and VME user-defined pins,
are also implemented.
The adapter also provides a 64-bit PCI-to-VME interface via 
a Tundra Universe II chip~\cite{U2}.

The L2$\beta$etas run a GNU/Linux system and all programs 
are written in C++.  Real time trigger performance is achieved by 
restricting system calls (including dynamic memory allocation) 
and promoting the L2 executable to a non-interruptible 
scheduling queue.  The second CPU provides full access to the 
SBC for runtime monitoring and control.

\subsubsection{L2 preprocessor subsystems}

\paragraph{L2Cal}
\label{sec:l2cal}
The calorimeter preprocessor system identifies jets and electrons/photons 
and calculates event \met\ for 
the global processor.  The worker code may be combined to 
run serially in a single processor or placed 
in separate processor cards to increase throughput as data rates and 
detector occupancies grow with luminosity.
Each processor uses the $E_T$ data from the 2560 calorimeter
trigger towers.  
The input data arrives on ten input links which together transport 
3~kB/event of tower transverse energy data, including both EM towers 
and the EM+H tower sums.
 
The jet algorithm operates by clustering $n\times n$ (currently $n=5$) 
groups of calorimeter trigger towers which are centered on seed towers.  
The seed towers are $E_T$-ordered with $E_T \ge 2$~GeV.  
Overlapping candidates may be reported as separate jets depending 
on programmable criteria based on the number of shared towers (otherwise the
highest-$E_T$ jet of the overlapping candidates is kept).  
The list of  jets is passed to L2Global which applies jet requirements 
as defined by the trigger menu.  

The electron/photon algorithm begins by defining an $E_T$-ordered list of 
EM towers with $E_T$ above 1~GeV.  For each seed tower, the neighboring 
tower with the largest $E_T$ is combined with the seed to make an EM cluster.  
The EM energy fraction of the leading- and sub-leading-$E_T$ trigger towers 
of the cluster and the amount of total $E_T$ in a $3 \times 3$ tower array 
surrounding the seed tower of the cluster are used to reduce background.  
The final list of electron candidates is sent to L2Global 
to apply the trigger requirements.  

The L2 calorimeter \met\ algorithm  
calculates the vector sum $E_T$ from the individual trigger tower total-$E_T$ 
energies passed from L1.  It is capable of calculating the \met\ for
different minimum tower $E_T$s and $\eta$ ranges.

\paragraph{L2Muon}
\label{sec:l2muon}
L2Muon uses calibration and more precise timing information to improve the 
quality of the muon candidates~\cite{l2mu-1}.  It receives the L1Muon output 
and data from approximately 150 front-end modules 
(from the PDTs, MDTs, and the scintillation counters).  The muon 
candidates contain the track $p_T$, $\eta$ and $\phi$ 
coordinates, and quality and timing information.  

\begin{figure}
\centerline{\includegraphics[width=3.in]{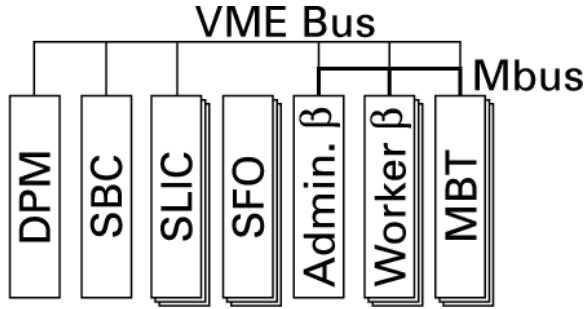}}
\caption{Layout of cards in an L2Muon crate.  Two stages of
processing are completed in a single crate using SLICs and 
L2$\beta$eta processors.  Central and forward muon regions are processed
in separate crates of similar configuration.}
\label{fig:slic_crate}
\end{figure}

L2Muon implements one extra level of preprocessing  in the 
stochastic pipeline sequence.  The first L2Muon stage incorporates eighty  
200-MHz DSPs in a parallel processing scheme.  Each DSP is 
responsible for finding track segments in a small region of the 
detector, so that the total execution time of the algorithms 
is independent of the number of hits.  The DSPs are geographically 
organized in eleven central and five forward 9U VME boards (second level input 
computers or SLICs), with each SLIC containing one administrator and four 
worker DSP chips.  The SLICs are programmable in C.  The layout of an 
L2Muon crate is shown in Figure~\ref{fig:slic_crate}.  The DSP algorithms, 
characterized by the detector region (central or forward) and the input or 
sub-detector plane (L1, A or BC muon layers), make use of 
detector symmetry to 
run the same basic processing code.  The MBT 
sends the stubs found by the SLICs to the L2$\beta$eta processor.  
The L2$\beta$eta board uses the track segments to construct
integrated muon candidates with an associated $p_T$ 
and quality.

\paragraph{L2PS}
Both the central and the forward preshower detectors are designed to provide 
high electron detection efficiency, electron-photon separation 
and high background (charged hadron) rejection at the trigger 
level.  This is accomplished by providing evidence 
for early shower development and by giving a good spatial point 
for comparison with calorimeter clusters or tracks.
At L2, the CPS and FPS are treated as separate detectors 
and their data are processed independently.

Upon L1Accept, CPS axial clusters from each of the eighty 
azimuthal trigger sectors are combined into 
quadrants in azimuth and transmitted to the L2 preprocessor.
Stereo clusters are sent directly to the L2 preshower preprocessor.  
Figure~\ref{fig:l2_connections} shows the L1 to L2 data transfer path into
the preshower subsystem.  
The L2CPS preprocessor receives axial ($x$) clusters over four 
G-Links, each serving one azimuthal quadrant.  The CPS stereo 
data are transmitted over four G-Links, two for each hemisphere
(positive/negative $\eta$), including both $u$ and $v$ layer clusters.
Axial clusters are tagged by L1 with the presence or 
absence of a CFT track.  Each of the G-Links can carry a maximum 
of 48 axial CPS clusters or 96 stereo clusters.

In the L2$\beta$eta processor, the CPS cluster centroids are 
compared to produce $\eta$ and $\phi$ coordinates for clusters
that match in three layers.  The presence or 
absence of CFT trigger tracks associated with CPS $x$ clusters is also 
provided, and output clusters are flagged as electrons (when there is a track 
associated with the cluster) or photons (no track).    The $\eta$ and $\phi$ 
coordinates are binned to correspond to the calorimeter trigger tower geometry 
of $\eta \times \phi = 0.2 \times 0.2$.  A window of width 0.05 is drawn 
around each calorimeter trigger tower, and any preshower hit in 
this $\eta,\phi$ region is designated a calorimeter match.
The FPS provides similar functionality and is the only source of forward 
tracking information available before the L3 trigger.

\paragraph{L2CTT}
The L2CTT preprocessor takes inputs from the L1CTT and the L2STT.  This 
preprocessor system has been designed to operate in two different modes: 
{\it i}) with input tracks straight from L1CTT and {\it ii}) with input tracks 
from L2STT which receives input from the L1CTT and SMT barrels.

In the first mode of operation, L2CTT reads in the track 
lists from different $\phi$ 
regions of the L1 tracking trigger system and
concatenates these into a single $p_T$-sorted list.  The $p_T$ measurements 
are refined using additional hit and tracking information than is 
available at L1.  For each track, the azimuthal angle with respect to the 
beamline, $\phi_0$, is determined.  The value of the azimuthal 
angle at the third layer of the EM calorimeter, $\phi_{em3}$, is also 
calculated ($\phi_{em3}$ is different from $\phi_0$ due to the bending of tracks
in the solenoidal magnetic field).  Finally, each track is evaluated according
to several isolation criteria to enhance the trigger capabilities for tau
leptons.  The $p_T$-sorted list of L2 tracks 
is reported to L2Global.  In the second mode of operation, input data are
provided by the L2STT along with refined L2 track $p_T$s.
Only $\phi_0$, $\phi_{em3}$, and isolation are calculated for these data.
However two separate lists of L2 tracks are passed on to L2Global, 
one sorted according to $p_T$ and another sorted according to impact parameter.

\paragraph{L2STT}
\label{sec:l2stt}
The L2STT performs online pattern 
recognition in the data from the SMT.  It reconstructs 
charged particle tracks found in the CFT at L1 
with increased precision by utilizing the much finer spatial resolution of 
the SMT.  

The L2STT improves the momentum measurement of charged particle tracks at the 
trigger level.  Requiring hits in the SMT helps reject spurious 
L1 triggers from accidental track patterns in the CFT.  The primary physics 
justification of the L2STT is its ability to measure the impact parameter of 
tracks precisely enough to tag the decays of long-lived particles, 
specifically $B$ hadrons.   

\begin{figure}
\centerline{\includegraphics[width=3.in]{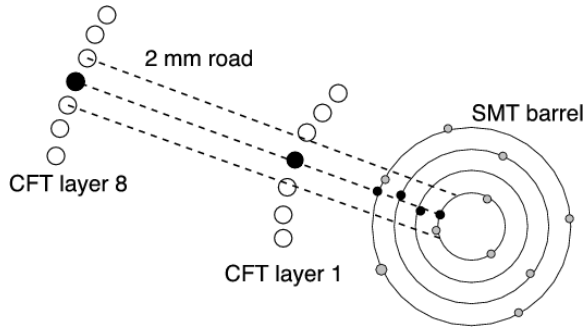}}
\caption{The definition of roads based on L1 tracks and SMT hit 
selection in L2STT.}
\label{fig:STT_idea}
\end{figure}

Figure \ref{fig:STT_idea} shows the basic principle of the L2STT.  For each 
event, the L1CTT sends a list of tracks to the L2STT.  
A road is defined around each track, and the SMT hits within the road are 
associated with the track.  The L2STT uses only the hits in the axial strips
of the silicon ladders, which define points in the $r-\phi$ plane.  The L2STT 
uses the hits in the innermost and 
outermost layers of the CFT and hits in at least three of the four layers 
of the SMT to fit the track parameters.  The results of the fits are sent to 
L2Global.  

The SMT barrel ladders are arranged in twelve sectors, each covering $30^\circ$ 
in azimuth.  The ladders of adjacent sectors overlap slightly such that more 
than 98\% of all tracks are contained in a single sector.  The L2STT therefore 
treats all $30^\circ$ sectors independently.  

The L2STT consists mainly of custom-designed digital electronics modules.  All 
custom modules are designed to plug into a motherboard, and a common
motherboard design is used throughout the system.  Data input from the 
SMT detector and the L1CTT is via optical fiber serial links into 
receiver cards (VTMs) located in the 
rear card cage of the VME64/VIPA crates that house the L2STT 
electronics.  The data are processed in large FPGAs and/or 
DSPs on a logic daughterboard that sits on the motherboard.  There are three 
different types of such daughterboards in the system.  Data are communicated 
between modules in an L2STT crate using a serial link transmitter and 
receiver cards.  Each module is equipped with a daughterboard 
that buffers data for readout through the data acquisition system once an 
event has been accepted by the trigger system.  The logic daughterboard is 
connected to the buffer cards, the serial links, and to the VME backplane by 
three PCI buses on the motherboard.  The VME bus is used for initialization 
and monitoring and to read data out of the buffer cards.  
 
The three types of logic daughterboards are the fiber road card (FRC), silicon 
trigger card (STC), and track fit card (TFC).  The FRC receives data from 
the L1CTT and the TFW that it fans out to the other 
modules.  The FRC also manages the storage of data in the buffer cards.  The
STC receives the SMT data, clusters hits in adjacent strips, and associates 
SMT clusters with roads.  The TFC 
performs the final hit selection and fits the tracks.

L2STT consists of six identical VIPA crates, each serving two $30^\circ$ 
sectors.  Each crate is equipped with one FRC module, nine STC modules, 
and two TFC modules.  Each crate also has a CPU board to program the FPGAs and 
DSPs, download constants, and monitor system performance.  An SBC 
is used to read data out of the buffer cards and
feed them into the data acquisition system.

\begin{figure}
\centerline{\includegraphics[width=3.5in]{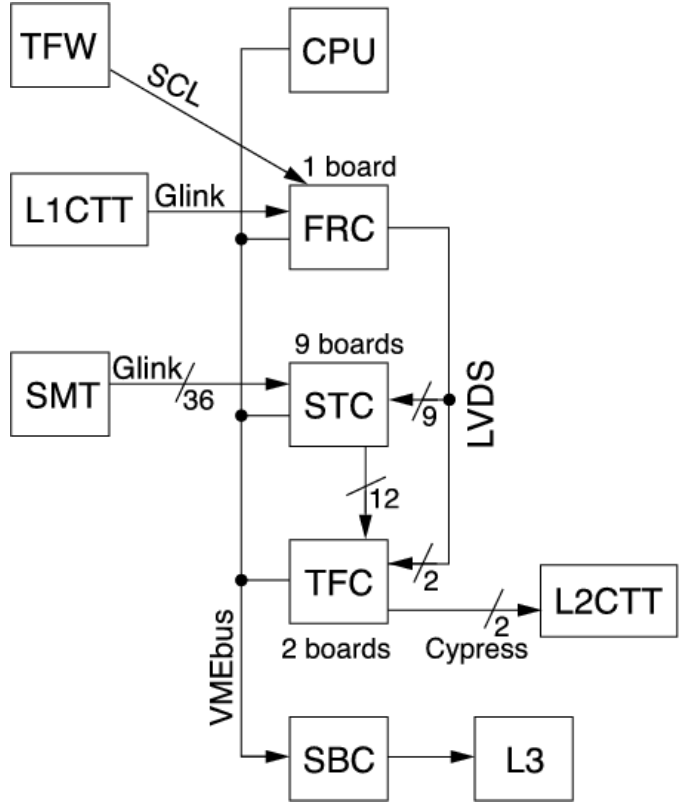}}
\caption{Data flow through an L2STT crate.}
\label{fig:STT-flow}
\end{figure}

Figure~\ref{fig:STT-flow} shows the flow of data through an L2STT crate.
Each of the twelve TFCs provides a list of tracks from its $30^\circ$ sector.  
The track information includes track parameters and the $\chi^{2}$ of the fit 
as well as additional information about the cluster selection and fit.
One TFC in each of the six crates also transmits a 
list of the initial L1CTT tracks.     
These data are transmitted to the L2CTT, where the 
tracks are sorted by $p_T$ and impact parameter and passed to L2Global
to be used in the trigger decision.
The data are also sent to a buffer card for readout to L3.

\begin{figure}
\centerline{\includegraphics[width=3.in]{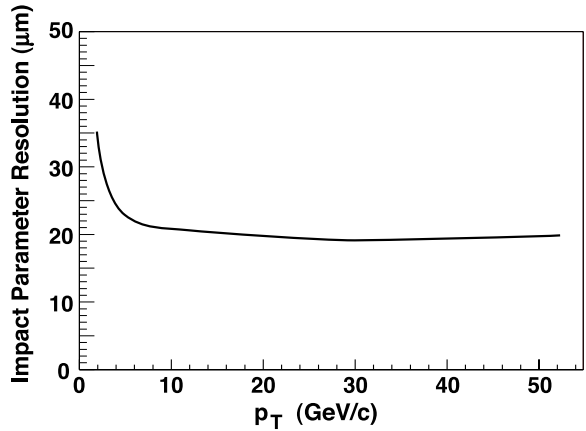}}
\caption{Impact parameter resolution as a function of $p_T$ determined
using simulated single muon events.}
\label{fig:mures}
\end{figure}

The L2STT helps to select events with an enhanced heavy-flavor content by 
measuring the impact parameter $b$ of reconstructed tracks with respect to 
the beam.  Figure~\ref{fig:mures} shows the impact 
parameter resolution $\sigma_b$ obtained from simulated single muon events.  
The impact parameter resolution has a $p_T$ dependence introduced by multiple 
scattering.  In the trigger, the effect of this
$p_T$ dependence can be reduced by using the impact parameter significance $S_b
\equiv b/\sigma_b$ instead of the impact parameter $b$.  The uncertainty
$\sigma_b = \sqrt{(18\, \mu\mathrm{m})^2 + 
[(50\, \mu{\mathrm{m\, GeV}/c})/p_T]^2}$ 
(for tracks with four SMT clusters in the fit) takes into account the effect 
of multiple scattering.  

\subsubsection{L2Global}

L2Global is the first level of the trigger to examine  
correlations across all detector systems.
The L2Global worker is responsible for making trigger decisions 
based on the objects identified by the L2 preprocessors.  Trigger 
decisions are made by creating global physics objects.  These 
objects can be based directly on the objects reported 
by the preprocessors or can be created by combining objects 
from different preprocessors.  The L2Global worker imposes requirements 
on the global physics objects according to configuration information 
it receives from the TCC based on 
the downloaded trigger menu.

\paragraph{Inputs to L2Global}
After an L1Accept is issued, the TFW sends a trigger decision mask 
to L2Global.  The SCL sends notification of an L1 accepted event to
every geographic sector.  In the case of L2 preprocessors, the receipt of 
the L1Accept SCL message means that the preprocessor
must send at least a header block to L2Global for this
event, and must prepare at least a header block for
eventual L3 readout in case the event passes L2.  When
L2Global receives the L1Accept, L2Global performs a decision cycle
and prepares at least a header block for L3 readout if
the event passes (or if it is an Unbiased Sample as described
below).

For each event, the L2Global worker uses the L1 trigger decision mask and
preprocessor inputs to 
decide which algorithms to run on the data from the preprocessors.  
It then makes a trigger decision and returns this 
decision to the TFW.  The trigger list specifies which trigger 
conditions L2 will impose for each run.  The trigger list can 
change as frequently as every run.  The trigger list is downloaded to the 
L2Global crate by the TCC, which receives its instructions from COOR.

A number of L1 qualifier bits may be sent along with a given L1Accept.
Of these L1 qualifiers, the ones of special interest to
L2Global are ``Unbiased Sample'' (UBS), ``Forced Write'',
and ``Collect Status.''  For a UBS event, L2Global sends the
event to L3 whether or not any of the L2 bits actually passed, marking 
the condition in the L2Global event header.  The actual L2 decision 
is also recorded.  Such events occur at a rate determined
by the trigger programming, for an independently 
adjustable fraction of events passing each L1 bit.  A
secondary effect of the UBS qualifier in
L2Global is that additional information is written to
L3 to assist in debugging the event.  In particular,
the output from L2Global is expanded to
include more information to allow detailed checking of
the processing.  Events marked with
the UBS qualifier are routed to a special data stream for system 
monitoring.

The Forced Write qualifier provides a mechanism to test
new trigger definitions.  Unlike the UBS qualifier, 
Forced Write produces detailed output at every firing of
an L1 trigger bit that is marked with this qualifier.  The Forced Write
qualifier has the same effect on L2 as the UBS qualifier, however
Forced Write events are routed to the standard output stream
for offline analysis.

After receipt of an event with a Collect Status 
qualifier, L2Global (and all preprocessors) capture internal
scaler and other monitoring information (including distributions of
processing times and pipeline occupancies) for readout by the TCC.
The monitoring blocks are tagged with the L1 crossing number 
of the event with the Collect Status qualifier, so the 
TCC can assemble a consistent set of statistics in the L1 and L2
systems.  Collect Status qualifiers are generated approximately 
once every five seconds in a typical data run.

\paragraph{Trigger scripts}
The L2Global worker begins the processing of every event by checking 
which of the L1 trigger bits fired.  The trigger list specifies 
which L2 script is associated with each L1 bit.  
The L2 script is the trigger condition that must be satisfied 
in order for the L2 trigger to fire for a given L1 bit.  
The L2 script is 
specified by a number of filters and a minimum number of objects 
required to pass each filter.  
An example of a script is an electromagnetic object filter 
and a minimum of two objects.  This script is satisfied if there are two 
electromagnetic objects in the event that satisfied the 
conditions of the filter.  If any script is satisfied, the 
event passes L2 and is sent to L3.

\paragraph{Tools and filters}
Tools and filters provide the main functions of the L2Global worker.  
The filters make up the scripts described above and  
in turn rely on tools.  Tools are C++ classes that build a 
specific type of L2Global object.  
L2Global objects are based on preprocessor objects.  
A tool starts with a list of preprocessor objects and applies 
selection criteria to decide which preprocessor objects 
should be made into global objects.  A tool can also 
correlate information from two preprocessors by combining 
two preprocessor objects to form one global object.  An example 
of this is the combination of a track object and an EM object 
that come from different preprocessors, but refer to the same  
electron candidate.  The tools are written 
to be flexible and are configurable through the trigger list.  
For the above example, one trigger list parameter specifies 
whether or not EM objects from the preprocessor should be matched to tracks 
from the tracking system.  

The tools produce lists of global objects.  
The filters then use these lists of global objects to make a trigger 
decision by imposing trigger requirements on the objects.  
For the case of an EM object, the electromagnetic fraction, 
transverse momentum, and isolation can be required to have values above or 
below specified thresholds to satisfy the filter.  
The filter generates its own list of objects that 
satisfy the trigger criteria.  Tools and filters can also operate on
global objects from earlier filters to construct additional global objects of 
greater complexity.  At the conclusion of processing, the script 
checks to see if there are at least the minimum number of objects 
required to satisfy the script requirements.

\subsection{The Level 3 trigger}
\label{sec:l3trigger}

The L3 trigger provides additional rejection both to enrich 
the physics samples and to maintain an acceptable throughput which can be 
recorded to tape. A high level, fully programmable software trigger, L3 
performs a limited reconstruction of events, reducing a nominal 1~kHz input 
rate to 50~Hz for data recorded for offline analysis. Its decisions are 
based on complete physics objects as well as on the relationships between such 
objects (such as the rapidity or azimuthal angle separating physics objects or 
their invariant mass).  Candidate physics objects, or relations between them, 
are generated by object-specific software algorithms (filter tools). 
Tools perform the bulk of the work: unpacking raw data, locating hits, forming 
clusters, applying calibration, and reconstructing electrons, muons, taus, 
jets, vertices, and \met.  Reference sets (refsets) 
of programmable algorithm parameters are input to the tools via the 
programmable trigger list. The refsets define the physics objects 
precisely (jet refsets specify cone size, for example, and electron refsets, 
the electromagnetic fraction, among other characteristics) for each invocation 
of the filter tool. All tools cache their results to expedite possible multiple 
calls within the same event, and if the event is accepted, add L3 object
parameters to the data block. 

Individual calls to the tools are made by filters that define the specific 
selection criteria employed by a tool or imposed on its results.  These criteria
include the refset used by the tool, as well as thresholds and other 
cuts applied by the filter on the results of a tool (for example, the 
requirement of two jets within a given pseudorapidity range above a fixed $E_T$ 
threshold).  Filter results are keyed for access by other filters, so in 
addition a key to the results of a previous filter can be included in these 
parameter sets.  Part of the trigger programming, this information can be 
changed with each trigger list download. 

The trigger list programming includes blocks of filter scripts that specify
one or more filters and that define the L3 trigger condition for each L3 
trigger or filter bit.  Each L3 filter script is associated with a L2 bit; 
multiple L3 scripts may be associated with each L2 bit.  Failure to 
pass an individual filter terminates execution of the script, calling no 
further tools, and skipping to the script for the next filter bit. Only when 
all filters in a script are satisfied, is the trigger satisfied and the event 
sent to the host cluster to be recorded.

\subsubsection{ScriptRunner} 

Each filter tool receives event data under the direction of ScriptRunner, 
the interface of the L3 framework to the tools. 
At the beginning of a run, ScriptRunner 
initiates parsing of the tool refsets and the filter scripts. L3-relevant 
pieces of the trigger list (and event data) are its input.  Needed tools 
initialize themselves with the necessary calibration constants 
and refsets.  ScriptRunner then processes any errors and parses the scripts to
build the execution tree.  The output of ScriptRunner is filtered event data 
dispatched to the data logger, monitoring information (individual and combined 
filter rates and timing of both filter scripts and individual tools collected 
in the event loop), and error messages.  At the end of a run (and upon request) 
ScriptRunner extracts and reports monitor information and rates. 

In the order specified by the trigger list, the execution path proceeds  
along the branches of each L2 trigger bit that has been set.  Each filter 
branch is traversed, with tools called in the given filter script order until 
an event fails a filter or passes all filters; execution then returns to the 
filter bit level and proceeds to a sister filter bit 
branch, if one exists, or continues to the next trigger bit. Control is 
returned to ScriptRunner as soon as an event fails a filter or after it passes 
all filters.  

\subsubsection{Available physics object tools}

\begin{figure}
\centerline{\includegraphics[width=3.in]{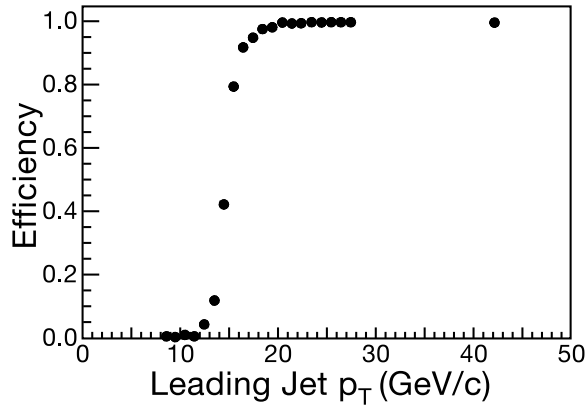}}
\caption{The L3 jet trigger turn-on.  Efficiency is plotted as a function 
of offline leading jet $p_T$.  A selection cut of 15~GeV/$c$ has been 
applied by the L3 filter.  Statistical errors are included, although small 
enough to be obscured by the symbol in most instances. }  
\label{fig:L3Jet}
\end{figure}

\paragraph{Level 3 jets and electrons}
The L3 jet tool relies on the high-precision calorimeter readout and primary 
vertex position available at L3 and the improved energy and position 
resolution this information makes possible.  
Implementing a simple cone algorithm and performing a suppression of hot 
calorimeter cells, the L3 jet tool is able to sharpen the turn-on curve 
dramatically.  Applying offline cleanup cuts on collected data shows sharp 
turn-ons to nearly 100\% efficiency for both jet and electron filters (see 
Figure~\ref{fig:L3Jet}). Rejection factors of 20--50 have been realized for 
the various jet triggers.

A simple $\sqrt{(\Delta\eta)^2 + (\Delta\phi)^2} = 0.25$ jet cone, with
requirements on $E_T$, electromagnetic fraction 
($>0.9$), and transverse shower shape, forms the basic electron tool.
An electron candidate match to a preshower detector signal can also be 
required.

\paragraph{Level 3 muons}
Wire and scintillator hits are used in the reconstruction of muon 
track segments inside and outside the toroid. 
Track-finding links these segments to identify muon tracks in three dimensions. 
Refsets determine the minimum number of hits in each detector layer required for
valid tracks.  In addition to unpacking data and determining tracks in the 
regions defined by L2, the L3 muon tool can call subordinate tools that 
utilize inner tracking and calorimetry. 

L3 improves on the ability of L2 to separate prompt and out-of-time hits 
by fitting the available scintillator hits along a track to the velocity of the
particle.  Remnant cosmic ray muons are recognized by the 
out-of-time signals in addition to their association with a 
penetrating track opposite their candidate track. 

The availability of vertex constraints and the ability to match central tracks 
makes possible improved momentum resolution in L3 compared with L2.
A track match tool extrapolates muon tracks to the central tracker, 
defining a region to be searched.  If more than 
one central track is available, the track that minimizes the $\chi^2$ 
calculated using the angular distances in $\theta$ between the central   
and extrapolated muon tracks is selected.  Additionally, matching local muon 
tracks to paths of minimum ionizing particles in the calorimeter distinguishes 
isolated from non-isolated muons. 

\paragraph{Missing transverse energy}
The L3 \met\ tool is based on calorimeter cells. Using corrected
calorimeter energy (assuming a nominal $(x,y,z) = (0,0,0)$ vertex), 
the \met\ is 
calculated through intermediate pseudorapidity ring sums (allowing quick 
recalculation by geometrically corrected ring sums when the primary vertex tool 
returns a vertex position).  The tool calculates the \met, the azimuthal angle 
of the \met, total scalar $E_T$, and a \met\  significance.

The filters that make up the L3 trigger are flexible and provide additional
rejection.  For
example, adding a \met\ filter to loose electron or muon filters, in parallel 
to single lepton scripts, provides redundancy for triggering on $W$ boson 
events; adding \met\ to lepton + jets top quark filters provides additional 
rejection factors (a factor of three for electron and eight for muon triggers). 

\paragraph{Level 3 tracking}
CFT track-finding can be performed either in specified regions or across the 
entire detector.  At initialization, the calculated $x,y,\phi$ fiber positions 
are stored in lookup tables.  Adjacent hits are merged into clusters with their 
$x,y,\phi$ position averaged.  Tracking is first done using a fast circle fit
through the axial layers, identifying candidates as arcs through the origin and
defining an arc length $S$ in the $x-y$ plane.
A straight line fit in the $S-z$ plane then determines the track helix 
parameters.  

A link-and-tree algorithm joins clustered hits from different layers.  
Starting from a link in the outer layer (and continuing recursively), 
candidate tracks are built by adding links from adjacent layers, extending the 
path length as long as the curvature is consistent with the preceding link.  
The longest extended path found starting from the initial link is kept as a 
track candidate.  The radius defined by the circle through clusters (and the 
origin) must exceed that corresponding to minimum trigger $p_T$.

SMT tracking modifies this method.  Segments connect neighboring hits between 
points within a specified $\Delta\phi$.  Segment paths are linked when their 
$z$-slope and $\Delta\phi$ are within (tunable) specifications. Unless 
seeded by a location set by the candidates from an earlier tool, the algorithm 
begins with the outermost SMT layer, looking for the longest paths toward the 
center, fitting these to a helix.  The path with the smallest $\chi^2$ is 
selected.

The CFT-track-based $z$-vertex tool offers substantially higher 
efficiency and purity than an SMT-hit-based algorithm.    
Its use sharpens the turn-on curve for jet triggers, and 
provides rejection against events with a primary vertex at large $|z|$.
Monte Carlo studies suggest that the tool achieves 0.5~mm $z$-vertex
position resolution.  The CFT-track-based vertex is used for jet 
identification and the \met\ calculation.

A global (SMT plus CFT) high-momentum-track finder starts from axial CFT tracks 
(with matched stereo clusters) propagated into the SMT.  If the CFT 
axial/stereo match fails, the CFT-SMT match is done in $x,y$ only.  
A stand-alone global track filter is a useful addition 
to single muon and electron triggers.  Starting from axial CFT tracks (with 
matched stereo clusters) propagated into the SMT, 
the L3 tracker makes an independent selection of individual charged high-$p_T$ 
tracks.  This algorithm (with 60 $\mu$m distance-of-closest-approach (DCA)
resolution for central tracks) runs in less than 200 ms.
Studies of off-line track-matched electrons in a $Z\rightarrow e^+e^-$ 
data sample show a 60\% overall efficiency (95\% within 
the CFT-axial acceptance).

The use of tracking information in the L3 filter reflects a strategy of 
parallel single-electron filters which increase efficiency at low $E_T$ and 
still provide redundancy at high $E_T$ (for cross checks).  These parallel 
triggers allow high-$E_T$ 
filters with loose cuts (where rejection is not critical) by introducing 
tighter cuts at lower $E_T$.  Similarly, L3 global track filters (run on 
muon triggers) are complemented by a suite of L3 muon filters running the 
stand-alone local muon filter for high-$p_T$ single muons.  

Online monitoring keeps track of current beam spot information (the 
mean position and spread in $x,y$ along with tilts in $x,z$ and $y,z$).  Using
this information, L3 can calculate a fully 3-dimensional primary vertex for 
each event.  
By recalculating L3 track parameters using the 3-d vertex, L3 is capable of 
triggering on the impact parameter of tracks.  The DCA resolution with respect 
to the primary vertex is 25~$\mu$m.  With input provided by a tracking, jet and 
vertex tool, event, jet and track $b$-quark-probabilities based on the signed 
impact parameters of tracks can be calculated, and $b$-tagging implemented in 
L3.

\paragraph{Relational filters}
Additional higher level selections, based on the relationship between physics 
tool candidates, are implemented at the filter level.  Examples include the 
invariant mass filter, an acoplanarity tool that selects events 
with the two leading-$p_T$ jet candidates separated by a polar angle between 
$\phi_{min}$ and $\phi_{max}$, and an $H_T$ tool that applies a 
selection cut to the scalar sum of the $E_T$s of all jet filter candidates.

\subsubsection{Special filter tools}
\label{sec:l3specialtools} 

For monitoring purposes, events can be written out regardless of the L3 trigger 
decision. Special ``mark and pass'' runs record every L2-accepted event and 
mark them with the results of all filter 
tools of every script under the L1/L2 bit(s) satisfied.  Additionally, a 
``pass one of $n$'' option for each script independently specifies a fraction 
of events to be forced through marked but unfiltered. Every event selected in 
this way passes into a special ``monitor'' stream, distinct from the physics 
stream. Monitor stream data are collected continuously during regular 
data collection.  

When necessary to prescale events, L3 can mark a filter bit as failed 
without running the filter script.  L3Prescale 
uses selection by random number generation (seeded uniquely by node) so that 
collectively, successive short runs still see events in the correct prescale 
fraction. This implementation by random numbers initialized differently in 
each node is the same means by which ``pass 1 of $n$'' events are generated. 

\subsubsection{Online monitoring}

Online monitoring of the L3 system is done in the control room during 
regular data collection. Quick analysis of the event record from randomly 
sampled online events produces distributions of 
physics quantities (for failed as well as passed candidate events).  Plots of 
electron, jet, muon, tau, and global track multiplicity, $E_T$, $\phi$, and 
$\eta$ are continuously reviewed.  In addition, a comparator 
package looks for discrepancies between online quantities and those computed 
when the data are run through the trigger simulator offline.  

\FloatBarrier

\section{Data acquisition system}
\label{sec:daq}

The data acquisition system (L3DAQ) transports detector
component data from the VME readout crates to the processing nodes of the 
L3 trigger filtering farm. The online host receives event data from the L3 
farm nodes for distribution to logging and monitoring tasks.  Overall
coordination and control of triggering and data acquisition is handled by the
COOR program running on the online host system.

\subsection{L3DAQ}

\begin{figure}
\centerline{\includegraphics[width=3.in]{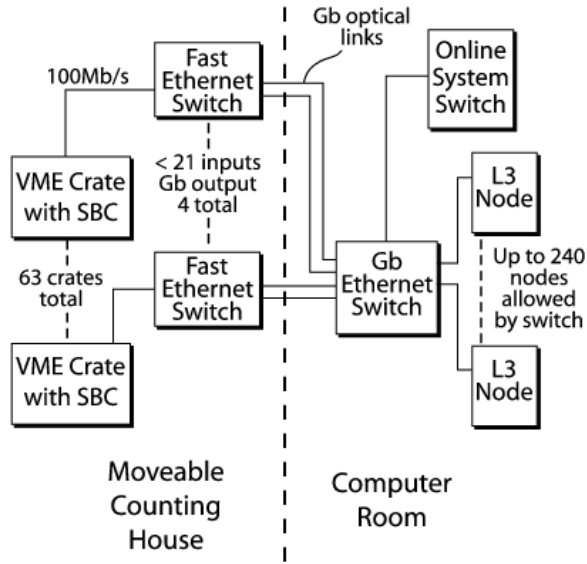}}
\caption{The physical network configuration of the L3DAQ system.  The moveable
counting house holds electronics for the detector and moves with the detector
between the assembly hall and the collision hall.}
\label{fig:l3daq-network}
\end{figure}

\begin{figure}
\centerline{\includegraphics[width=3.in]{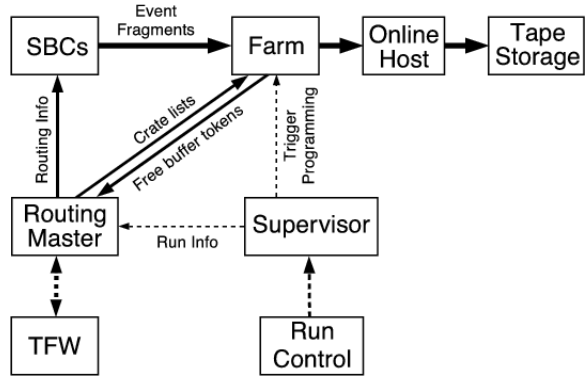}}
\caption{Schematic illustration of the information and 
data flow through the L3DAQ system.}
\label{fig:dataflow}
\end{figure}

The L3DAQ system's designed bandwidth is 250~MB/s,
corresponding to an average event size of about 200~kB at an L2
trigger accept rate of 1~kHz.  
As shown in Figure~\ref{fig:l3daq-network}, the system is built around 
a single Cisco 6509~\cite{CISCO} ethernet switch.
A schematic diagram of the communication and data flow in the system is shown
in Figure~\ref{fig:dataflow}. All nodes in the system are based on 
commodity computers (SBCs) and run the Linux operating system.
TCP/IP sockets implemented via the ACE~\cite{ACE} C++ network and
utility library are used for all communication and data transfers.

Up to sixty-three VME crates are 
read out for each event, each containing 1--20~kB of data distributed among VME
modules. An SBC (single board computer) in each VME crate reads out the VME
modules and sends the data to one or more farm nodes
specified by routing instructions received from the routing master (RM)
process. An event builder (EVB) process on each farm node builds a complete
event from the event fragments and makes it available to 
L3 trigger filter processes.

The supervisor process provides the interface between the 
main D\O\ run control program (COOR) and the L3DAQ system. 
When a new run is configured, the supervisor passes run and general
trigger information to the RM and passes the COOR-provided L3 filter 
configuration to the EVB process on relevant farm nodes, 
where it is cached and passed on to the L3 filter processes.

The SBCs are single-board computers with
dual 100~Mbit/s Ethernet interfaces and a VME-to-PCI interface.
An expansion slot is occupied by a digital-I/O (DIO) module, 
used to coordinate the readout of VME modules over the VME user (J3) backplane.
A custom kernel driver on the SBC handles
interrupt requests from the DIO module that are triggered by
readout requests from the crate-specific electronics.
On each readout request, the kernel module 
performs the VME data transfers and stores the event
fragment in one 
of several buffers in kernel memory.

A user-level process on the SBC receives route information from the RM 
in the form of route tags that contain a
unique event identifier (L3 transfer number) and the indices of
the farm nodes to which that event should be sent. 
If the L3 transfer number of the route tag matches that of the transfer number 
embedded within the head event fragment in the kernel buffers, the event
fragment is sent to the specified farm nodes. 

The EVB process on each farm node collates the event fragments
received from SBCs into complete events, keyed by L3 transfer
number. For each event, the EVB receives an expected-crate list
from the RM in order to determine when an event is complete.
Complete events are placed in shared memory buffers for
processing by the L3 filtering processes (Section~\ref{sec:l3trigger}).
The EVB process routinely informs the RM of the number
of free memory buffers that it has available.

The RM program executes on an SBC in a special VME crate
containing a hardware interface to the TFW.
The TFW provides trigger information and the L3 transfer number upon each L2
accept decision and allows the RM to asynchronously disable the firing
of L1 triggers.  For each event, the RM program chooses a node for processing 
based on the run configuration, 
the trigger information, and the number of available buffers 
in the set of nodes configured to process the type of event.  
A node is chosen in a round-robin fashion from among the set of nodes
with the most free buffers.
If the number of available free buffers is too few, the RM instructs the
TFW to disable triggers so that the farm nodes have time to catch up.

\subsection{COOR and the online host}
\label{sec:coor}

Run control and detector configuration are handled by the central
coordination program COOR.  COOR receives requests from users
to configure the detector or to start and stop runs and
sends the necessary commands to the rest of the system
to carry out those requests.  COOR sends commands to
L1, L2, and L3 and to the manager processes for the SDAQ (see below) 
and data logging
subsystems.  COOR can also configure EPICS (Section~\ref{sec:controls}) 
devices via a connection to the COMICS (Section~\ref{sec:controls-config}) 
program.  It also maintains a database of
name/value pairs accessible to the online system for recording the current 
network addresses of various online services and for access by 
trigger configurations; this allows for communicating time-dependent
values (such as the current beam spot position) to the trigger systems.

The online host system receives event data from the L3 farm nodes at a 
combined rate of approximately 10~MB/s (50~Hz L3 accept rate of 200~kB 
events) and distributes that data to logging and monitoring tasks. 
The host system is capable of supporting multiple simultaneous 
data runs.
Events that pass the L3 filters are tagged with a data stream 
identification that is a function of the satisfied hardware and software 
trigger components.  Different streams are recorded independently; 
events are assigned to only one stream, excepting some events which are
additionally assigned to a special monitoring stream.  

The final repository for the raw event data is tape, maintained in a robotic 
tape system located about 3~km from the detector.  Data must be transmitted 
to each tape drive at approximately 10~MB/s to keep the drive operating in 
streaming mode, since the remote tape drive nodes have no intermediate disk 
buffer.  The online system is capable of simultaneous output to multiple tape 
streams and of buffering in case of tape robot unavailability.  
In addition to logging data, the online host system must supply between ten 
and  twenty data monitoring clients at anywhere from 1\% to 100\% of the full 
data rate.  

\begin{figure}
\centerline{\includegraphics[width=3.in]{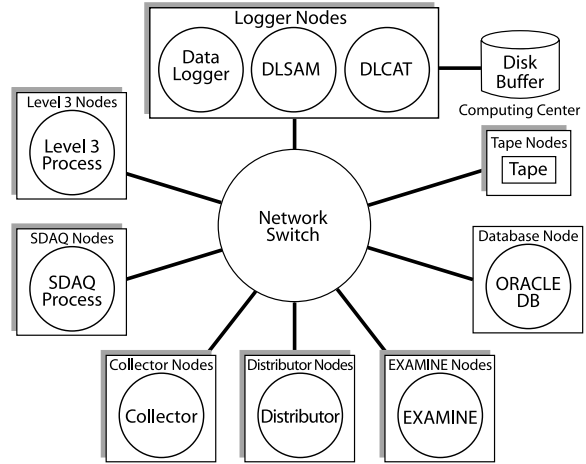}}
\caption{Schematic of the physical architecture of the online host system.
We are using commodity computers running Linux connected to a Cisco 6509 
switch.}
\label{fig:host-hardware}
\end{figure}

\begin{figure}
\centerline{\includegraphics[width=3.in]{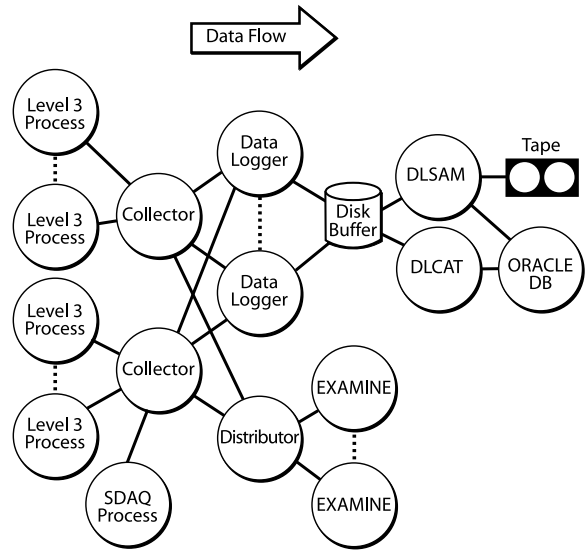}}
\caption{Schematic of the software architecture of the online host system.}
\label{fig:host-software}
\end{figure}

Figures~\ref{fig:host-hardware} and \ref{fig:host-software} illustrate the 
physical and software architecture of the online host system.
Event data arrive from the L3 trigger nodes at collector processes.
The collector directs each event to the data logger appropriate for the 
stream identifier determined for the event.  
The collector also sends, on a 
best-effort basis (there is no flow control backpressure to the L3 
nodes), a copy of each event to a distributor process,
which is an event queueing system that provides event data in near real-time
to online analysis and monitoring clients (EXAMINE programs).

A diagnostic secondary data path, SDAQ, makes possible the 
processor-based readout of digitizing electronics.  Information 
from SDAQ is also routed via the collector processes, allowing all 
downstream DAQ components to be shared.  The SDAQ mechanism bypasses the 
L3DAQ and L3 trigger and is valuable in the commissioning 
and calibration of detector components.  The detector-specific components of 
an SDAQ application have access to a library of
SDAQ functions that handle queuing of data messages between components, 
interrupt management with callbacks, run synchronization, and priority-based 
scheduling.  Several of the subdetectors use the SDAQ system: 
the SMT for calibration and to monitor the performance of individual silicon 
detector channels during a run, 
and the CFT, CPS, FPS, and FPD to calibrate scintillator response. 

The data logger writes data to files, grouped according to stream 
classification tags.  Each data logger is responsible for a set of streams. 
The data logger also generates metadata information in file format for 
storage in a database to enhance the offline access of data.  The DLSAM 
processes are the interfaces to the SAM/ENSTORE mass storage facility 
(Section~\ref{sec:SAM} and the associated database descriptions).  DLSAM 
monitors the local data buffer disks and requests that files be stored in 
the mass storage system (ENSTORE).  This request is made through 
the database interface (SAM) which negotiates the request for file storage 
with the ENSTORE system. 

All of the online computing systems are connected to 
a single high-capacity network switch. The architecture 
provides for parallel operation of multiple instances of the 
bandwidth-critical components. 

The high-level online software applications are predominately 
constructed using the Python scripting language.  
Network communication between the components is implemented with
the InterTask Communication (ITC) package, a multi-platform, 
multi-threaded client/server messaging system developed at Fermilab and 
based on the ACE~\cite{ACE} network and utility libary. 

\FloatBarrier

\section{Controls and monitoring}
\label{sec:controls}

\begin{figure}
\centerline{\includegraphics[width=3.in]{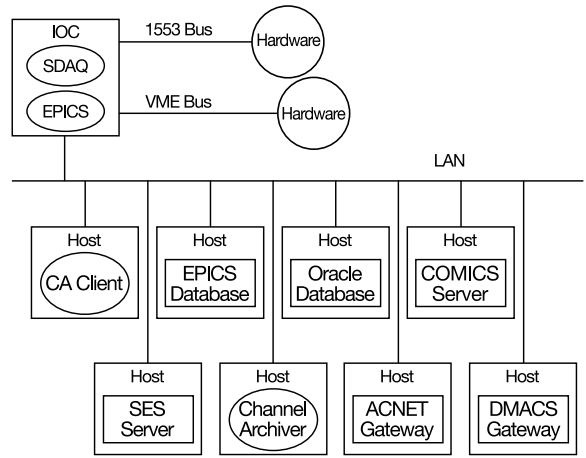}}
\caption{Organization of the control system components.  The Hosts are 
computers
running Linux or Tru64 Unix and the IOCs are embedded computers in VME crates. 
Only one IOC is shown, and only a few of the online hosts.}
\label{fig:controls}
\end{figure}

The D\O\ experiment has extended EPICS (Experimental Physics and 
Industrial Control System) \cite{epics} to meet the control and monitoring 
requirements of a large high energy physics detector.  EPICS, an integrated 
set of software building blocks for implementing a distributed control system, 
has been adapted to satisfy the slow controls needs of the detector by {\it i}) 
extending the support for new device types and an additional field bus, 
{\it ii}) the addition of a global event reporting system that augments the 
existing EPICS alarm support, and {\it iii}) the addition of a centralized 
database with supporting tools for defining the configuration of 
the control system.  Figure~\ref{fig:controls} shows the architecture and 
components of the D\O\ controls and monitoring system.  

EPICS uses a distributed client-server architecture consisting of 
host-level nodes that run application programs (clients) and input/output 
controller (IOC) nodes (servers) that interface directly with the detector 
hardware.  The two classes of nodes are connected by a local area network.  
Clients access process variable (PV) objects on the servers using the 
EPICS channel access protocol.

One of the unique properties of the D\O\ detector interface is the use of the 
MIL-STD-1553B \cite{mil-1553} serial bus for control and monitoring operations 
of the 
electronics components located in the collision hall.  Since the detector 
is inaccessible for extended periods of time, a robust, high-reliability 
communication field bus is essential.  EPICS was extended by providing a 
queuing driver for MIL-STD-1553B controllers and a set of device support 
routines that provide the adaptive interface between the driver and the 
standard EPICS PV support records.  With these elements in place, all of 
the features of EPICS are available for use with D\O's remote devices.

\subsection{Global event reporting}

\begin{figure}
\centerline{\includegraphics[width=3.in]{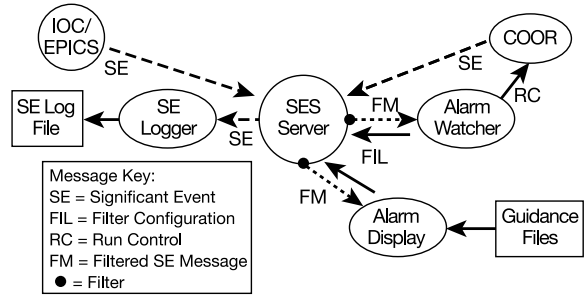}}
\caption{The major components of the SES are the server, logger, watcher, and
alarm display.  COOR is a coordinating process that provides run control to the
experiment.  Significant events originate in the detector hardware, run
control system, or online applications.}
\label{fig:ses}
\end{figure}

To process significant events from all experiment sources, a separate facility, 
the significant event system (SES), collects and distributes all 
changes of state.  The SES has a central server that collects event messages  
from sender clients and filters them, via a Boolean expression, for routing to 
receiving clients.  Sender clients, including the IOCs, connect to the server 
and all state changes on those clients, including alarm transitions, are sent 
to the server.  The architecture of the SES and the flow of messages within 
the system are illustrated in Figure~\ref{fig:ses}.

The alarm class of SES messages receives special handling in the server.  The 
SES server maintains the current alarm state of the entire experiment so that 
receiving clients can obtain the current state when they first connect 
to the server.  In addition to specialized receiving clients that may connect 
to the server, there are three standard clients: the SES logger, the SES alarm 
display, and the SES alarm watcher.  The logger has a pass-all filter so that 
it receives all SES messages sent to the server and writes the messages 
received to a disk file.  The current state of the detector stored in the server 
is relayed to users through the alarm display.  For alarms that compromise 
data quality, the alarm watcher automatically pauses the current run.  In 
addition to its monitoring and logging functions, the SES system provides the 
means for distributing synchronizing messages to other components of the online 
software system.  Tools have been developed for mining data from the SES log 
files.  Hardware experts review the log files to understand which hardware 
devices are unstable and collaborators performing data analysis can be sure 
that the event they are examining is real and not caused by a fault in the 
detector.

\subsection{Centralized device database}

The EPICS databases that configure the individual IOCs are flat ASCII files 
containing record definitions, the database equivalent of a PV, that 
are read by the 
IOCs during startup.  The EPICS system additionally provides a higher-level 
construct, called a template, which is a parameterized collection of record 
definitions.  Generator files, which reference the templates, supply the 
parameter values to produce instances of these templated devices.  While these 
collections of files are adequate for EPICS initialization, they are not easily 
accessible to host-level processes, which require the same information.

To address this problem, we have centralized the relevant information
in a relational database (Oracle \cite{oracle}) and provided a family of 
scripts to manage the transformation between the relational database and the 
EPICS ASCII-format files.  At the time this document was prepared, the 
database contained approximately 6200 templated devices, corresponding to about 
137000 process variables, and that number is constantly expanding.

In addition to the database management scripts, a WWW browser interface to the 
relational database is available for the initial definition, modification, and 
viewing of the relational database entries.  With control system device 
specifications centralized in the relational database, they are easily 
accessible to other host-level processes.  

\subsection{Detector configuration management}
\label{sec:controls-config}

One of the most complex tasks performed by the control system is the 
configuration of the detector for specific run conditions.  The set of distinct 
configurations, both for physics data collection and for calibration runs, is 
very large; the usual technique of uploading a specific detector 
configuration and saving it as a file for subsequent downloading is impractical.

For ease of configuration management, the detector is represented as a tree 
with nodes at successively deeper levels corresponding to smaller, more 
specialized units of the detector.  The terminal nodes of the 
tree, called action nodes, each manage the configuration of a specific,
high-level device.  They are instances of the high-level devices discussed 
in the preceding database section.  The intermediate nodes of the tree 
primarily serve to 
organize the traversal order of the subordinate nodes since the detector is, 
in general, sensitive to the order in which devices are initialized.  

A single server program, COMICS, manages the configuration of the 
EPICS-accessible part of the detector.  The tree nodes, both intermediate and 
action, are all specialized instances of a base node class that defines the 
majority of the methods that characterize node behavior.  The detector tree 
structure is defined by a set of configuration files that are Python program 
segments that instantiate instances of these nodes.

\subsection{Operator interfaces and applications}

An application framework, in the Python scripting language, 
assists in developing operator interfaces and provides a consistent ``look 
and feel'' for all visual displays.  This framework includes a collection of 
specialized, graphical objects that construct updating displays of PV 
values using a Python interface to the EPICS channel access application 
program interface (APS).  The experiment uses more than forty instances of 
these monitoring displays in the control room to manage the detector 
components.

\subsection{Archiving EPICS data}

While using EPICS records for control and monitoring tasks, 
almost every detector group needs to maintain structured access 
to archived PV values.  There are two major archiving tools employed by D\O: 
{\it i}) the channel archiver \cite{epics-archive}, for needs that require 
sampling rates of 1~Hz or faster, but do not require frequent access 
to historical data; and {\it ii}) the EPICS/Oracle Archiver, for long-term 
studies that require slower sampling rates (once per minute or less 
frequently), easy access to data at any moment, and minimal maintenance.

Many channel archivers are running concurrently, monitoring several thousand 
PVs.  About once a week, collected archives are sent to the central Fermilab 
robotic tape storage via the SAM data management system 
(Section~\ref{sec:SAM}).  The channel archiver toolset has 
interfaces, including web-based tools, that enable retrieval from an 
archive of data in different formats and generation of time plots with 
various options.

\subsection{ACNET gateway}

In the operation of the detector, it is vital to have a fast and reliable 
messaging connection between D\O\ and accelerator operations to exchange 
control and monitoring information.  The D\O\ control system supplies cryogenic 
and magnet data, the luminosity determined using the luminosity monitor
(Section~\ref{sec:lum-monitor}), as well as FPD pot positions and counter 
rates.  The accelerator control system (ACNET), in turn, sends information 
about critical accelerator devices.  A gateway between the D\O\ and ACNET 
control systems, based on the XML-RPC protocol \cite{xml-rpc}, provides this 
interconnection.

\FloatBarrier

\section{Computing and software}
\label{sec:software}

A large amount of software has been developed for data acquisition, 
monitoring and controlling hardware, Monte Carlo event simulation, and  
data and Monte Carlo event reconstruction.  Early in the development
of software for Run II, we made the decision that all new software would be
written using the C++ programming language.  Legacy Run I Fortran software and
programs from other sources (e.g.  Monte Carlo event simulation programs) are
wrapped in C++ code.  In this section, we give an overview of the computing and
software in use during Run II.

\subsection{Event data model}
\label{sec:edm}

The D\O\ event data model (EDM) is a library of C++ classes and
templates whose purpose is to support the implementation of
reconstruction and analysis software. 

The central feature of the EDM is the event, a class that represents
the results of a single beam crossing. The event acts as a container
to manage all of the data associated with a single crossing: the raw
output of the detector, the results of trigger processing, and
the results of many different reconstruction tasks.
Each of the items in this collection contains both the
data describing the crossing in question as well as metadata that
describe the configuration of the program that constructed these
results. This allows us to run multiple instances of single algorithms
with different configuration information (for example, several
cone-based jet algorithms with different cone radii), and to
distinguish between the output of these different algorithms. The EDM
also provides a mechanism for access to the collected reconstruction
results, relieving the users from the burden of understanding the
somewhat complex internal organization required for the management of
the event data and the corresponding metadata.

\subsection{Data persistency}
\label{sec:d0om}

\begin{figure}
\centerline{\includegraphics[width=5.in]{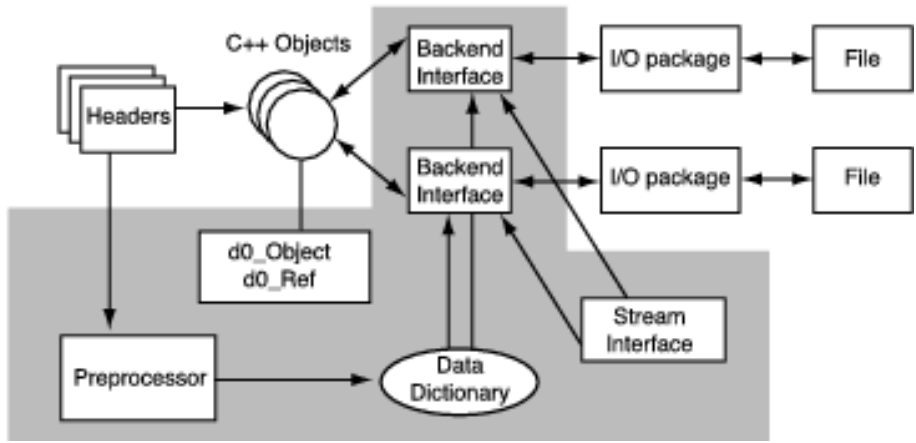}}
\caption{The structure of D\O OM.  The shaded area shows the
components which are part of D\O OM proper.}
\label{fig:d0om}
\end{figure}

The conversion of the C++ objects used in the reconstruction
program to a persistent format is handled by the
D\O\ object model (D\O OM) \cite{d0om-chep}.  This has several parts
(Figure~\ref{fig:d0om}).  First, D\O OM maintains a dictionary
describing the layout of the C++ classes that are to be used
persistently, which is generated by running
a preprocessor over the C++ headers defining the classes.
This preprocessor is based on a modified version of the
CINT C/C++ interpreter (which is also used in the
ROOT system \cite{root}).

The actual translation between C++ objects and the persistent
format is handled by one of several I/O packages.  User code
has no dependence on the I/O packages, so that new formats
can be added without changing any reconstruction code.
The I/O packages are typically built on top of external software
packages to do the actual I/O.

Finally, the external interface to the package is provided
by a set of stream classes.  A set of D\O\ framework
packages that use these classes to read and write events
within the framework is also provided; in most cases,
a framework user need only to include these packages
to read and write events.

D\O OM includes numerous features to assist with schema evolution
and versioning of data.  The dictionary information is maintained
along with the saved data, so the layout of the saved data
is always known.  During reading, class members are matched
between the C++ and persistent forms based on name.
This allows D\O OM to handle the common cases of adding and
deleting data members with no explicit action on the part of the programmer.
For more complicated cases, it is possible to provide
conversion code that is automatically run when needed.
The D\O OM dictionary information may also be queried at run
time.  The system can also make use of the persistent dictionary
information to build objects at run time for which no dictionary
information was compiled into the program.  This is useful
for programs to dump or browse arbitrary D\O OM data files.

One I/O package is based on the DSPACK library, which originated with 
the NA49 experiment \cite{dspack}.  
DSPACK handles the conversion from C-like structures
to a serial data format; the D\O OM I/O package converts from
the C++ objects to the DSPACK structures.  At D\O, DSPACK
data are usually encapsulated inside another, lightweight, format
called EVPACK, which provides data compression and random access
within files with keyed lookup.  EVPACK-encapsulated DSPACK
records may also be sent over the network; this is used to
distribute data within the online host system.
All event data are stored in (EVPACK-encapsulated) DSPACK format.
In addition, this format is used for several static, structured
data files used by the reconstruction program, such as the
description of the detector geometry.

\subsection{Calibration databases}
\label{sec:databases}

\begin{figure}
\centerline{\includegraphics[width=8cm]{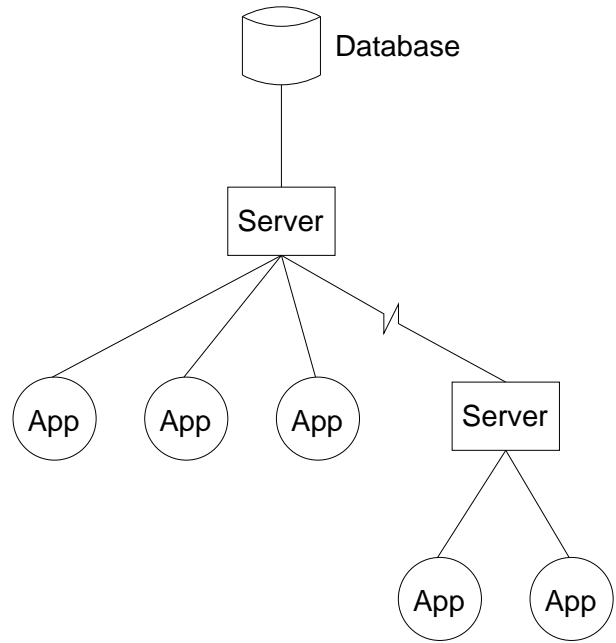}}
\caption{The hierachy of the database, middle tier server, and client
applications.  Each server can cache data, and the servers can be chained to 
provide higher-speed access for clients located at remote processing sites.}
\label{fig:dbserver}
\end{figure}

Run-dependent information needed for reconstruction, such as
magnetic field polarities and calibration constants, is stored in a
database. To decouple the reconstruction program from any
particular proprietary database implementation, D\O\ makes use of a
three-tier architecture \cite{dbserver}, where the database and
the client code make up the top and bottom tiers
(Figure~\ref{fig:dbserver}).  In between is a middle tier server which
accepts requests from the clients, makes the query to the database,
and returns the data to the client. Data can be cached by the server,
thus speeding up the response to the client and reducing the load on
the database.

This caching is especially effective for the large-scale reconstruction
of raw data. If runs that require the same calibration constants are
processed together, only one query to the database is made for
each calibration set; all other requests for this set are fulfilled
directly from the cache. 

Servers can be chained together so that processing sites outside
Fermilab may run their own local servers which request their data from
the servers at the laboratory. This allows a remote processing site to
have its own cache to speed up queries without requiring the site to
access the database directly.

The particular software used for storing the calibration data is the
Oracle relational database.  The middle tier servers are mostly
implemented in the Python programming language with some parts written
in C++ to improve performance. CORBA distributed object technology
\cite{CORBA} is used for the communication between the client and the
servers. The core part of the server is written to be independent of
the particular application it is serving data for; the customized
code which queries specific database tables and returns the results to
the client is automatically generated from the database schema for
each particular application. Each subsystem of the detector (SMT,
CFT, preshower detectors, calorimeter, and muon
detectors) has its own database schema and a separate server.

The client end of the system is an I/O package for D\O OM. The
necessary C++ class definitions are also automatically generated from
the database schema. Accessing the data is then a matter of navigating
a C++ object hierarchy: the client application need have no knowledge
of CORBA, Oracle, or the method by which it obtained the data. This
decoupling makes it relatively easy to replace the specific
implementation of the middle tier, if this becomes either necessary or
desirable, without requiring the whole of the client implementation to
change.

\subsection{Event simulation}
\label{sec:MC}

The generation of Monte Carlo (MC) events involves multiple stages and many 
executables.  To integrate all processes, all 
programs use the EDM to carry data in memory and 
D\O OM to store persistent event data.  All code is 
organized in independent packages running in a standard D\O\  
framework and is written in C++ or embedded in C++ driving routines.

The first step in MC event generation is the simulation of a 
physical process, a \ppbar\ collision producing a particular final 
state.  Nearly all existing event generator programs are written 
in Fortran, but the StdHep code from the FNAL Computing Division can be used to 
store the output in a standard common block format.  This allowed us to write 
a C++ wrapper that converts the StdHep Fortran format to C++ classes  
satisfying the EDM requirements.  

To trace the particles through the D\O\ detector, determine where their paths
intersect active areas, and simulate their energy deposition and secondary 
interactions, we use the CERN program GEANT v3.21 \cite{geant}, which is also 
written in Fortran.  
A C++ wrapper is used to read files produced by the event generators and to 
write the output of GEANT in D\O OM format.  This executable is 
called D\O GSTAR. All subsequent steps in the event simulation are 
handled by programs written almost entirely in C++.

The D\O SIM program modifies the generated Monte Carlo data to account for
various detector-related effects. 
After the particles from the simulated reaction have been traced through the 
detector, the generated energy depositions must be converted to the form that 
the real data takes when processed through the D\O\ electronics.  Detector 
inefficiencies and noise (from the detector and electronic readout) must be
taken into account, and more than one interaction may 
occur during a beam crossing.  In addition, some portions of the detector 
(such as the calorimeter) remain sensitive to interactions over a period of 
time that includes more than one beam crossing. 
Simulation of the trigger electronics and the 
effects of the trigger on data selection is performed by a separate 
program, D\O TRIGSIM.  D\O TRIGSIM contains simulation code only 
for the L1 trigger.  The L2 and L3 triggers are based on
filtering code, and exactly the same software runs in D\O TRIGSIM.
The output of D\O SIM and D\O TRIGSIM is in the same format as 
the data recorded by the D\O\ data acquisition system, but contains additional 
MC information to make it possible to correlate detector data with 
the original generator output.

\subsection{Reconstruction}
\label{sec:reco}

The D\O\ offline reconstruction program D\O RECO is responsible for 
reconstructing objects used for physics analysis.  It 
is a CPU-intensive program that processes events recorded 
during data collection and simulated MC events.  The executable is run on 
the offline production farms and the results are placed into the central data 
storage system (Section~\ref{sec:SAM}) for further
analysis.  Information and results for each event are organized using the EDM. 
The EDM manages information within the event in blocks called chunks. 
The raw data chunk (RDC), created either by an L3 processor node
or the MC, contains the raw detector signals and is the primary 
input to D\O RECO.  The output from D\O RECO consists of many additional 
chunks associated with each type of reconstructed object.  

D\O RECO reconstructs events in several hierarchical steps.  The 
first involves detector-specific processing.  Detector unpackers process 
individual detector data blocks within the RDC, decoding the raw 
information, associating electronics channels with physical detector elements, 
and applying detector-specific calibration constants.  For many of the 
detectors, this information is then used to reconstruct cluster (for example, 
from the calorimeter and preshower detectors) or hit (from the tracking 
detectors) objects.  These objects use geometry constants to associate detector 
elements (energies and positions) with physical positions in space.  The second 
step in D\O RECO focuses on the output of the tracking detectors,
reconstructing global tracks from the hits in the SMT and CFT.  
This process, involving several different tracking algorithms, 
is the most CPU-intensive activity of D\O RECO.  
The results are stored in corresponding track chunks, which are 
used as input to the third step of D\O RECO, vertexing.   First, primary 
vertex candidates are found.  These vertices indicate the locations of \pbarp\ 
interactions and are used in the calculation of various kinematical quantities 
(e.g. $E_T$).  Next, displaced secondary vertex candidates are 
identified.  Such vertices are associated with the decays of long-lived 
particles.  The results of the above algorithms are stored in vertex chunks, 
and are then available for the final step of D\O RECO --- particle 
identification. 
Using a wide variety of algorithms, information from each of the preceding 
reconstruction steps is combined and physics object candidates are 
created.  RECO first finds electron, photon, muon, neutrino (\met), and 
jet candidates, after which it identifies candidates for heavy-quark and tau 
decays.

\subsection{Data handling and storage}
\label{sec:SAM}

The sequential access via metadata \cite{SAM} (SAM) data handling system 
gives users access to all the data created by the D\O\ experiment (both 
detector data and simulation data), in a flexible and transparent manner. 
The user does not need to know where the files are physically stored, nor 
worry about exactly how they are delivered to her/his process. 
SAM oversees 
the functions of cataloging data (files and events, and associated metadata 
regarding production conditions), transferring data in and out of mass storage 
systems, transferring data among different computer systems (whether connected 
via local or wide area network), allocating and monitoring computing resources 
(batch slots, tape mounts, network bandwidth, disk cache space), and maintaining
file delivery status at the user process level.  The bookkeeping 
functions of the SAM system are provided by an Oracle \cite{oracle} database, 
which is accessed via a client-server model utilizing CORBA technology.   
Files are stored in SAM using interfaces that require appropriate metadata for 
each file.   The files are organized, according to the metadata provided, by 
data tier, and by 
production information (program version which produces the data, etc.). The 
SAM system also provides file storage, file delivery, and file caching 
protocols that permit the experiment to control and allocate the computing 
resources.  Tape resources can be guaranteed to high priority activities 
(data acquisition and farm reconstruction), high usage files can be required 
to remain in the disk cache, and different priorities and allocations for 
resource usage can be granted to groups of users.

\ack

We would like to thank Ken Ford and Scott Baxter for their work on the
illustrations and to acknowledge in-kind contributions by
the Altera and Xilinx Corporations.
We thank the staffs at Fermilab and collaborating institutions, 
and acknowledge support from the 
DOE and NSF (USA);
CEA and CNRS/IN2P3 (France);
FASI, Rosatom and RFBR (Russia);
CAPES, CNPq, FAPERJ, FAPESP and FUNDUNESP (Brazil);
DAE and DST (India);
Colciencias (Colombia);
CONACyT (Mexico);
KRF (Korea);
CONICET and UBACyT (Argentina);
FOM (The Netherlands);
PPARC (United Kingdom);
MSMT (Czech Republic);
CRC Program, CFI, NSERC and WestGrid Project (Canada);
BMBF and DFG (Germany);
SFI (Ireland);
Research Corporation,
Alexander von Humboldt Foundation,
and the Marie Curie Program.

\bibliography{run2_nim}

\begin{thebibliography}{100}
\bibitem{l2-mbus}
The Mbus was originally designed for use by the CDF experiment. D\O\ uses a
  similar physical backplane, but implements a re-engineered communications
  protocol to support effective arbitration in an environment with numerous
  master and target devices.
\end{thebibliography}

\end{document}